\documentclass[10pt]{iopart}

\expandafter\let\csname equation*\endcsname=\relax
\expandafter\let\csname endequation*\endcsname=\relax
\usepackage{color}
\usepackage{amsmath}
\usepackage{amssymb}
\usepackage{enumerate}
\usepackage{graphicx}
\usepackage{mathbbol}
\usepackage{amsfonts}
\usepackage{show labels}
\usepackage{url}

\newcommand{\Prob}{{\rm Prob}}
\newcommand{\bea}{\begin{eqnarray}}
\newcommand{\eea}{\end{eqnarray}}
\newcommand{\beq}{\begin{equation}}
\newcommand{\eeq}{\end{equation}}

\newcommand{\PDF}{{\rm pdf }}
\newcommand{\GF}{{\rm generating\ function }}

\def\XXint#1#2#3{{\setbox0=\hbox{$#1{#2#3}{\int}$}
 \vcenter{\hbox{$#2#3$}}\kern-.5\wd0}}

\begin{document}
\title[Record statistics of a strongly correlated time series]{Record statistics of a strongly correlated time series: random walks and L\'evy flights}

\author{Claude Godr\`eche}
\address{Institut de Physique Th\'eorique, Universit\'e Paris-Saclay, CEA and CNRS, 91191 Gif-sur-Yvette, France}

\author{Satya N. Majumdar}
\address{LPTMS, CNRS, Univ. Paris-Sud, Universit\'e Paris-Saclay, 91405 Orsay, France}

\author{Gr\'egory Schehr}
\address{LPTMS, CNRS, Univ. Paris-Sud, Universit\'e Paris-Saclay, 91405 Orsay, France}

\begin{abstract}
We review recent advances on the record statistics of
strongly correlated time series, whose entries denote the positions of a 
random walk or a L\'evy flight on a line. After a brief survey of the theory of records for independent and identically distributed random variables, we focus on random walks. During the last few years, it was indeed realized that random walks are a very useful ``laboratory'' to test the effects of correlations on the record statistics. We start with the simple one-dimensional random walk with symmetric jumps (both continuous and discrete) and discuss in detail the statistics of the number of records, as well as of the ages of the records, i.e., the lapses of time between two successive record breaking events. Then we review the results that were obtained for a wide variety of random walk models, including random walks with a linear drift, continuous time random walks, constrained random walks (like the random walk bridge) and the case of multiple independent random walkers. Finally, we discuss further observables related to records, like the record increments, as well as some questions raised by physical applications of record statistics, like the effects of measurement error and noise. 
\end{abstract}

\maketitle

\section{Introduction}

The statistics of extreme and rare events have recently generated a lot of interest in various areas of science. 
In particular, the study of the statistics of {\it records} in a discrete time series, initiated in the early fifties \cite{Cha1952}, has become fundamental and important in a wide variety of systems, including climate studies \cite{hoyt,basset,SZ1999,benestad, RP2006,WK2010,AB2010,WHK2013}, finance and economics \cite{records_finance,WBK2011,SL2014}, hydrology \cite{records_hydrology}, sports \cite{Gembris,sports}, in detecting heavy tails in statistical distributions \cite{FWK2012}, and others \cite{glick,gregor_review}. 


Consider any generic time 
series of $N$ entries $\{X_1, X_2, \ldots, X_N\}$ where $X_i$ may 
represent the daily temperature in a given place, the price of a stock or 
the yearly average water level in a river. A record happens at step $k$ if 
the $k$-th entry exceeds all previous entries.
Questions related to records are obviously intimately connected to extreme value
statistics \cite{gumbel,Gal87}. For instance the actual record value at step $k$ is just the maximal value of the entries after $k$ steps,
which is a key observable in extreme value statistics. On the other hand, record statistics has deep connections with
first-passage problems \cite{Redner_book,Satya_review,Bray_review}. For instance, the record rate, i.e., the probability that a record
is broken at step $k$, is related to the survival probability, i.e., the probability that the time series remains below a certain level up to step $k$, which is a key quantity in first-passage problems. 

However, despite its connections with extreme value statistics and first-passage problems, the statistics of records of a time series raises specific new questions which require new tools and techniques. In this paper, we focus on a class of observables associated to the record statistics. This includes, for instance, 
 the number of records in a given sequence of size $N$ as well as the ages of the records. The age of a record is defined as the time up to which the record survives, i.e., before it gets broken by the next record. We will also study 
the record values as well as the increments of the record values. The statistics of these observables can not be understood from extreme value statistics or first-passage problems solely and they require new techniques that will be discussed in this review.

Remarkably, the study of records have found a renewed interest and 
applications in diverse complex physical systems such as the evolution of the 
thermo-remanent magnetisation in spin-glasses \cite{jensen,sibani}, evolution of the vortex 
density with increasing magnetic field in type-II disordered 
superconductors \cite{jensen,oliveira}, avalanches of elastic lines in a disordered medium \cite{fisher,sibani_littlewood,ABBM,LDW09}, the evolution of fitness in biological populations \cite{sibani_fitness,krugjain,franke}, jamming in colloids \cite{RBSY2016}, in the study of failure events in porous materials \cite{PRLKM2016}, in models of growing networks \cite{GL2008}, and in quantum chaos \cite{SLJ2013,SL2015} amongst others. The 
common feature in all these systems is a {\em staircase} type temporal evolution of physical observables (see figure \ref{fig_record}). For instance, when a domain wall 
in a disordered ferromagnet is driven by an increasing
external magnetic field, its center of 
mass remains immobile (pinned by disorder) for a while and then, as the 
field increases further, an extended part of the wall gets depinned, 
giving rise to an avalanche and, consequently, the center of mass jumps 
over a certain distance \cite{fisher,sibani_littlewood,ABBM,LDW09}. The position of the center of 
mass as a function of time (or increasing drive), displays a 
staircase structure as in figure \ref{fig_record}. Some useful insights on such a staircase evolution in these 
various systems can be gained by studying the dynamics of records
in a time series \cite{sibani,sibani_littlewood,oliveira}, where the 
record value remains fixed for a while until it gets broken by the next 
record and jumps by a certain {\it increment} (see figure~\ref{fig_record}). For instance, 
in the case where the positions $X_i$ are the positions of a random walker after $i$ steps, this 
``record process'' is at the heart of the so called ABBM model \cite{ABBM,LDW09} which has been extensively used to model 
the so-called Barkhausen noise in disordered ferromagnets \cite{ZCDS1998}. 

%
%
The record statistics for independent and identically distributed (i.i.d.) random variables have been extensively studied in the past, both
in the mathematics \cite{Nevzorov,ABN1992,Res1987} and also more recently in the physics literature (for a recent review on the i.i.d.~case see \cite{gregor_review}). 
Many aspects of these studies are now theoretically well understood and, here, we will briefly recall the main useful results, with a special emphasis on the statistics of the ages, which is
somehow less well known. Another class of time series for which record statistics has been studied recently corresponds to independent but non identically distributed random variables. This is quite relevant in sports, where with time the average performance of a sportsman/woman typically increases with time due either to increased nutrition or technologically advanced sports equipments used for the preparation. Similarly in the context of climate studied, there can be a typical linear trend in time of the average temperature. Various interesting results have been derived for this independent but non identically distributed time series \cite{Kru2007, FWK2010,WFK2011}. As these results have already been reviewed in ref \cite{gregor_review}, we will not repeat them here. 

In many realistic time series, the entries $X_i$
are however correlated. So, the question naturally arises: what can we say
about the record statistics for {\em correlated} sequences? For a {\em weakly} correlated time series, i.e., with a finite correlation 
time, one would expect the record statistics for a large sequence to 
be asymptotically similar to the uncorrelated case. This, however, is no 
longer true when the entries are {\em strongly} correlated. It turns out that in this strongly correlated case, the study of record statistics is technically challenging. 
The difficulty of the task can be estimated by considering the aforementioned connections with extreme value statistics and first-passage problems, which are notoriously hard to solve for strongly correlated time series.
As a consequence, there exist very few results in the literature and in fact all the classical textbooks on records \cite{Nevzorov,ABN1992,Res1987} deal essentially, if not exclusively, with the case of i.i.d.~random variables.

One of the 
simplest and most natural strongly correlated time series is the random 
walk sequence on a line, where the entry $X_i$ corresponds to the position 
of a random walker at discrete time step $i$, starting from the origin 
$X_0=0$, and undergoing random jumps at each time step. Despite the 
striking importance and abundance of random walks in various areas of 
research, the record statistics of such a single, discrete-time random 
walk with a symmetric jump distribution on a line was not computed and 
understood until only a few years ago \cite{MZ2008}. Indeed, while the positions of a random
walker are strongly correlated, the random walk itself is a Markov
process. Thanks to this key Markov property, it was recently realized that the random
walk and its variants is an ideal laboratory to test analytically 
the effects of strong correlations on the record statistics of time series.

Indeed, recently, there have been much 
progress in understanding the record statistics for such a random walk 
sequence, both with and without drift and also for the case of multiple random 
walkers, and many interesting analytical results were derived -- some of 
them rather surprisingly universal. This random walk sequence is thus 
useful as it provides an exactly solvable example for the record 
statistics of strongly correlated time series. These results for random walks have been briefly reviewed
in refs \cite{gregor_review,satya_leuven,SM_review}. Since then, however, the subject has rapidly evolved and a 
detailed account of these recent progresses is still lacking. The purpose of this review is to provide
an updated survey of the known results both for i.i.d. and for strongly correlated time series, like random walks and L\'evy flights.  

The review is organised as follows. We start, in section 2, by a brief survey of the theory
of records for i.i.d.~random variables. In section 3, we develop the basic theory of record statistics for random walks, which is the cornerstone of this review. These results are based on a general renewal structure which is then exploited to obtain detailed information about the statistics of both the number and ages of the records for several models of random walks, including symmetric random walks -- with both continuous and discrete jumps --, random walks with a linear drift and continuous time random walks. In section 4, we focus on the record statistics of constrained random walks, with a special focus on the (symmetric) random walk bridge -- i.e., a random walk conditioned to start and end at the origin after $N$ steps. In section~5, we discuss the record statistics for $K$ independent random walkers and in section~6, we present several generalisations of these results, emphasizing in particular the similarities between the ages of records and the size of excursions between consecutive zeros in the {lattice random walk and Brownian motion} and more generally in renewal processes. Finally, in section 7, we present some related issues that have been recently discussed in the literature -- like the effects of measurement error and noise -- before we conclude in section~8.

\section{Record statistics for i.i.d.~random variables}

\begin{figure}[ht]
\centering
\includegraphics[width = 0.8\linewidth]{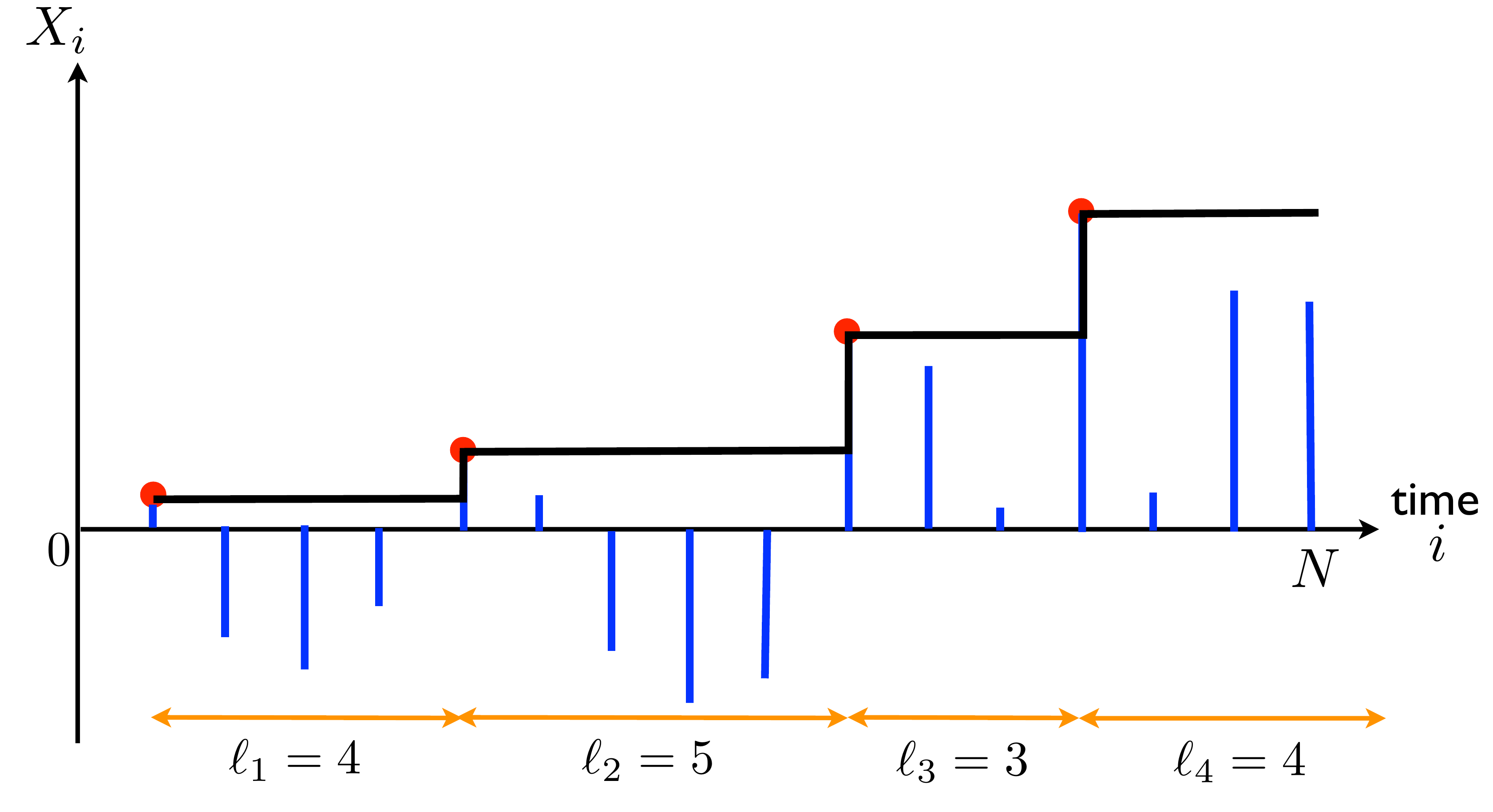}
\caption{ 
For any time series $\{X_1,\ldots, X_N \}$, the solid black line representing the current record value as a function of time
exhibits a generic staircase evolution. 
In this realization the number $N$ of random variables $X_i$ is equal to $16$ and the number of records (the red dots) is $M = 4$.
The {\it increments} in record values are the jumps in the staircase.
They occur at the {\it record times} $N_1=1$, $N_2=5$, $N_3=10$ and $N_4=13$.
The {\it ages} of the records are the lapses of time during which a record survives before it gets broken by the next one.
Thus $\ell_1=N_2-N_1=4$, $\ell_2=N_3-N_2=5$, $\ell_3=N_4-N_3=3$.
The last age $\ell_4$ (denoted by $\ell_M$ in general) has a different status.
It is equal to the difference $N-N_4$ shifted by one unit, i.e., $\ell_4=4$ in the present example.
With such a choice, the sum of the ages $\ell_1+\cdots+\ell_M$ is equal to $N$.
}
\label{fig_record}
\end{figure}

We begin by reviewing the main results for the record statistics of i.i.d.~random variables. We consider a collection of $N$ random
variables $X_1, X_2, \ldots, X_N$ which are drawn from a continuous probability density function ($\PDF$) $p(X)$. These random variables being
i.i.d., their joint $\PDF$ $P(X_1, X_2, \ldots, X_N)$ simply factorizes as 
\begin{eqnarray}\label{eq:def_iid}
P(X_1, X_2, \ldots, X_N) = \prod_{i=1}^N p(X_i) \;. 
\end{eqnarray}
By definition, $X_k$ is an upper record if and only if it is larger than all previous entries,
\begin{eqnarray}\label{def_record}
X_k > \max\, \{X_1, \dots, X_{k-1} \} \;.
\end{eqnarray} 
For instance in figure \ref{fig_record}, $X_1$ is, by definition, a record, then $X_{5}$ is a record, on so on (see the caption of the figure for details).
One can similarly define a lower record which is such that $X_k < {\rm min} \{X_1, \dots, X_{k-1} \}$. For now, we will mainly focus on upper records, which we will simply call ``records''. 
%
%
%
Let $M$ be the number of records among these $N$ random variables. We first discuss a straightforward method, based on indicator variables, to investigate the statistics of $M$. Then we discuss more complicated joint probability distributions of the number and the ages of the records. This second method is not only useful to investigate the age of the longest and shortest record but can be generalised, with some appropriate modifications, to the study of the records of random walks (see section \ref{section:RW}), constrained random walks (see section \ref{sec:bridge}) as well as multiple random walker systems (see section \ref{sec:multi}). 
\subsection{Distribution of the number of records}

To study the distribution of the number of records $M$ it is useful to introduce indicator variables $\sigma_k$ which take the value $0$ or $1$:
\begin{eqnarray}\label{def_sigma}
\sigma_k = 
\begin{cases}
& 1 \; {\rm if \;} X_k \; {\rm is \; a \; record} \;, \\
& 0 \; {\rm if \;} X_k \; {\rm is \; {\it not} \; a \; record} \;
\end{cases}\;, \;
M = \sum_{k=1}^N \sigma_k \;.
\end{eqnarray} 
For i.i.d.~random variables, these indicator functions $\sigma_k$ are independent [see (\ref{final_2p}) below]. We define 
\begin{eqnarray}\label{def_rate}
\langle \sigma_k \rangle = r_k \;,
\end{eqnarray}
where the average is taken over the different realizations of the random variables $X_1, \dots, X_N$.
Thus $r_k$ is the probability that $X_k$ is a record, i.e., that the event in (\ref{def_record}) happens. 
In other words, $r_k$ represents the rate at which a record is broken at ``time'' $k$, or equivalently the {\it probability of record breaking} at time $k$ for the sequence $X_1, \dots, X_N$.
For i.i.d.~random variables
this probability can be easily computed from the joint distribution in (\ref{eq:def_iid}) and this yields
\begin{eqnarray}\label{rate_iid}
r_k = \int_{-\infty}^\infty p(x) \left[\int_{-\infty}^x p(y) dy \right]^{k-1} \, dx = \int_0^1 u^{k-1} du = \frac{1}{k} \;,
\end{eqnarray} 
where we have used the change of variable $u = \int_{-\infty}^x p(y) dy$. Interestingly, this result $r_k = 1/k$ (\ref{rate_iid}) is {\it universal}, i.e., it is independent of the parent distribution $p(x)$. 
This can be easily understood since the probability that $X_k$ is the maximum among $X_1, \dots, X_k$ is indeed equal to $1/k$ as the maximal value can be realized with equal probability by any of these $k$ i.i.d.~random variables. From the record rate in (\ref{rate_iid}), we get the mean number of records as
\begin{eqnarray}\label{exact_mean_iid}
\langle M \rangle = \sum_{k=1}^N r_k = \sum_{k=1}^N \frac{1}{k} = H_{N} \;,
\end{eqnarray}
where $H_{N}$ denotes the $N$-th harmonic number. 
For large $N$, it behaves as
\begin{eqnarray}\label{exact_mean_iid_asympt}
\langle M \rangle = \ln{N} + \gamma_E + {\cal O}(N^{-1}) \;,
\end{eqnarray}
where $\gamma_E = 0.57721\ldots$ is the Euler constant. 
By a similar calculation, one can compute the variance of the number of records, $\langle M^2 \rangle - \langle M \rangle^2$. 
This computation involves the two-point correlations $\langle \sigma_j \sigma_k \rangle$. 
From the joint distribution (\ref{eq:def_iid}) it is easy to show that $\sigma_j$ and $\sigma_k$ are linearly independent for $j \neq k$ \cite{SM_review}.
Indeed, by a simple generalisation of the reasoning made above for~(\ref{rate_iid}), one has
\begin{eqnarray}\label{final_2p}
\langle \sigma_j \sigma_k \rangle 
&=&{\rm Prob}(X_j=\max(X_1,\dots,X_j),X_k=\max(X_j,\dots,X_k))
\nonumber\\
&=&\int_0^1 du\, u^{k-j-1}\int_0^u
dv\, v^{j-1}=\frac{1}{jk}
= \langle \sigma_j \rangle \langle \sigma_k \rangle \;, \; j \neq k \;,
\end{eqnarray}
while $\langle \sigma_k^2 \rangle = \langle \sigma_k \rangle = r_k$. 
Hence, using~(\ref{final_2p}), one obtains
\begin{eqnarray}\label{second_moment}
\langle M^2 \rangle - \langle M \rangle^2 = \sum_{k=1}^N \langle \sigma_k^2 \rangle - \langle \sigma_k \rangle^2 &=& \sum_{k=1}^N \left[ \frac{1}{k} - \frac{1}{k^2} \right] \nonumber \\
&=& \ln{N} + \gamma_E - \frac{\pi^2}{6} + {\cal O}(1/N) \;.
\end{eqnarray} 
Similarly, one can compute the generating function of the probability distribution of the number of records $P(M|N)$ using (for $N \geq 1$)
\begin{eqnarray}\label{GF_rising}
\sum_{M\ge1} P(M|N) x^M &=& \langle x^{M} \rangle = \prod_{k=1}^N \langle x^{\sigma_k} \rangle = \prod_{k=1}^N \left(\frac{x-1}{k}+1 \right) \nonumber \\
&=& \frac{x(x+1)\dots(x+N-1)}{N!} = \frac{1}{N!} \frac{\Gamma(x+N)}{\Gamma(x)} \;,
\end{eqnarray}
where $\Gamma(z)$ is the Gamma function. One also recognizes that the ratio of Gamma functions $\Gamma(x+N)/\Gamma(x)$ appearing in (\ref{GF_rising}) is the $\GF$ of the unsigned Stirling numbers of the first kind \cite{Riordan}, i.e., 
\begin{eqnarray}
x(x+1)\dots(x+N-1) = \sum_{M=1}^N {N \brack M} x^M \;,
\end{eqnarray}
where the unsigned Stirling numbers ${N \brack M}$ enumerate the number of permutations of $N$ elements with exactly $M$ disjoint cycles. Hence one has
\begin{eqnarray}\label{exact_dist_number}
P(M|N) = \frac{1}{N!}{N \brack M} \;,
\end{eqnarray} 
which thus shows that the number of records of $N$ i.i.d.~random variables is distributed like the number of cycles in random permutations of $N$ objects with uniform measure. 
Finally, using the asymptotic behaviour of Stirling numbers, one can show that the distribution of $M$ approaches, when $N \to \infty$, a Gaussian distribution 
\begin{eqnarray}\label{pdf_iid}
P(M|N) \approx \frac{1}{\sqrt{2 \pi \ln N}} \exp{\left[-\frac{(M-\ln N)^2}{2 \ln N} \right]} \;.
\end{eqnarray}

Here we have discussed the case where the random variables $X_i$ are continuous random variables. We refer the reader to ref \cite{rounding} for a discussion of the effects of discreteness, in particular when continuous random variables are subsequently discretized by rounding to integer multiples of a discretization scale (see also section \ref{sec:other} for related issues). 

\subsection{Joint distribution of the ages of records and of their number}

Apart from the number of records, other important observables are the ages of the records, which we now focus on. For a realization of the sequence of $N$ random variables $X_i$ with $M$ records, we denote by ${\vec \ell} = ({\ell_1, \ell_2, \dots, \ell_M})$ the time intervals between successive records as depicted in figure \ref{fig_record}. 
Thus $\ell_k$ is the age of the $k$-th record, i.e., it denotes the time up to which the $k$-th record survives (in the mathematical literature the ages are called ``inter-record times'' \cite{ABN1992,Schorrock}). 
Note that the last record, the $M$-th record in this sequence, is still a record at ``time'' $N$. 
Its age $\ell_M$ is defined as its lifetime $N-N_M$ shifted by one unit, where $N_M$ is the time of occurrence of this last record (see figure \ref{fig_record}).
This definition simplifies the computations that follows.

We first compute the joint probability distribution $P(\vec \ell, M | N)$ of the ages $\vec \ell$ and the number $M$ of records, given the length $N$ of the sequence. This distribution can be computed from the joint distribution of the $X_i$ in (\ref{eq:def_iid}) as
\begin{eqnarray}\label{gen_formula_iid}
&&P(\vec \ell, M | N) = \int_{-\infty}^\infty dy_M p(y_M) \left[\int_{-\infty}^{y_M} p(x) dx \right]^{\ell_M-1} \nonumber \\
&&\times \prod_{k=1}^{M-1} \int_{-\infty}^{y_{k+1}} dy_{k} p(y_k) \left[\int_{-\infty}^{y_{k}} p(x) dx \right]^{\ell_{k}-1} \, 
\delta\left({\sum_{k=1}^M \ell_k, N}\right) \;,
\end{eqnarray}
where the Kronecker delta, $\delta(i,j) = 1$ if $i=j$ and 0 otherwise,
ensures that the size of the sample is $N$. If one performs the change of variables $u_k = \int_{-\infty}^{y_k} p(x) dx$, the distribution $P(\vec \ell, M | N)$ in (\ref{gen_formula_iid}) can be written as
\begin{eqnarray}\label{gen_formula_iid_2}
P(\vec \ell, M | N) = \int_0^1 du_M u_M^{\ell_M-1} \prod_{k=1}^{M-1} \int_0^{u_{k+1}} du_{k} u_{k}^{\ell_{k}-1}
\delta\left({\sum_{k=1}^M \ell_k, N}\right)\,.
\end{eqnarray}
This multiple integral in (\ref{gen_formula_iid_2}) can be performed straightforwardly to obtain
\begin{eqnarray}\label{full_jpdf}
P(\vec \ell, M | N) = \frac{1}{\ell_1(\ell_1+\ell_2)\dots(\ell_1+\ell_2+\cdots+\ell_{M})} \delta\left({\sum_{k=1}^M \ell_k, N}\right)\;. 
\end{eqnarray}
It is important to stress that this joint distribution is completely universal, i.e., independent of the parent distribution $p(x)$. This means that any observables depending only on the ages of the records is totally universal. Quite interestingly, although the variables $X_i$ are independent, we see on (\ref{full_jpdf}) that the ages $\ell_k$ are correlated, which yields a non trivial statistics of the ages in this i.i.d.~case. 
In the next section we discuss the marginal distribution of the age of the $k$-th record as well as the statistics of the longest or shortest records and refer the reader to ref \cite{Ahsanullah} -- chapter 1 -- for further details and references on the ages of records for i.i.d.~random variables in the mathematical literature. 

We conclude this section by a remark on the record times, which are the times at which the records occur,
\beq\label{eq:Nk}
N_k = 1+\sum_{j=1}^{k-1} \ell_j,\quad (2\le k\le M),
\eeq 
with $N_1=1$.
Elements of the study of these record times can be found in \cite{GL2008}.
In particular, in the continuum limit of large times, these record times, now real variables denoted by $t_k$, are generated recursively.
The successive ratios 
\beq\label{eq:ratios}
\frac{t_{k-1}}{t_k}=U_k,
\eeq are i.i.d.~random variables uniform on $(0,1)$.
This property is instrumental in the derivation of the asymptotic distribution of the duration of the longest lasting record \cite{GL2008} [see (\ref{pdf_lmax_iid}-\ref{eq:fVhat}) in section \ref{sec:fVR} below].


\subsection{Marginal probability distribution of the age of the $k$-th record}

The marginal probability distribution of $\ell _k$ can be obtained by summing the full joint distribution $P(\vec \ell, M | N)$ in (\ref{full_jpdf}) over all the ages $\ell_j$ with $j \neq k$ and then summing over the number of records:
\begin{eqnarray}
P(\ell_k | N) = \sum_{M\geq 1} \sum_{\ell_1 \geq 1} \ldots \sum_{\ell_{k-1} \geq 1} \sum_{\ell_{k+1} \geq 1} \ldots \sum_{\ell_M \geq 1} P(\vec \ell, M | N) \;.
\end{eqnarray}
The full distribution $P(\vec \ell, M | N)$, given in (\ref{full_jpdf}), is obviously not invariant under the permutation of the ages $\ell_j$ and therefore $P(\ell_k|N)$ depends explicitly on $k$. Its $\GF$ with respect to $N$ can be computed exactly -- using the integral representation of the full joint distribution (\ref{gen_formula_iid_2}) -- with the result \cite{Lorenzo}
\begin{eqnarray}\label{GF_Pellk}
\sum_{N\geq 1} P(\ell_k | N) z^N = \frac{1}{1-z} \int_0^z dx (1-x) \frac{\left[-\ln(1-x)\right]^{k-1}}{(k-1)!} x^{\ell_k-1} \;.
\end{eqnarray}
We see that the right hand side of (\ref{GF_Pellk}) behaves like $\propto (1-z)^{-1}$ when $z \to 1$, from which we conclude that $P(\ell_k| N)$ tends to a stationary distribution as $N \to \infty$, which is given by \cite{Schorrock,Neuts}
\begin{eqnarray}\label{Pell_explicit}
P(\ell_k) = \lim_{N \to \infty} P(\ell_k | N) &=& \int_0^1 dx (1-x) \frac{\left[-\ln(1-x)\right]^{k-1}}{(k-1)!} x^{\ell_k-1} \; \\
&=& \sum_{m=0}^{\ell_k-1} (-1)^m {\ell_{k}-1 \choose m} \frac{1}{(2+m)^k} \;, \nonumber
\end{eqnarray} 
where the second line is obtained by performing the change of variable $u = - \ln(1-x)$ in the integral in (\ref{Pell_explicit}) and using the binomial formula to expand $x^{\ell_k - 1} = (1-\e^{-u})^{\ell_k - 1}$ in order to perform the integral over $u$. 
The probability $P(\ell_k)$ is a monotonically decreasing function, starting from $P(\ell_k = 1) = 2^{-k}$. For large $\ell_k$, its asymptotic behaviour is more conveniently obtained from the integral representation (\ref{Pell_explicit}) which can be analysed in the large $\ell_k$ limit by performing the change of variable $v=(1-x) \ell_k$ which yields
\begin{eqnarray}
P(\ell_k) \sim \dfrac{1}{(k-1)!} \dfrac{\left[\ln \ell_k\right]^{k-1}}{\ell_k^2} \;, \; \ell_k \to \infty \;.
\end{eqnarray}
This indicates that the first moment of $\ell_k$ is diverging when $N \to \infty$. In fact, one can show from (\ref{GF_Pellk}) that
\begin{eqnarray}\label{average_lk_iid}
\langle \ell_k \rangle \sim \frac{\left[ \ln N \right]^k}{k!} \;, \; N \to \infty \;.
\end{eqnarray}
Interestingly, by using the Stirling formula $k!\approx \sqrt{2 \pi k} \, \e^{k \ln k - k}$, one sees that the average $\langle \ell_k \rangle$, as a function of $k$, admits a maximum for $k_{\max} \sim \ln{N}$, for which $\langle \ell_{k_{\max}} \rangle \sim N$ (up to possible logarithmic corrections). Hence, $k_{\max}$ coincides with the typical number of records $\langle M \rangle \sim \ln N$, see (\ref{exact_mean_iid_asympt}). 
This indicates that the longest lasting record is rather likely to be the last one, which happens with a rather high probability $\approx 0.62433$ [see~(\ref{c1}) and (\ref{eq:Qinfty}) below], or close to it \cite{Lorenzo}.
The statistical properties of the longest lasting record will be discussed in the next section.

\subsection{Distribution of the age of the longest lasting record}
\label{sec:fVR}

We have seen in the previous section that the mean age of the $k$-th record, $\langle \ell_k \rangle$, depends rather strongly on $k$, see (\ref{average_lk_iid}). It behaves typically as $(\ln N)^k/k!$ as a function of $k$ 
and reaches its maximum for $k_{\max} \sim {\cal O}(\ln N)$ where it is of order $ {\cal O}(N)$. In this section, we characterize this extreme behaviour and focus on the age of the longest record, denoted by $\ell_{\max, N}$, which is defined as
\begin{eqnarray}\label{def_lmax}
\ell_{\max,N} = \max \{\ell_1, \ell_2, \dots, \ell_M \} \;.
\end{eqnarray}
Its cumulative distribution $F(\ell |N) = {\rm Prob}(\ell_{\max,N} \leq \ell)$, for $\ell \geq 1$, is obtained from the full joint distribution in (\ref{full_jpdf}) by summing over $M$ and $\ell_1, \dots,\ell_M$ with the constraint that $\ell_1 \leq \ell$, $\dots$, $\ell_M \leq \ell$. It reads, for $N \geq 1$, 
\begin{eqnarray}\label{cumulative distribution function_lmax_iid}
F(\ell | N) = \sum_{M\geq 1} \sum_{\ell_1 =1}^\ell \dots \sum_{\ell_M = 1}^\ell \frac{\delta\left({\sum_{k=1}^M \ell_k, N}\right)}{\ell_1(\ell_1 + \ell_2)\dots(\ell_1+\ell_2+\cdots+\ell_M)} \;,
\end{eqnarray}
while $F(\ell | N = 0) = 1$. The $\GF$ of $F(\ell | N)$ with respect to $N$ is conveniently written using the integral representation of the distribution in (\ref{gen_formula_iid_2}). After some manipulations, it can be written as \cite{SM_review}
%
%
\begin{eqnarray}\label{gf_distlmax}
\sum_{N\ge0} z^N F(\ell | N) = \exp{\left(\sum_{k=1}^\ell \frac{z^k}{k} \right)} \;.
\end{eqnarray}
From the $\GF$ of the full distribution of $\ell_{\max,N}$ (\ref{gf_distlmax}) one obtains the $\GF$ of the average value $\langle \ell_{\max,N} \rangle = \sum_{\ell\ge1} (1 - F(\ell | N))$ as
\begin{eqnarray}\label{gf_av_lmax}
\sum_{N\ge0} \langle \ell_{\max,N} \rangle z^N = \frac{1}{1-z} \sum_{\ell\ge1}\left[ 1 - \exp{\left(-\sum_{k\ge\ell+1} \frac{z^k}{k} \right)}\right] \;.
\end{eqnarray}
By analysing this expression (\ref{gf_av_lmax}) in the limit $z \to 1$, where the discrete sums can be replaced by integrals (setting $z = \e^{-s}$) one obtains
the large $N$ behaviour of $\langle \ell_{\max,N} \rangle$ as
\begin{eqnarray}\label{c1}
\langle \ell_{\max,N} \rangle = \lambda N + {\cal O}(1) \;, \; \lambda = \int_0^\infty ds \, \e^{-s - E(s)} = 0.62433\ldots \;,
\end{eqnarray}
where $E(s) = \int_s^\infty dy\, \e^{-y}/y$. In (\ref{c1}), $\lambda$ is known as the Golomb-Dickman or Goncharov constant \cite{Finch_book}.
This constant $\lambda$ also describes the linear growth of the longest cycle of a random permutation \cite{Finch_book}. It also appeared in a model of growing network \cite{GL2008} and in a one-dimensional ballistic aggregation model \cite{MMS09}. The complete asymptotic expansion of $\langle \ell_{\max,N} \rangle$, beyond the leading order, was established in ref \cite{Gou96}.

\begin{figure}[t]
\centering
\includegraphics[width = 0.8\linewidth]{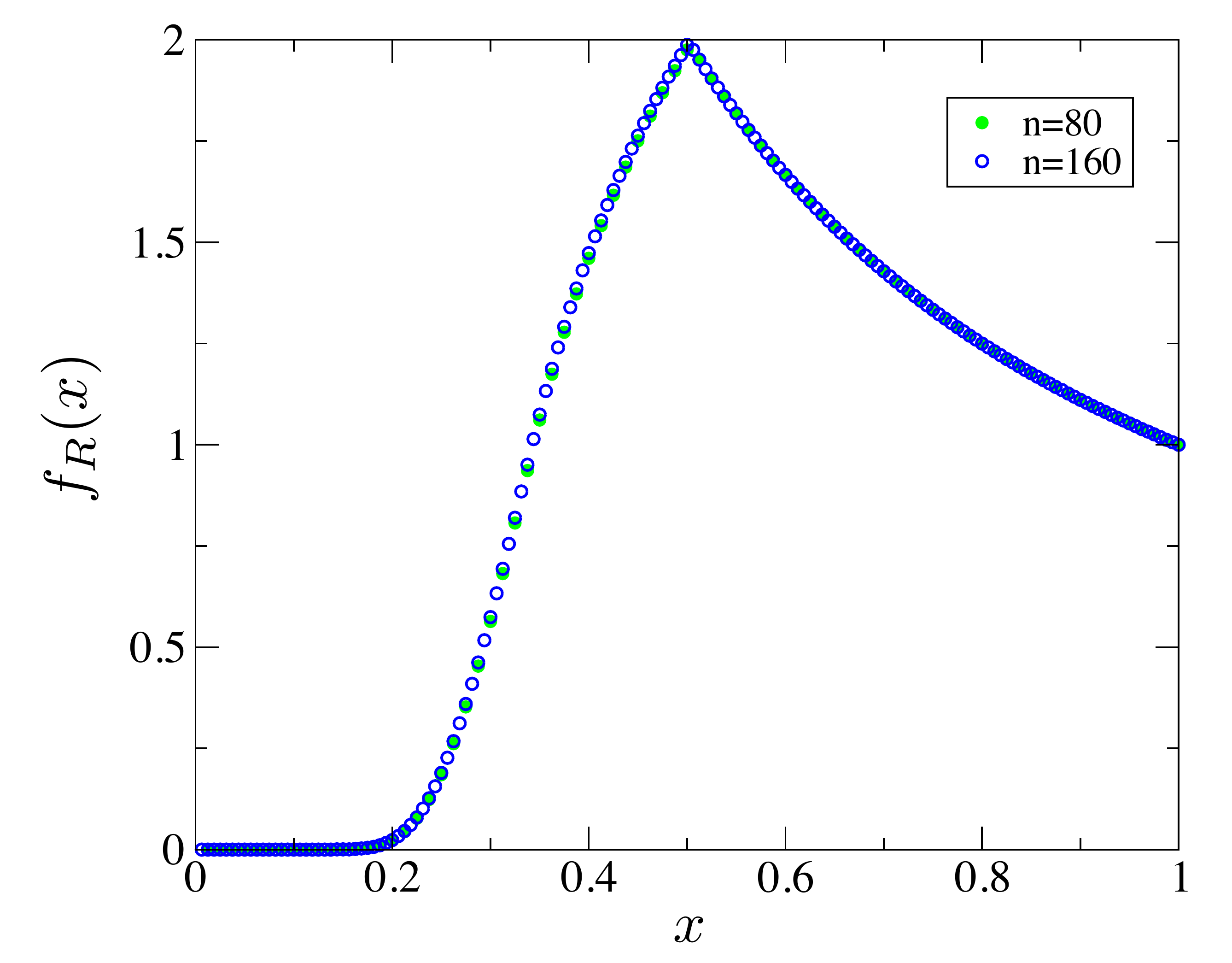}
\caption{Limiting distribution of the scaled random variable $R = \ell_{\max,N}/N$, see (\ref{pdf_lmax_iid}). It was obtained from the analytical expression (\ref{gf_distlmax}) of the generating function of $\ell_{\max,N}$ for $N=80$ (green full circles) and for $N=160$ (blue empty circles). The good collapse of the data confirms the scaling form in (\ref{pdf_lmax_iid}).}\label{fig:fRiid}
\end{figure}
A complementary approach to the statistics of $\ell_{\max,N}$ can be found, e.g., in refs \cite{GL2008,PY97}. 
In particular, in the regime of long times 
the scaled random variable $R = \ell_{\max,N}/N$ has a limiting density denoted by $f_R$,
\begin{eqnarray}\label{pdf_lmax_iid}
{\rm Prob}(\ell_{\max,N} = \ell) \underset{N \to \infty}\longrightarrow \frac{1}{N}f_{R}\left(\frac{\ell}{N} \right).
\end{eqnarray}
To compute this limiting distribution $f_R(x)$, it is convenient to study the inverse variable $V=1/R$, which has a limiting density $f_V(x)$. It turns out that 
the Laplace transform of the inverse variable $V=1/R$ has an explicit expression
\beq\label{eq:fVhat}
\widehat f_V(s)=\langle \e^{-sV}\rangle= \int_0^\infty d x \, f_V(x) \e^{-sx} = 1-\e^{-E(s)},
\eeq
where we recall that $E(s) = \int_s^\infty dy\, \e^{-y}/y$. Note that from (\ref{eq:fVhat}) one can straightforwardly compute the average $\langle R \rangle$ as 
\beq
\lambda=\langle R\rangle=\left\langle \frac{1}{V}\right\rangle=\int_0^\infty ds\,\widehat f_V(s),
\eeq
which, after a simple integration by parts, yields back the Golomb-Dickman constant in~(\ref{c1}), i.e., $\langle R \rangle = \lambda$. Furthermore, from (\ref{eq:fVhat}), and using $f_R(x) = x^{-2} f_V(1/x)$, one can show that the function $f_{R}(x)$ is a piecewise continuous function on the interval $[0,1]$, continuous on each interval of the form $[1,1/2]$, $[1/2,1/3]$, $\ldots$, and 
exhibiting singularities at the points $x_k = 1/k$, with $k = 2, 3, \ldots$. It has a maximum at $x=x_2 = 1/2$ and its asymptotic leading behaviours are given by \cite{GL2008}
\begin{eqnarray}
f_{R}(x) \sim
\begin{cases}
& \exp\left(\dfrac{1}{x} \ln x \right) \;, \; x \to 0 \;, \\
& \\
&1 \;, \; x \to 1 \;. 
\end{cases}
\end{eqnarray} 
We refer the reader to ref \cite{GL2008} for further details on this limiting distribution.
Figure~\ref{fig:fRiid} depicts $f_R(x)$ obtained
from the analytical expression of the generating function of $\ell_{\max,N}$ given by~(\ref{gf_distlmax}) for $N=80$ (green full circles) and $N=160$ (blue empty circles). The good collapse of the data confirms the scaling form~(\ref{pdf_lmax_iid}).

Another, related, quantity of interest is the probability that the longest lasting record is the last one, or probability of record breaking for the sequence of ages, namely
\beq\label{eq:QN}
Q_N=\Prob(\ell_M>\max(\ell_1,\dots,\ell_{M-1}))=\Prob(\ell_{\max,N}=\ell_M).
\eeq
This sequence converges, at large $N$, to the Golomb-Dickman constant $\lambda$ \cite{GL2008},
\beq\label{eq:Qinfty}
\lim_{N\to\infty} Q_N=\lambda,
\eeq
which means that, for a very long sequence, the fraction of records with longest duration is equal to $\lambda$.


\subsection{Distribution of the age of the shortest record}

We now focus on the age of the shortest record, denoted by $\ell_{\min,N}$, which is defined~as
\begin{eqnarray}\label{def_lmin}
\ell_{\min,N} = \min \{\ell_1, \ell_2, \dots, \ell_M \} \;.
\end{eqnarray}
We define $G(\ell | N) = {\rm Prob}(\ell_{\min,N} \geq \ell)$, $\ell \geq 1$, and $G(\ell | N=0) = 0$. Using the same reasoning as above for $\ell_{\max,N}$ we find the $\GF$ of $G(\ell | N) = \Prob(\ell_{\min,N} \geq \ell)$ with respect to $N$ as
\begin{eqnarray}\label{dist_min_GF}
\sum_{N\ge0} G(\ell |N)z ^N = \exp{\left[\sum_{k\ge \ell} \frac{z^k}{k} \right]} - 1 \;.
\end{eqnarray}
The $\GF$ of the average value $\langle \ell_{\min,N}\rangle = \sum_{\ell\ge1} G(\ell | N)$ can be obtained from~(\ref{dist_min_GF}) which yields the asymptotic result for large $N$ \cite{SP1966}
\begin{eqnarray}\label{lmin_iid}
\langle \ell_{\min,N} \rangle = \e^{-\gamma_E} \ln{N} + o(\ln{N}) \;,
\end{eqnarray}
with the numerical value $\e^{-\gamma_E} = 0.56145\ldots$, where $\gamma_E = 0.57721\ldots$ is the Euler constant. 

\begin{figure}
\centering
\includegraphics[width=0.8\linewidth]{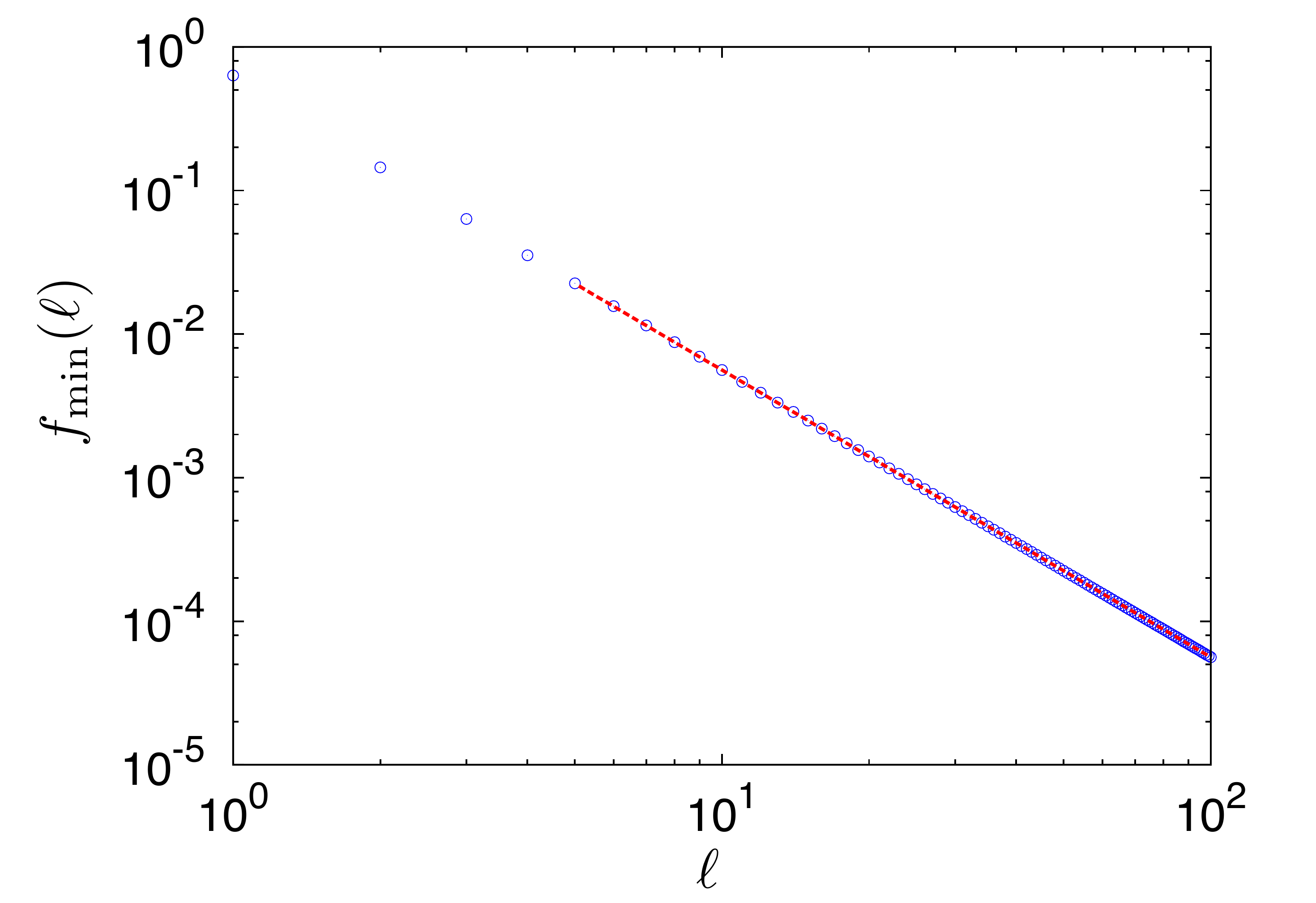}
\caption{Plot of the limiting distribution $f_{\min}(\ell)$ given in (\ref{stationary_min}) (blue circles). The dotted red line corresponds to the large $\ell$ asymptotic behaviour given in (\ref{eq:fmin_asympt}).}\label{fig:lmin}
\end{figure}
On the other hand, when $N \to \infty$, one can easily show that $G(\ell | N)$ converges to a stationary cumulative distribution function, from which one obtains the limiting distribution
\begin{eqnarray}\label{stationary_min}
\fl{\rm Prob}(\ell_{\min,N} = \ell) \underset{N \to \infty}{\longrightarrow} f_{\min}(\ell) = \exp{\left[-\sum_{k=1}^{\ell-1} \frac{1}{k}\right]}\left(1 - \e^{-1/\ell} \right) \;, \; \ell \geq 2 \;,
\end{eqnarray}
while $f_{\min}(\ell = 1) = 1-\e^{-1}$. The limiting distribution $f_{\min}(\ell)$ is a monotonously decreasing function of $\ell$ and its asymptotic behaviours are given by
\begin{eqnarray}\label{eq:fmin_asympt}
f_{\min}(\ell) \approx
\begin{cases}
&1-\e^{-1} \;, \; \ell \to 1 \;, \\
&\dfrac{\e^{-\gamma_E}}{\ell^2} \; , \; \ell \to \infty \;,
\end{cases}
\end{eqnarray}
Remembering that this asymptotic behaviour for large $\ell$, $f_{\min}(\ell) \approx \e^{-\gamma_E}/\ell^2$, is valid for $\ell \leq N$, this yields the large $N$ estimate for $\langle \ell_{\min,N} \rangle$ as given in (\ref{lmin_iid}). In figure \ref{fig:lmin} we show a plot of this limiting distribution $f_{\min}(\ell)$, where we see in particular that the asymptotic large $\ell$ behaviour $\sim \e^{-\gamma_E}/\ell^2$ (\ref{eq:fmin_asympt}) gives a quite accurate description of the exact distribution $f_{\rm min}(\ell)$ already for $\ell \gtrsim 10$.

\section{Record statistics for correlated sequences: Random Walk model} \label{section:RW}

We have seen in the previous section that for an uncorrelated time series
$\{X_1,X_2,\ldots, X_N\}$ of length $N$, the statistics of the number of 
records $M$, as well as the statistics of the ages of records can be
computed analytically. In many realistic time series, the entries $X_i$ 
are however correlated. So, the question naturally arises: what can we say
about the record statistics for {\em correlated} sequences? 
%
%
We review below the recent results that have been obtained for the random walk sequence.


We start with the simple case of a discrete-time random walk on a line.
This will include both short-ranged random walks as well as long-ranged 
L\'evy walks as explained below. In addition, it may include
random walks in the presence of a constant drift.
The walker starts at the origin $X_0=0$ and its position evolves in 
discrete-time via the Markov rule
\begin{equation}
X_i= X_{i-1} + \eta_i
\label{evolrw.1}
\end{equation}
where $\eta_i$ represents the random jump length at step $i$. These noise
variables $\eta_i$ are i.i.d.~random variables, each drawn from the
jump distribution $\phi(\eta)$. The jump distribution may be
symmetric (no drift) or asymmetric (e.g., in the presence of a constant 
drift). 

Few examples
of symmetric jump distributions are:

\begin{itemize}

\item[(i)] $\phi(\eta)= \frac{1}{2}\, \e^{-|\eta|}$ (exponential),

\item[(ii)] $\phi(\eta)= \frac{1}{\sigma_0 \sqrt{2\pi}}\, 
\e^{-\eta^2/{2\sigma_0^2}}$ (Gaussian),

\item[(iii)] $\phi(\eta)= \frac{1}{2}\, 
\left[\Theta(\eta+1)-\Theta(\eta-1)\right]$ (uniform in $[-1,1]$),

\item[(iv)] $\phi(\eta)\sim |\eta|^{-1-\mu}$ for large $|\eta|$ with 
$0\le \mu < 2$ (L\'evy flights),

\item[(v)] $\phi(\eta)= \frac{1}{2}\left[\delta(\eta-1)+ \delta(\eta+1)\right]$ 
(lattice random walk).
\end{itemize}
In the first four examples, the jump distribution
is continuous. In the last example, the jump distribution is not 
continuous, and the walker is restricted to move on a one-dimensional lattice with unit lattice spacing. For the
first three examples, the variance of the step length $\sigma^2= 
\int_{-\infty}^{\infty} \eta^2\, \phi(\eta)\, d\eta$ is finite, while
in the L\'evy case, $\sigma^2$ is infinite.

Note that even though the noise variables $\eta_i$ are uncorrelated, the
positions $X_i$ are strongly correlated.
We consider such a sequence of $N$ entries $\{X_1,X_2,\ldots, X_N\}$
with $M$ records. 
For an illustration, see figure \ref{fig.rw}.
Our first goal is to compute the distribution $P(M|N)$ of the number of records $M$.
We will also be interested in the statistics of the ages of the records.
The random variable $\ell_k$ denotes the age of the $k$-th record, i.e., the length of
time between the $k$-th record and the $(k+1)$-th record (see
figure \ref{fig.rw}). 
The ages are thus defined as in the i.i.d.~case (see figure \ref{fig_record}), except for the last one.
In both cases one sets $\ell_M=N-\sum_{k=1}^{M-1}\ell_k$.
However the origins of time are different in the two cases, namely
for i.i.d.~variables the first record starts at time 1, while for the random walk it starts at time zero.
Hence there is a shift of one unit between the two ages $\ell_M$.
In figure \ref{fig_record} one has $\ell_4 = 4$, while in figure \ref{fig.rw} one has $\ell_4 = 3$.

Hence our main observables are the number of records 
$M$, and the ages $\{\ell_1,\ell_2,\ldots, \ell_M\}$ of the records.
Following the i.i.d.~case investigated in the previous section (see (\ref{def_sigma}) and below), we can still write 
$M=\sum_{k=1}^N \sigma_k$, where $\sigma_k$ is a binary variable:
$\sigma_k=1$ if a record occurs at step $k$ and $\sigma_k=0$ otherwise.
However, unlike in the i.i.d.~case, the variables $\sigma_k$ are now
correlated in the random walk case. Hence, it is hard to compute
directly the distribution of $P(M|N)$. So, how does one proceed to
compute $P(M|N)$ in this case?

We will see below that one can make progress in
calculating $P(M|N)$ by using
the renewal property of the random walk. 
Indeed this approach
was used in ref \cite{MZ2008} to compute exactly
$P(M|N)$ for symmetric jump distributions. But the renewal
property is more general, and can be used even for random walk
sequence with a drift \cite{PLDW2009,MSW2012}, as we will see below.
For introductions to renewal processes, see, e.g., \cite{Feller,Cox1962,GL2001}.

\subsection{The general renewal property}\label{sec:renewal}

Following ref \cite{MZ2008}, we note that instead of trying to
compute $P(M|N)$ directly, it is convenient to first consider a 
bigger collection of random variables in a given sequence, namely the
number of records $M$ as well as the collection of their ages 
denoted by the vector $\vec \ell=\{\ell_1,\ell_2,\ldots, \ell_M\}$. 
The joint 
distribution of these random variables will be denoted by $P(\vec \ell, M|N)$
as in the i.i.d.~case. The main point 
is that this apparently more 
complicated joint distribution actually has a rather simple
structure, due to the renewal property (as explained below). Consequently, 
by integrating out
the age variables $\vec \ell$ from the joint distribution, one can
exactly obtain the marginal distribution $P(M|N)$ of the record number
only.

\begin{figure}
\centering
\includegraphics[width=0.8\textwidth]{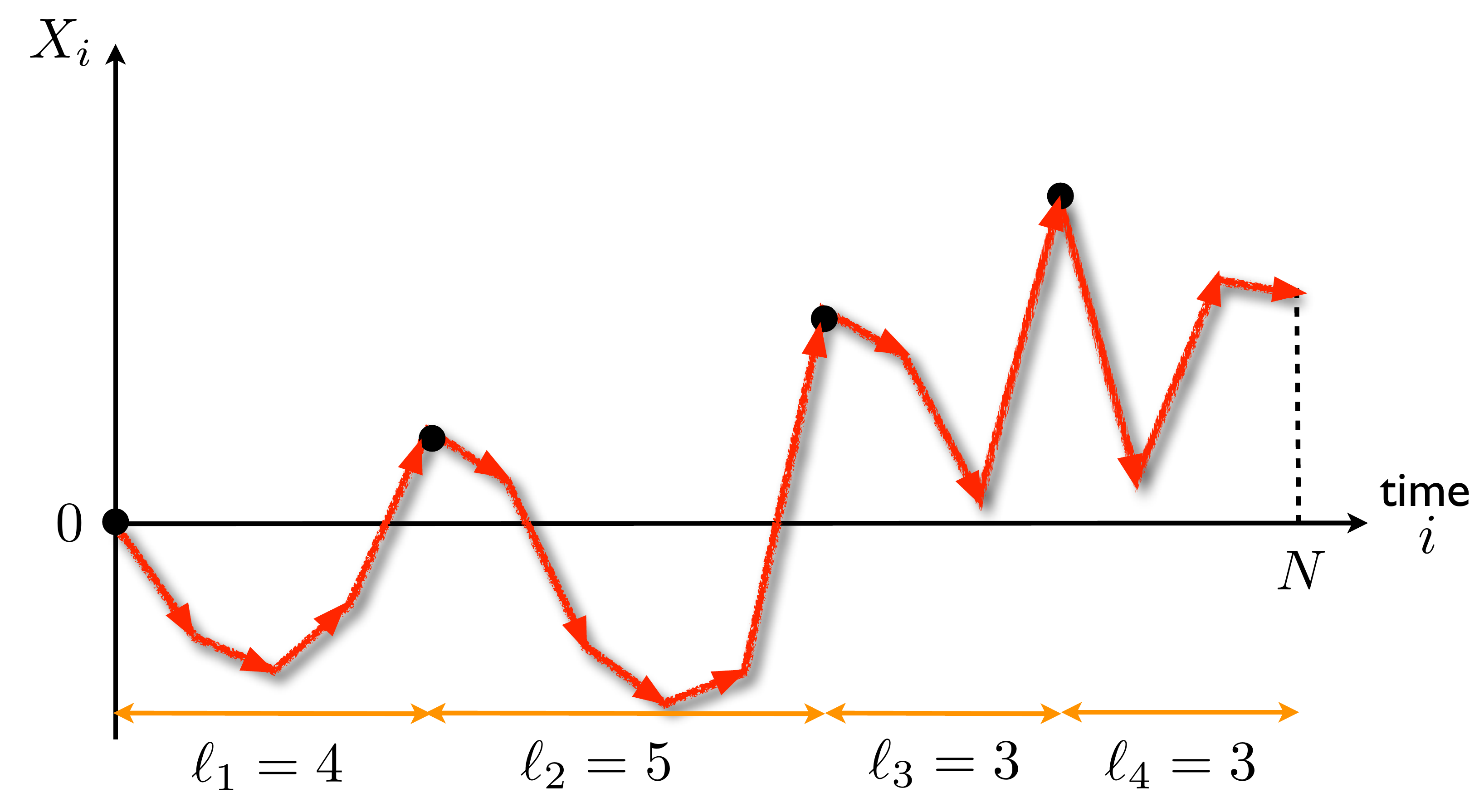}
\caption{ A typical realization of the random walk sequence
$\{X_0=0,X_1,X_2,\ldots, X_N\}$ of $N=15$ steps with $M=4$ records. Each 
record
is represented by a filled circle. The set 
$\{\ell_1,\ell_2,\ell_{3}\}$ represents
the time intervals between the successive records and $\ell_M = \ell_4$ is the age of
the last record which is still a record at time $N$.}
\label{fig.rw}
\end{figure}

Our goal now is to first compute the joint distribution $P(\vec \ell,M|N)$ 
for the generic random walk sequence. 
For this, we will need two crucial quantities as building 
blocks~\cite{MZ2008}. 

\begin{itemize}

\item The first quantity is the so called persistence or 
survival probability 
$q(\ell)$.
It is the probability that a random walk, starting at the 
initial position $X_0$, stays below $X_0$ up to step $\ell$
\begin{eqnarray}
\hspace*{-0.5cm}q(\ell)&= & {\rm Prob}\left(X_1<X_0,\, X_2<X_0,\,X_3<X_0,\dots,X_\ell<X_0\big| X_0\right) \nonumber \\
\hspace*{-0.5cm}&=& {\rm Prob}\left(X_1<0,\, X_2<0,\,X_3<0,\dots,X_\ell<0\big| X_0=0\right),
\label{ql_def}
\end{eqnarray}
with $q(0)=1$ by definition. In going from the first to the second line 
in (\ref{ql_def}), we have used the translation 
invariance of the
process with respect to the starting point. Evidently, 
$q(\ell)$ does not depend on $X_0$ and we can set $X_0=0$. 
For later purposes, let us also define its generating function
\begin{equation}
{\tilde q}(z)= \sum_{\ell\ge 0} q(\ell)\, z^\ell \, .
\label{ql_gf}
\end{equation}

\item The second ingredient
is the related first-passage probability $f(\ell)$ (starting at $X_0=0$)
and defined as
\begin{equation}
f(\ell)= {\rm Prob}\left(X_1<0,\, X_2<0,\,\ldots, X_{\ell-1}<0,\, 
X_\ell>0\Big|X_0=0 \right) \;.
\label{fl_def}
\end{equation}
It is clear that $f(\ell)$ is simply related to $q(\ell)$ via 
\begin{equation}
f(\ell)= q(\ell-1)-q(\ell) \, .
\label{fl_ql}
\end{equation}
Consequently, the generating function of $f(\ell)$ is simply related to that
of $q(\ell)$ as
\begin{equation}
{\tilde f}(z)= \sum_{\ell\ge 1} f(\ell)\, z^{\ell} = 1- (1-z)\, {\tilde q}(z)\, .
\label{fz_qz}
\end{equation}
\end{itemize}

We will see later that both probabilities $q(\ell)$ and $f(\ell)$ for a random 
walk can be computed exactly. But for now, we can proceed even 
without
the explicit knowledge of the two. In fact, the discussion below
will hold for any arbitrary renewal process, not necessarily restricted to 
the random walk sequence.

Armed with these two probabilities $q(\ell)$ and $f(\ell)$, and using the fact
that the
successive intervals
between records are statistically independent due to the Markov nature of 
the process (also called the renewal property), it follows immediately
that (see figure \ref{fig.rw})
\begin{equation}
P(\vec \ell,M|N)= f(\ell_1)\, f(\ell_2)\, \ldots f(\ell_{M-1})\, q(\ell_M)\, 
\delta\left({\sum_{k=1}^M \ell_k,N}\right) \;,
\label{renewal.1}
\end{equation}
where the Kronecker delta enforces the global constraint that the sum of 
the
time intervals equals $N$. The fact that the last record, i.e., the $M$-th 
one,
is still surviving as a record at step $N$ indicates that the distribution
$q(\ell_M)$ of the last interval is different from the preceding ones. It is 
easy to check 
that $P(\vec \ell, M|N)$ is normalised to unity when summed over $\vec \ell$ and 
$M$.

The record number distribution $P(M|N)= \sum_{\vec \ell}P(\vec \ell,M|N)$ is 
just the
marginal of the joint
distribution when one sums over the interval lengths. 
Due to the global constraint, this sum is most easily carried out by considering
the generating function with respect to $N$. Multiplying (\ref{renewal.1})
by $z^N$ and summing over $\vec \ell$ and $N$, one arrives at the fundamental 
relation
\begin{equation}
\sum_{N\ge0} P(M|N)\, z^N = \left[{\tilde f}(z)\right]^{M-1}\, 
{\tilde 
q}(z)=\left[1-(1-z) {\tilde q}(z)\right]^{M-1}\, {\tilde q}(z) \;,
\label{renewal.genf}
\end{equation}
where we used (\ref{fz_qz}). Thus the knowledge
of ${\tilde q}(z)$ enables one to determine the distribution $P(M|N)$
and all its moments. For instance, multiplying (\ref{renewal.genf}) by $M$ and summing over all $M$, one obtains
the exact generating function of the average 
number of records ${\langle M\rangle}$ in $N$ steps
\begin{equation}
\sum_{N\ge0} {\langle M\rangle}\, z^N= \frac{1}{(1-z)^2\, {\tilde 
q}(z)}\, .
\label{avg_genf.1}
\end{equation}
Similarly, the higher moments can also be computed in principle, once one
knows $q(\ell)$.

Let us emphasize again that the result~(\ref{renewal.genf}),
and consequently~(\ref{avg_genf.1}) are 
rather general, and hold for
any renewal process. So, we only need to know $q(\ell)$.
This, however,
can be computed explicitly for any random walk process on a line using an 
elegant theorem due to Sparre 
Andersen \cite{SA54}. According to this theorem, 
the generating function ${\tilde q}(z)$ satisfies
a nontrivial combinatorial identity \cite{Feller,SA54}
\begin{equation}
{\tilde q}(z) = \sum_{\ell\ge0} q(\ell)\, z^\ell=
\exp\left[\sum_{n\ge1}\frac{z^n}{n}\, p_{-}(n)\right] \;,
\label{SA.1}
\end{equation}
where $p_{-}(n)= {\rm Prob}\,[X_n\le 0]$. Note that $q(\ell)$ involves
a non-local property of the trajectory from the $0$-th to the $\ell$-th step, 
namely it is the probability
that $X_i$ stays negative up to step $\ell$, starting at the origin. In 
contrast, $p_{-}(n)$ is a local quantity: it is the probability that 
exactly
at step $n$, the walker is on the negative side of the origin.

In the next subsections, we will consider
several cases where $q(\ell)$, or equivalently ${\tilde q}(z)$ can
be computed explicitly using this theorem, leading to exact results for~$P(M|N)$.

\subsection{Statistics of the record number}\label{sec:record_nber}

In this subsection, we will apply the general renewal theory developed 
above to compute explicitly the distribution of the record number $M$ 
for a random walk sequence for a variety of jump distributions, with and 
without drift.

\vskip 0.5cm

\subsubsection{Symmetric and continuous jump distribution.} 

For symmetric and 
continuous jump distributions (see examples (i)-(iv) discussed in the introduction of section \ref{section:RW}), 
clearly 
$p_{-}(n)=1/2$ for 
all $n\ge 1$ (by 
symmetry). Consequently, (\ref{SA.1}) gives
\begin{equation}
{\tilde q}(z) = \sum_{\ell\ge0} q(\ell)\, z^\ell= \frac{1}{\sqrt{1-z}}\, ,
\label{qz_exact}
\end{equation} 
a completely {\em universal} result, i.e., independent of the jump 
distribution $\phi(\eta)$, as long as it is symmetric and continuous.
Expanding in $z$, it gives the universal result 
\begin{equation}
q(\ell)= {{2\ell} \choose {\ell}}\, 2^{-2\,\ell} \approx \frac{1}{\sqrt{\pi \,\ell}} \;, \; \ell \to \infty \;.
\label{ql.1}
\end{equation}
Let us remark that this result for $q(\ell)$ is universal {\em for all} $\, \ell\ge 0$. Consequently, from (\ref{renewal.genf}), the 
record number distribution $P(M|N)$ also becomes universal for all 
$N$ \cite{MZ2008}. For instance, substituting (\ref{qz_exact}) in 
 (\ref{renewal.genf}) and inverting with respect to $z$ gives
the exact distribution, universal for all $N$ (equations (\ref{nore1})-(\ref{scaling_record_dist.1}) were first derived in \cite{MZ2008})
\begin{equation}
P(M|N)= {{2N-M+1}\choose N}\, 2^{-2N+M-1} \, .
\label{nore1}
\end{equation}
From this exact result in (\ref{nore1}) all moments of $M$ can be 
computed as well.
For example, the average number of records is 
given by
\begin{equation}
{\langle M\rangle} = (2N+1)\,{2N \choose N}\, 2^{-2N} \;.
\label{avg_rec.1}
\end{equation}
In particular, for large $N$, the mean number of records grows as
\begin{equation}
{\langle M\rangle} \approx \frac{2}{\sqrt{\pi}}\, \sqrt{N} \, ,
\label{avg_rec.2}
\end{equation}
much faster than the logarithmic growth for i.i.d.~sequences discussed in the
previous section [see (\ref{exact_mean_iid_asympt})]. It is easy to show from
the exact distribution that the variance grows linearly 
for large $N$
\begin{equation}
{\langle M^2\rangle}-{\langle M\rangle}^2
\approx 2\left(1-\frac{2}{\pi}\right)\, N \,.
\label{var_rec}
\end{equation}
Thus, both the mean and the standard deviation grow as $\sqrt{N}$ for 
large $N$, indicating that the fluctuations are large. This is also
vindicated by the scaling analysis of the distribution $P(M|N)$
in (\ref{nore1}) in the scaling limit, by setting $M\sim {\cal O}(\sqrt{N})$
and taking the $N\to \infty$ limit. In this scaling limit, one 
obtains
\begin{equation}
P(M|N)\approx \frac{1}{\sqrt{N}}\,g\left(\frac{M}{\sqrt{N}}\right) \;, \,\, 
{\rm with}\,\, g(x)= \frac{1}{\sqrt{\pi}}\, \e^{-x^2/4}\, \Theta(x)\, ,
\label{scaling_record_dist.1}
\end{equation}
where $\Theta(x)$ is the Heaviside step function (i.e., $\Theta(x) = 1$ if $x\geq 0$ and $\Theta(x) = 0$ if $x<0$). Indeed, this scaling 
behaviour of $P(M|N)$ for the random walk case is markedly different from
the i.i.d.~case discussed before in (\ref{pdf_iid}), where $P(M|N)$ approaches a Gaussian distribution
$P(M|N)\sim \exp[-(M-\ln N)^2/{2\ln N}]$, with
mean ${\langle M\rangle} = \ln N$ and
standard deviation $\sigma= \sqrt{\ln N}$.

\begin{figure}
\centering
\includegraphics[width = 0.8\linewidth]{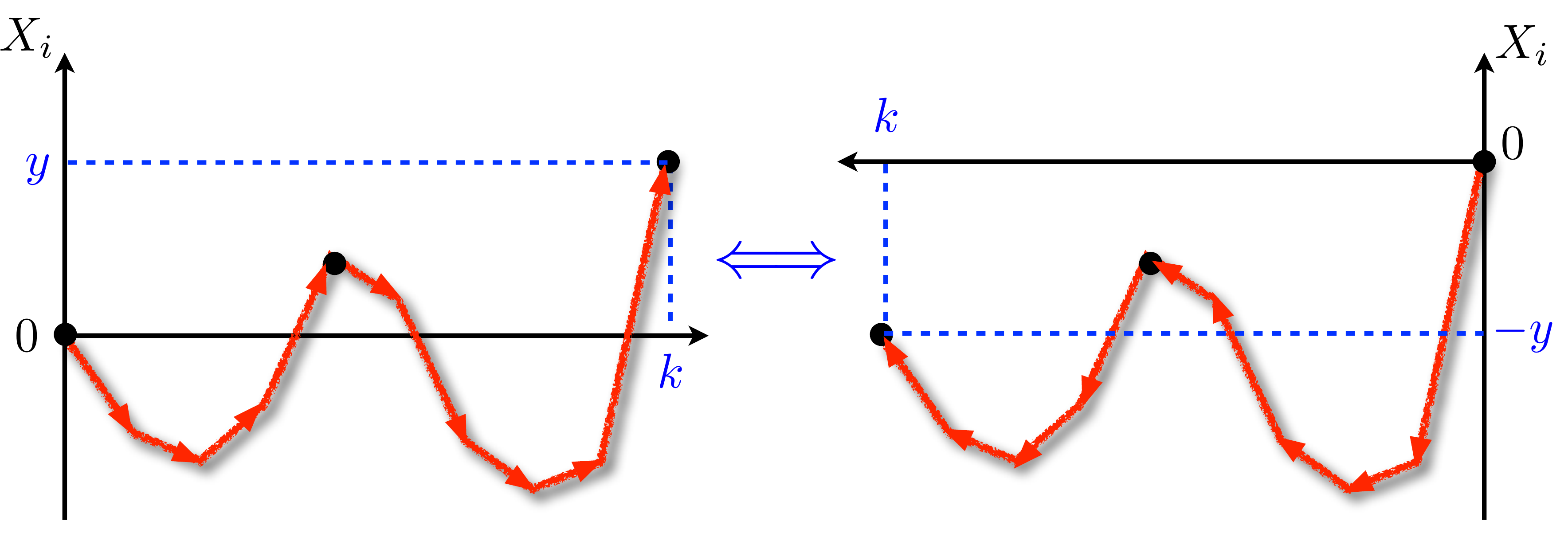}
\caption{Schematic proof of the fact that the record rate $r_k$ is (for a symmetric random walk) precisely the survival probability $q(k)$, defined in (\ref{ql_def}).}\label{fig:record_survival}
\end{figure}

We conclude this subsection by noting that the mean record number $\langle M \rangle$ can easily be computed following the rationale presented in the i.i.d.~case (\ref{def_sigma})--(\ref{exact_mean_iid}), and using the Sparre Andersen theorem.
\begin{figure}
\centering
\includegraphics[width = 0.8\linewidth]{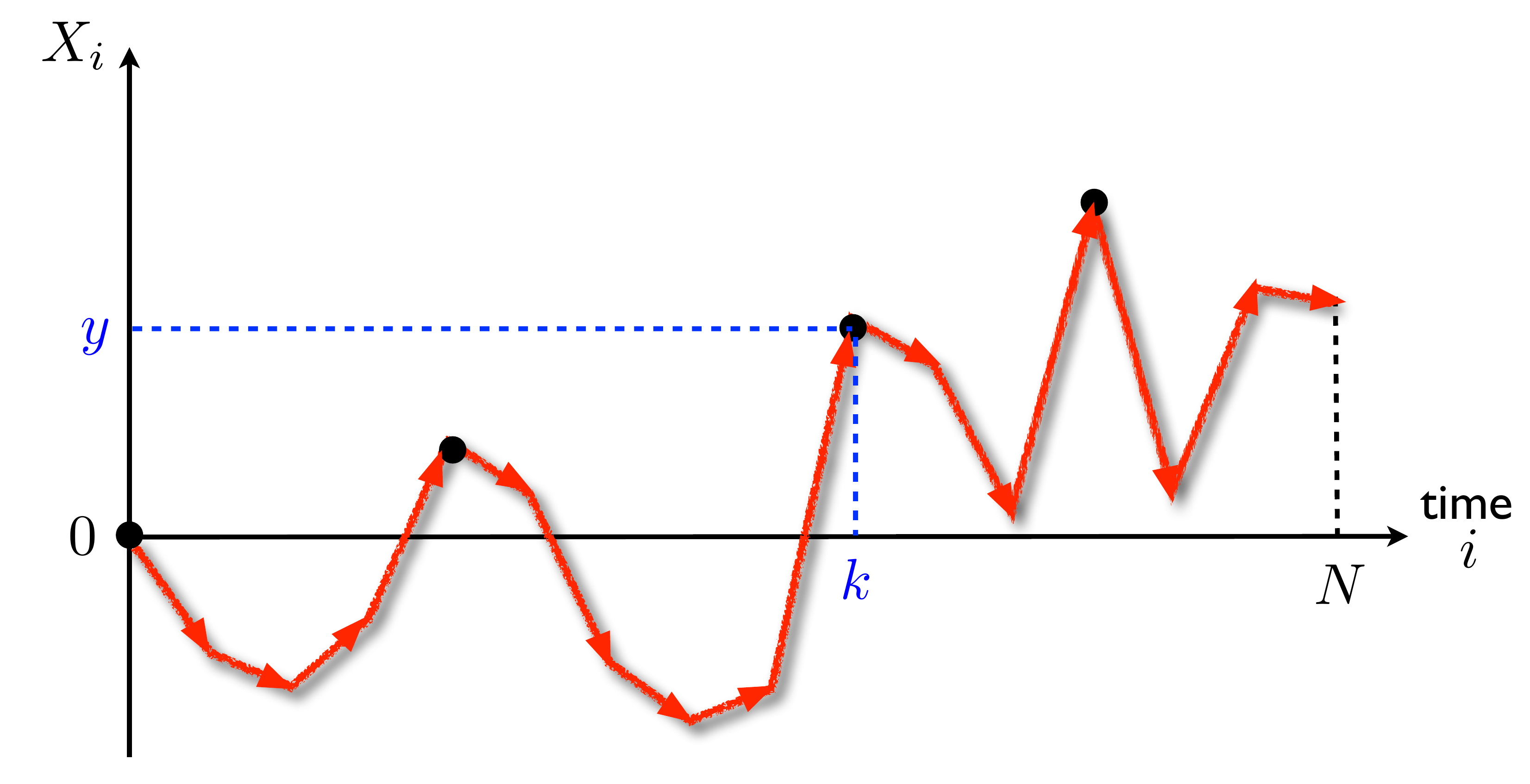}
\caption{Realization of a random walk for which a record is broken at step $k$, at which the record has value $y$. The rate $r_k$ at which a record is broken at step $k$ is obviously independent of the piece of the trajectory of the random walk after step $k$.}\label{fig:record_rate1}
\end{figure}
Indeed, as in the i.i.d.~case, the mean record number can be computed as
\beq\label{average_rw_simple}
\langle M \rangle = \sum_{k=0}^N r_k \;,
\eeq 
where $r_k$ is the record rate at step $k$, i.e., the probability that a record occurs at step $k$, as in figure \ref{fig:record_rate1}. To compute the probability
of such an event, we isolate the first $k$ steps of the trajectories (as $r_k$ does not depend on the positions of the random walker $X_i$ with $i>k$), as in figure \ref{fig:record_survival}, and we denote by $y>0$ the actual value of the record at step $k$. 
Next, we choose as a new origin of the random walk the last point, with coordinates $(k,y)$, and we change the direction of the time axis (see figure \ref{fig:record_survival}). 
In this new frame, we see that the event depicted in figure \ref{fig:record_survival} contributes to the probability that the random walk starts from the origin and arrives at $-y<0$ after $k$ time steps, staying negative in between. 
By integrating over the final position $y$, one obtains that the rate $r_k$ is precisely identical to the survival probability $q(k)$ as defined in (\ref{ql_def}). Thus, from (\ref{average_rw_simple}), and using the Sparre Andersen theorem (\ref{ql.1}), one obtains
\beq
\langle M \rangle = \sum_{k=0}^N q(k) = \sum_{k=0}^N \frac{1}{2^{2k}} {2k \choose k} = (2N+1) {2N \choose N} 2^{-2N} \;,
\eeq
recovering the result obtained in (\ref{avg_rec.1}) in a different manner. We will see that this way of computing the average number of records (\ref{average_rw_simple}) can be generalised to the case of a random walk bridge (see section \ref{sec:bridge}) and multiple random walks (see section \ref{sec:multi}). 

\vskip 0.5cm

\subsubsection{Symmetric random walk on a lattice.}

This case corresponds to the random walk sequence with a non-continuous
jump distribution, $\phi(\eta)= [\delta(\eta-1)+\delta(\eta+1)]/2$.
Since the walker starts at $X_0=0$, the walk stays on the one-dimensional lattice
with unit lattice spacing. For such a sequence $\{X_1,X_2,\ldots,X_N\}$, 
there are evidently a lot of degeneracies. We will count an entry $X_k$
as a record if it is strictly bigger than all previous entries, i.e., 
if $X_k>{\rm max}\left \{X_1,X_2,\ldots,X_{k-1}\right\}$. The general
renewal result for the record number distribution in (\ref{renewal.genf}) still holds for this case, but ${\tilde q}(z)$
is no longer given by the simple form ${\tilde q}(z)=1/\sqrt{1-z}$ 
that is only valid for symmetric and continuous jump distributions.
However for the lattice walk ${\tilde q}(z)$ can be computed
explicitly either by standard generating function techniques \cite{Feller} 
or via the 
Sparre Andersen 
theorem in (\ref{SA.1}). We illustrate below both methods for 
completeness.

To compute $q(\ell)$ for a lattice random walk starting at the origin, it is 
convenient first to consider $Q(X,\ell)$, which denotes the probability that
starting at $X$, the walker does not go to the negative side (but can come 
back to the origin) up to step $\ell$. Clearly, $q(\ell)= Q(0,\ell)$. It is easy to
see that $Q(X,\ell)$ satisfies a backward recurrence 
equation \cite{Bray_review,Feller}, for $\ell \geq 1$
\begin{equation}
Q(X,\ell) = \frac{1}{2}\left[Q(X+1,\ell-1) + Q(X-1,\ell-1)\right],\,\, X\ge 0,
\label{lw_recur.1}
\end{equation}
with the boundary condition $Q(-1,\ell)=0$ and $Q(\infty,\ell)$ non-divergent for all $\ell \geq 1$, and 
the initial condition 
$Q(X,0)=1$ for all $X\ge 0$. The generating function ${\tilde Q}(X,z)=
\sum_{\ell\ge0} Q(X,\ell)\, z^\ell$ then satisfies the recursion relation
\begin{equation}
{\tilde Q}(X,z)= 1+ \frac{z}{2}\left[{\tilde Q}(X+1,z)+ {\tilde 
Q}(X-1,z)\right], \,\, X\ge 0 \;.
\label{lw_recur.2}
\end{equation}
This linear recursion relation can be trivially solved for the appropriate
boundary conditions given above, yielding, for all $X\ge 0$
\begin{equation}
{\tilde Q}(X,z)= \frac{1}{1-z}\,\left[1- [\lambda(z)]^{X+1}\right] \;,\,\, 
{\rm where}\,\, \lambda(z)= \frac{1}{z}\left[1-\sqrt{1-z^2}\right]\, .
\label{lw_recur.sol1}
\end{equation}
In particular, we get
\begin{equation}
{\tilde q}(z)= \sum_{\ell\ge0} q(\ell)\, z^\ell= {\tilde 
Q}(0,z)=\frac{1-\lambda(z)}{1-z}= 
\frac{\sqrt{1+z}-\sqrt{1-z}}{z\sqrt{1-z}}\, .
\label{lw_recur.sol2}
\end{equation}
In particular, when $z \to 1$, $\tilde q(z) \approx \sqrt{2}/\sqrt{1-z}$, yielding 
\bea
q(\ell) \approx \frac{\sqrt{2}}{\sqrt{\pi \ell}} \;, \; \ell \to \infty \label{qrw_largel} \;,
\eea
which differs by a factor $\sqrt{2}$ from the result in (\ref{ql.1}) obtained for continuous jump distributions. 

It is amusing to see how the same result in (\ref{lw_recur.sol2}) can 
also be derived from the
Sparre Andersen theorem in (\ref{SA.1}). For this, we need to
compute $p_{-}(\ell)= {\rm Prob}(X_\ell\le 0)$. Note that
for lattice walks, ${\rm Prob}(X_\ell\le 0)= {\rm Prob}(X_\ell<0) + {\rm 
Prob}(X_\ell=0)$. Using the symmetry ${\rm Prob}(X_\ell>0)={\rm Prob}(X_\ell<0)$
and the fact that the total probability adds up to unity, we have
$2\, {\rm Prob}(X_\ell<0)+ {\rm Prob}(X_\ell=0)=1$. Hence, we get
\begin{equation}
p_{-}(\ell)= \frac{1}{2}\left[1+ {\rm Prob}(X_\ell=0)\right]\,.
\label{lw_SA.1}
\end{equation}
But the $\GF$ of the return probability to the origin can be trivially 
computed \cite{Feller}
\begin{equation}
\sum_{\ell\ge0} {\rm Prob}(X_\ell=0)\, z^\ell= \frac{1}{\sqrt{1-z^2}}\, .
\label{lw_SA.2}
\end{equation}
Using this result on the right hand side of (\ref{SA.1}) and a few steps of 
straightforward algebra gives us the desired result in (\ref{lw_recur.sol2}). 

Substituting the exact ${\tilde q}(z)$ from (\ref{lw_recur.sol2}) 
in (\ref{renewal.genf}) then
gives us the exact $P(M|N)$ for lattice walks. One can also compute
all the moments of $M$ exactly. For instance, substituting
${\tilde q}(z)$ in (\ref{avg_genf.1}), we get \cite{MZ2008}
\begin{equation}
\sum_{N\ge0} {\langle M \rangle} z^N = 
\frac{\sqrt{1+z}+\sqrt{1-z}}{2(1-z)^{3/2}},
\end{equation}
which, when inverted, gives \cite{MZ2008}
\begin{equation}
{\langle M \rangle} =\frac{1}{2}\left[1+ \frac{ (-1)^{N+1}
\Gamma(N-\frac{1}{2}) _2F_1(\frac{3}{2},-N,\frac{3}{2}-N,-1)}{2\sqrt{\pi}\Gamma(N+1)}\right],
\label{discrete}
\end{equation}
where $_2F_1(a,b,c,z)$ is the standard hypergeometric function.
For $N=0$, $1$, $2$, $3$, $4$, one gets
${\langle M \rangle} = 1, 3/2, 7/4, 2, 35/16$ respectively. In particular, for large $N$, one has
\beq\label{avg_rec_discrete}
{\langle M \rangle} \approx \sqrt{\frac{2}{\pi}} \sqrt{N}\;,
\eeq
which is smaller by a factor $1/\sqrt{2}$ than the expression for the mean number of records in 
the continuous case given in (\ref{avg_rec.2}).
 
The full distribution $P(M|N)$ of the record number $M$ can also be obtained using the general formula in (\ref{renewal.genf}) and the appropriate result for the survival probability for the discrete random walk in (\ref{lw_recur.sol2}). However, this distribution can be obtained more directly for the lattice random walk by noticing the connection between the number of records and the maximal displacement of the random walk $X_{\max,N}$ up to step $N$, i.e., $X_{\max,N} = \max \left \{X_0, X_1, \dots, X_N\right\}$. This relation reads \cite{WMS2012}
 \beq\label{eq:record_max}
 M = X_{\max,N} + 1 \;.
 \eeq
To derive this relation (\ref{eq:record_max}) it is useful to consider the time evolution of the two processes $M$ and $X_{\max,N}$ as $N$ increases. At the next time step $N+1$, if a new site on the positive axis is visited for the first time, the process $X_{\max,N}$ increases by $1$, otherwise its value remains unchanged. On the other hand, when this event happens, then the record number $M$ is also increased by one, and otherwise it remains unchanged. Therefore we see that the two processes are locked with each other at all steps.
Since, by convention, the first position is a record implying that initially $M=1$, while $X_{\max,0} = X_0 = 0$, one obtains immediately the relation in (\ref{eq:record_max}). From this relation it is possible to obtain the statistics of $M$ from the one of $X_{\max,N}$, which can be computed easily, e.g., using the method of images. This yields \cite{WMS2012}, for $1\leq M \leq N+1$
\begin{eqnarray}\label{exact_Bernoulli_N1}
P(M|N) = \frac{1}{2^N} {N \choose \lceil \frac{N+M-1}{2}\rceil } \;,
\end{eqnarray} 
where $\lceil x \rceil$ denotes the smallest integer not less than $x$. One can check that this exact formula for the distribution (\ref{exact_Bernoulli_N1}) yields back the result in (\ref{discrete}) for the first moment. In addition, in the large $N$ limit the probability distribution $P(M|N)$ takes the scaling form
\begin{eqnarray}\label{eq:scaling_lattice_RW}
P(M|N) \approx \sqrt{\frac{{2}}{{N}}} \; g\left(\frac{\sqrt{2}\;M}{\sqrt{N}} \right) \;, \,\, 
{\rm where}\,\, g(x)= \frac{1}{\sqrt{\pi}}\, \e^{-x^2/4}\, \Theta(x) \;,
\end{eqnarray} 
which is similar, up to a factor $\sqrt{2}$ as already noticed below (\ref{discrete}), to the result obtained for continuous jump distributions in (\ref{scaling_record_dist.1}). This scaling form (\ref{eq:scaling_lattice_RW}) can be understood by reminding that the lattice random walk properly scaled, $X_{\lceil \tau N\rceil}/\sqrt{N}$ with $\tau \in [0,1]$, converges for large $N$ to the standard Brownian motion $x(\tau)$ with diffusion coefficient $D=1/2$ on the unit time interval $0 \leq \tau \leq 1$. Hence one expects from the identity in (\ref{eq:record_max}) that $M/\sqrt{N}$ converges for large $N$ to the maximum of the Brownian motion on the unit time interval, which is indeed given by the half-Gaussian in (\ref{eq:scaling_lattice_RW}). As we will see later, this identity (\ref{eq:record_max}) can be used to compute the record statistics of constrained random walks, like random walk bridges (see section \ref{sec:bridge}) or the one of multiple random walks (see section \ref{sec:multi}).

\vskip 0.5cm

\subsubsection{Random walk in the presence of a constant drift.}

In this subsection we will study the record statistics for
a sequence $\{X_0=0, X_1, X_2,\dots, X_N\}$
where $X_i$ denotes the position of a one-dimensional random
walker at step $i$ (discrete time and continuous space), in the presence
of a constant drift $c$. The position $X_i$ evolves in time $i$ via
the Markov rule (starting from $X_0=0$)
\begin{equation}
X_i= X_{i-1} + c+ \eta_i,
\label{evol_drift.1}
\end{equation}
where $\eta_i$ are i.i.d.~jump variables as before.

To keep the discussion simple, we will restrict ourselves to
the case of a symmetric and continuous jump distribution $\phi(\eta)$.
We will see later that in the presence of a nonzero drift $c$, the 
asymptotic tail
of the symmetric jump distribution $\phi(\eta)$ for large $|\eta|$ plays
a rather crucial role. These tails can be nicely characterized in terms
of the Fourier transform ${\hat \phi}(q)= \int_{-\infty}^{\infty} 
\phi(\eta)\, \e^{iq \eta}\, d\eta$ of the jump distribution. We will focus below
on a large class of jump distributions whose Fourier transform
has the following small $k$ behaviour
\begin{equation}
{\hat \phi}(q)= 1- (l_\mu\,|q|)^{\mu}+\ldots
\label{smallk.1}
\end{equation}
where $0< \mu\le 2$ and $l_\mu$ represents a typical length scale 
associated
with the jump. The exponent $0<\mu\le 2$ dictates
the large $|\eta|$ tail of $\phi(\eta)$. For jump densities with a finite
second moment $\sigma^2= \int_{-\infty}^{\infty} \eta^2\, \phi(\eta)\,d\eta$,
such as Gaussian, exponential, uniform etc.,
one evidently has $\mu=2$ and $l_2=\sigma/\sqrt{2}$. In contrast, 
$0<\mu<2$
corresponds to jump densities with fat tails
$\phi(\eta)\sim |\eta|^{-1-\mu}$ as $|\eta|\to \infty$.
A typical example is ${\hat \phi}(q)=\exp[-|q|^\mu]$ where $\mu=2$
corresponds to the Gaussian jump distribution while $0<\mu<2$ corresponds 
to L\'evy flights (for reviews on these jump processes see 
\cite{BG1990,MK2000}). 

In this subsection, we are interested in computing the statistics of the 
record number $M$ for the sequence in (\ref{evol_drift.1}), for
a nonzero constant drift $c$. In fact, this problem was first studied
in ref \cite{PLDW2009} for the special case of
Cauchy jump distribution $\phi_{\rm Cauchy}(\eta)=1/[\pi (1+\eta^2)]$
[which belongs to the $\mu=1$ family of jump densities in (\ref{smallk.1})]. 
By using the renewal approach mentioned above,
it was found that the mean number of records in 
this Cauchy case grows
asymptotically for large $N$ as \cite{PLDW2009}
\begin{equation}
{\langle M\rangle} \approx 
\frac{1}{\Gamma(1+\theta(c))}\, N^{\theta(c)},\quad {\rm
where}\quad \theta(c)= \frac{1}{2}+\frac{1}{\pi}\,\arctan(c)\,.
\label{cauchy_mean.1}
\end{equation}
In addition, the asymptotic record number distribution $P(M|N)$ for large 
$N$ was
found~\cite{PLDW2009} to have a scaling
distribution, $P(M|N)\approx N^{-\theta(c)}\, g_c\left(M\,
N^{-\theta(c)}\right)$ with a nontrivial scaling function $g_c(x)$ which 
reduces, for $c=0$, to the
half-Gaussian in~(\ref{scaling_record_dist.1}).
 
For jump densities with a finite second moment $\sigma^2$ and in the presence 
of a nonzero positive drift $c>0$, the mean number of records ${\langle 
M\rangle}$ was analysed in ref \cite{WBK2011} and was found to grow 
linearly with $N$ for large $N$, ${\langle M\rangle} \approx a_2(c)\, N$,
where the prefactor $a_2(c)$ was computed approximately for the Gaussian 
jump distribution. However, an exact expression of the prefactor for 
arbitrary jump densities with a finite $\sigma^2$ was still missing. These 
results for the mean record number were then compared to the stock prices 
data from the Standard and Poors 500 \cite{WBK2011}.

Finally, in ref \cite{MSW2012}, the full distribution of the record 
number $P(M|N)$ for large $N$ was analysed in detail for the whole family 
of continuous and symmetric jump distributions with Fourier transforms as 
in (\ref{smallk.1}), for all $0<\mu\le 2$ and all $c$.
An extremely rich behaviour for the record statistics was 
found \cite{MSW2012}
for varying $0<\mu\le 2$ and $c$ (see below).
Here we just summarise the main steps behind this analysis and
the main results (for details we refer the reader to 
ref \cite{MSW2012}). There are three main steps for the computation 
of record statistics that are described as follows.

\begin{itemize}

\item To use the general renewal approach outlined in the previous 
subsection, which is valid for arbitrary $c$. The only requirement is
the knowledge of the persistence probability $q(\ell)$.

\item The persistence probability $q(\ell)$ can be estimated from the 
Sparre Andersen identity in (\ref{SA.1}), which is also valid for 
arbitrary $c$. One needs to just evaluate the local quantity
$p_{-}(n)={\rm Prob}(X_n\le 0)$. For this one needs to know the
probability distribution $P(X_n)$ at step $n$ of the walker evolving via 
 (\ref{evol_drift.1}). In fact, for the asymptotic analysis of record 
number, it suffices to know the behaviour of $q(\ell)$ for large $\ell$, which
in turn requires the knowledge of $P(X_n)$ for large $n$. This latter
quantity has been well studied in the literature and one has a rather
complete knowledge of this distribution \cite{BG1990,MK2000}. Using
this, one can estimate $p_{-}(n)$ and hence $q(\ell)$ via the Sparre Andersen
identity. This was carried out in detail in ref \cite{MSW2012} for
all $0<\mu\le 2$ and all $c$. A summary is provided in the table \ref{tab:cmu} below.

\item Once $q(\ell)$ is known for large $\ell$, one can then use the renewal
results in equations (\ref{avg_genf.1}) and (\ref{renewal.genf}) to
estimate respectively the mean number of records ${\langle 
M\rangle}$
and the record number distribution $P(M|N)$, asymptotically for large $N$ \cite{MSW2012}.
 
\end{itemize}

\begin{figure}
\centering
\includegraphics[width=0.8\textwidth]{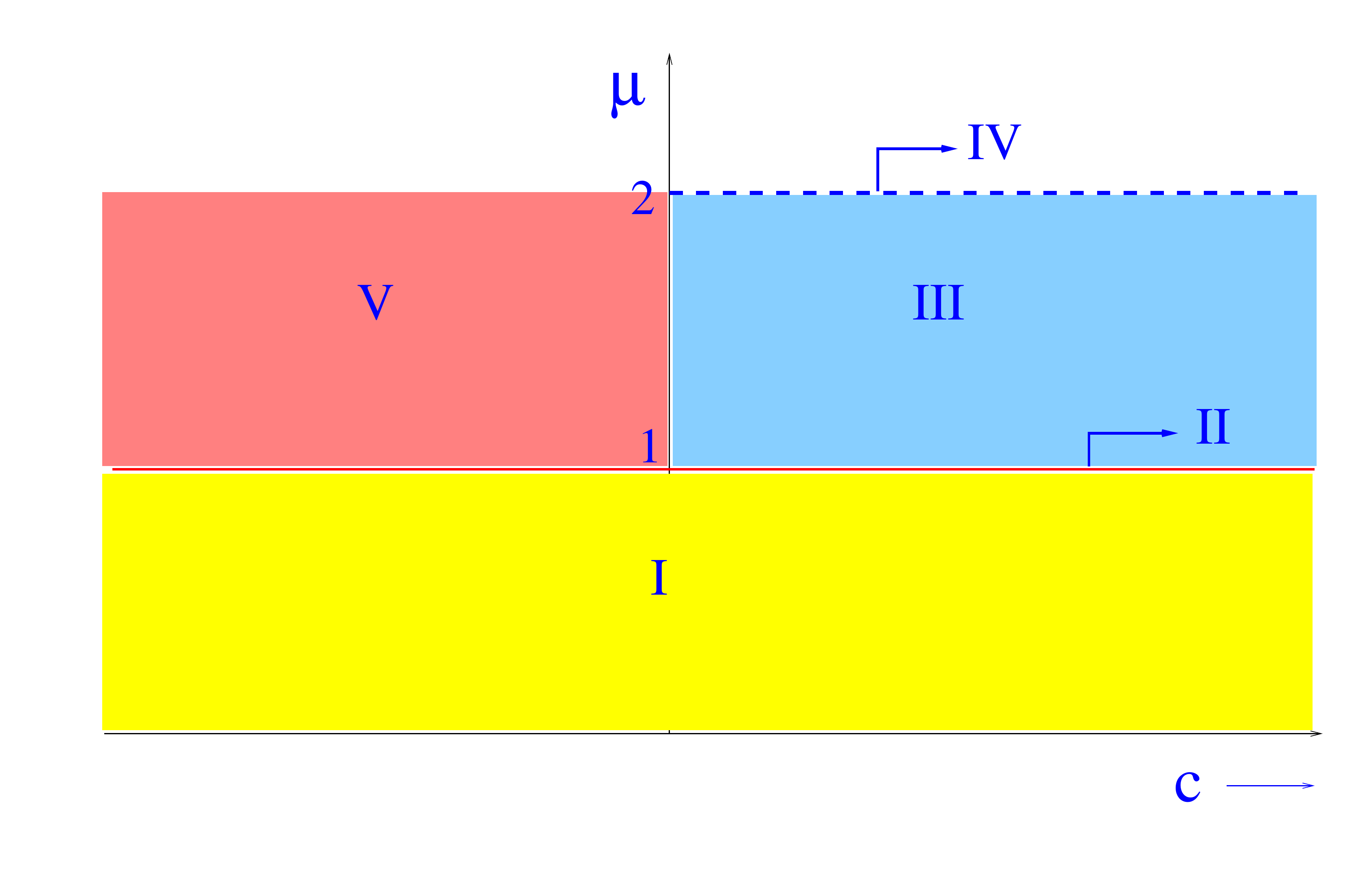}
\caption{Phase diagram in the $(c,0<\mu\le 2)$ strip
depicting $5$ regimes: (I) $0<\mu<1$ and $c$ arbitrary (II) the line
$\mu=1$ and $c$ arbitrary (III) $1<\mu<2$ and $c>0$ (IV) the semi-infinite 
line
$\mu=2$ and $c>0$ and (V) $1<\mu\le 2$ and $c<0$.
The line $\mu=1$ (regime II above) is a critical line on which the 
persistence and the record
statistics exhibits
marginal behaviour.
} 
\label{fig.phd}
\end{figure}

As mentioned above, both the persistence $q(\ell)$ and the mean record 
number ${\langle M \rangle}$ (as well as the distribution $P(M|N)$) 
display a rather rich and varied 
behaviour as functions of $\mu$ and $c$ \cite{MSW2012}. 
On the strip $(c,0<\mu\le 2)$ (see figure \ref{fig.phd}), it turns out
that there are five 
distinct
regimes: (I) when $0<\mu<1$ with $c$ arbitrary
(II) when $\mu=1$ and $c$ arbitrary (III) when $1<\mu<2$ and $c>0$
(IV) when $\mu=2$ and $c>0$ and (V)
when $1<\mu\le 2$ and $c<0$. 

In each of these five regimes 
the persistence $q(\ell)$ and the mean record number
${\langle M \rangle}$ have different asymptotic
dependence on the sequence size $N$ for large $N$ (see table 1).
Consequently, $P(M|N)$
also displays different scaling distributions in the five phases. 
The line $\mu=1$ (regime II above) is a critical line on which both
the persistence and the mean record number 
exhibits
marginal behaviour, in the sense that the exponents characterizing the
asymptotic behaviours of these quantities depend continuously on the
drift $c$.

Without giving further details, we just summarise in table~\ref{tab:cmu} the
asymptotic behaviour of $q(\ell)$ and $\langle M\rangle$
in these five regimes in the strip $(c, 0<\mu\le 2)$ in figure \ref{fig.phd}.
Let us make few remarks concerning the asymptotic results presented in 
table 1. The prefactors $B$ for the persistence $q(\ell)$ in the second 
column of table 1 can be computed exactly \cite{MSW2012}. In the marginal
case (regime II, i.e., along the line $\mu=1$ in figure \ref{fig.phd}),
the persistence exponent $\theta(c)= 
\frac{1}{2}+\frac{1}{\pi}\,\arctan(c)$ was first computed in 
ref \cite{BGL1999}. The constant $\alpha_\mu(c)$ was computed
in ref \cite{MSW2012}. Finally, the constant prefactors appearing
in the asymptotic expressions for ${\langle M\rangle}$ in the 
third column of table 1 are also explicitly computable \cite{MSW2012}. 
Finally, the full scaling forms of the record number distribution $P(M|N)$
for large $N$ are also computed explicitly in ref \cite{MSW2012} in all
the five regimes. Note that the results obtained in the region IV (corresponding to $\mu=2$ and arbitrary $c$) were
extended in ref \cite{GLS2015} to a wider class of random walks with stationary correlated jumps (and no assumption on the symmetry and/or continuity of the jump distributions). 

\begin{table}[h]
\begin{center}
\begin{tabular}{|c||c|c|}
\hline
${\rm regimes\,\, in\,\, figure \;} {\ref{fig.phd}}$&$q(\ell)$&${\langle 
M\rangle}$\\
\hline
I&$\approx B_I\, \ell^{-1/2}$&$\approx A_I\, \sqrt{N}$\\
II&$\approx B_{II}\, \ell^{-\theta(c)}$&$\approx A_{II}\, N^{\theta(c)}$\\
III&$\approx B_{III}\, \ell^{-\mu}$&$ a_\mu(c)\, N$\\
IV& $\approx B_{IV}\, \ell^{-3/2}\,\exp\left[-(c^2/2\sigma^2)\, \ell\right] 
$&$\approx a_2(c)\, N $\\
V&$\approx \alpha_\mu(c) $&$ {\rm const.} $\\
\hline
\end{tabular} 
\end{center} 
\caption{Asymptotic results for the persistence $q(\ell)$ for large $\ell$ and
the mean record number ${\langle M\rangle}$ in the five regimes
in the $(c,0<\mu\le 2)$ strip in figure \ref{fig.phd}.}\label{tab:cmu}
\end{table}

We conclude this section by mentioning that these results for the record statistics of random walk with a drift were used in the context of finance, in \cite{Cha15a,Cha15b}, to demonstrate that records provide a useful unbiased estimators of the so-called Sharpe ratios -- which characterize the signal-to-noise ratio of a financial time series, like the one generated by the time evolution of a price return. We refer the reader to \cite{Cha15a, Cha15b} for more detail on this question as well as to \cite{Cha_url} for an implementation (an R-package) of this estimator.

\subsubsection{Continuous-time random walk.}

Another model
where the general renewal approach outlined above can be exploited
to compute exactly the record number statistics \cite{Sanjib2011} is the 
so called
continuous-time random walk model, introduced by Montroll and Weiss \cite{MW1965}.
In the continuous-time random walk model, both space and time are
continuous. The walker 
moves 
on a continuous line by successive jumps as before, 
with jump lengths drawn independently from the distribution $\phi(\eta)$.
However, between
two jumps, the walker waits for a random amount of time $\tau$ drawn, 
independently
for each jump instance, from a waiting time distribution $\Psi(\tau)$. 
One considers waiting time distributions with a power law tail
$\Psi(\tau) \sim \tau^{-1-\gamma}$ for large $\tau$. If $\gamma>1$, the
mean waiting time is finite and in this case the walker essentially 
behaves like a discrete-time random walk as discussed earlier. However,
interesting new behaviour emerges when the mean waiting time is divergent,
i.e., in the case when $0<\gamma\le 1$ (see the reviews 
\cite{BG1990} and \cite{MK2000} for
detailed discussions). In this case, the Laplace transform
of the waiting time distribution behaves as
\begin{equation}
{\tilde \Psi}(s)= \int_0^{\infty} \e^{-s\,\tau}\, \Psi(\tau)\, d\tau 
\approx 
1- 
(\tau_0\, s)^{\gamma}+\cdots \quad {\rm as}\quad s\to 0 \;,
\label{wtdist.1}
\end{equation}
where $\tau_0$ is a microscopic time scale.

To compute the record statistics, one can again use the general renewal 
approach outlined before, except that now one considers a continuous-time
analogue of (\ref{renewal.1}) that reads
\begin{equation}
P_c(\vec \ell, M|t)= f_c(\ell_1)f_c(\ell_2)\ldots f_c(\ell_{M-1})\, 
q_c(\ell_M)\, \delta\left(\sum_{k=1}^M \ell_k-t\right)\; ,
\label{ctrw_renewal}
\end{equation}
where the subscript $c$ stands for the {\em continuous} time.
Here, $\vec \ell\equiv \{\ell_1,\ell_2,\ldots,\ell_M\}$ denotes
the collection of ages of records and $P_c(\vec \ell, M|t)$ is
the joint distribution of the ages and the number $M$ of records in time 
$t$. Note that in this expression (\ref{ctrw_renewal}), the variables $\ell_k$ (as well as $t$) are continuous
and, consequently, the $\delta$ function is a Dirac delta function and no longer a Kronecker delta as for the discrete-time random walk~(\ref{renewal.1}).

The function $q_c(\ell)$ denotes the probability
that the walker stays below $0$ up to time $\ell$.
Similarly, $f_c(\ell)= - dq_c(\ell)/d\ell$ denotes
the first-passage probability density, i.e., $f_c(\ell)d\ell$
denotes the probability that the process, starting at the origin,
crosses to the positive side for the first time in the time interval
$[\ell,\ell+d\ell]$. Taking the Laplace transform of (\ref{ctrw_renewal}) with respect to $t$ and 
integrating over $\ell_k$ gives
\begin{equation}
\int_0^{\infty} dt\, \e^{-s\,t}\, P_c(M|t) = \left[{\tilde 
f}_c(s)\right]^{M-1}\, {\tilde q_c}(s)= \left[{\tilde
f}_c(s)\right]^{M-1}\, \frac{(1- {\tilde f}_c(s))}{s}\; ,
\label{ctrw_renewal.2}
\end{equation}
where ${\tilde f}_c(s)= \int_0^{\infty} d\ell\, \e^{-s\,\ell}\, f_c(\ell)$. 
In deriving the last equality in (\ref{ctrw_renewal.2}) we have used 
${\tilde q}_c(s)= (1-{\tilde 
f}_c(s))/s$ which follows by taking the Laplace transform of the relation 
$f_c(\ell)= - 
dq_c(\ell)/d\ell$. In (\ref{ctrw_renewal.2}), $P_c(M|t)$ is just
the probability of having $M$ records in time $t$. Hence (\ref{ctrw_renewal.2}) is the exact
continuous-time analogue of (\ref{renewal.genf}) derived earlier.

To make further progress, we need to determine ${\tilde f}_c(s)$ in
terms of the waiting time distribution $\Psi(\tau)$. This can be
easily done as follows. Consider a time interval $t$ between
two successive zero crossings that contains
exactly $n$ jump events. 
For fixed $t$, clearly the number of possible steps $n$ is a random
variable. Its distribution $p_n(t)$ can be easily computed
using the fact that successive waiting time intervals are
statistically independent, i.e., 
\begin{equation}
p_n(t) = \int_0^{\infty}d\tau_1\int_0^{\infty} d\tau_2\ldots 
\int_0^{\infty} d\tau_n \Psi(\tau_1)\Psi(\tau_2)\ldots \Psi(\tau_n)\, 
\delta\left(t- \sum_{i=1}^n \tau_i\right)\, .
\label{ctrw_pnt.1}
\end{equation}
Taking Laplace transform with respect to $t$ gives
\begin{equation}
{\tilde p}_n (s)= \left[{\tilde \Psi}(s)\right]^n\; .
\label{ctrw_pnt_laplace.1}
\end{equation}
Using $p_n(\tau)$, one observes immediately that
\begin{equation}
f_c(\tau)= \sum_{n\ge1} f(n) p_n(\tau),
\label{ctrw_fct.1}
\end{equation}
where $f(n)$ is precisely the first-passage probability in discrete step 
$n$, defined before in (\ref{fl_def}). Taking Laplace transform of
 (\ref{ctrw_fct.1}) and using (\ref{ctrw_pnt_laplace.1}) then gives
\begin{equation}
{\tilde f}_c(s)= \sum_{n\ge1} f(n)\, \left[{\tilde 
\Psi}(s)\right]^n = {\tilde f}(z= {\tilde \Psi}(s)),
\label{ctrw_fct.2}
\end{equation}
where ${\tilde f}(z)= \sum_{n\ge1} f(n)\, z^n$ is the generating
function of the first-passage probability of the discrete-time random 
walk. For example, for symmetric and continuous jump distribution
$\phi(\eta)$, we have ${\tilde f}(z)= 1-\sqrt{1-z}$ from (\ref{fz_qz}). 
Hence, in this case, plugging (\ref{ctrw_fct.2})
in (\ref{ctrw_renewal.2}) gives the following main 
result \cite{Sanjib2011}
\begin{equation}
\int_0^{\infty} dt\, \e^{-s\,t}\, P_c(M|t) =
\frac{\sqrt{1-{\tilde \Psi}(s)}}{s}\, \left[1- \sqrt{1-{\tilde 
\Psi}(s)}\right]^{M-1} \; .
\label{ctrw_record.1}
\end{equation}

While the Laplace transform in (\ref{ctrw_record.1}) is not easy to 
invert for arbitrary $t$, one can make progress in the scaling limit
for large $M$, large $t$ but keeping the product $M\, 
(t/\tau_0)^{-\gamma/2}$ fixed. In this limit, using the small
$ $ behaviour of ${\tilde \Psi}(s)$ in (\ref{wtdist.1}) one obtains
a limiting scaling distribution \cite{Sanjib2011}
\begin{equation}
P_c(M|t) \approx \left(\frac{t}{\tau_0}\right)^{-\gamma/2}\, 
g_{\gamma}\left(M\, \left(\frac{t}{\tau_0}\right)^{-\gamma/2} \right)\, ,
\label{ctrw_record.2}
\end{equation}
where the scaling function $g_\gamma(x)$ is given by
\begin{equation}
g_{\gamma}(x)= \frac{2}{\gamma}\, x^{-1+2/\gamma}\, 
L_{\gamma/2}(x^{-2/\gamma})\;, \quad 0< \gamma\le 1 \;.
\label{ctrw_record.3}
\end{equation}
The function $L_{\mu}(x)$ is the standard one-sided L\'evy stable density.
Note that for $\gamma=1$, one can show \cite{Sanjib2011} that the result 
in 
 (\ref{ctrw_record.2}) reduces to the half-Gaussian result in (\ref{scaling_record_dist.1}), as one would expect.

Thus, to summarise, for the continuous-time random walk with waiting time distribution 
$\Psi(\tau)$
and jump length distribution $\phi(\eta)$, the distribution $P_c(M|t)$
of the record number $M$ in time $t$ is independent of the jump 
distribution $\phi(\eta)$ (for symmetric and continuous $\phi(\eta)$),
but does depend on the waiting time distribution $\Psi(\tau)$.
For power-law waiting time distribution, $\Psi(\tau)\sim \tau^{-1-\gamma}$
as $\tau\to \infty$ with divergent mean, i.e., $0<\gamma\le 1$, 
$P_c(M|t)$
has a scaling form as in (\ref{ctrw_record.2}) and the typical
number of records grows with time as, $M\sim t^{\gamma/2}$ for large $t$.
In the borderline case $\gamma=1$, one recovers the discrete-time result
discussed earlier.


\subsection{Statistics of the ages of records for random walk models}\label{sec:age_rw}

Apart from the number of records, other interesting observables are the ages
of the records of a random walk sequence.
As defined in the introduction of section \ref{section:RW},
the age $\ell_k$ of the $k$-th record is the number of
steps between the $k$-th and ($k + 1$)-th records, i.e.,
the time up to which the $k$-th record survives (see figure \ref{fig.rw}). Note that the last record is still
a record at step $N$ and hence the last age $\ell_M$ is not on the same footing as the other ones. 

Thanks to the (spatial) translational invariance of the random walk (see e.g., (\ref{ql_def})),
the sets of the ages $\ell_k$ behave similarly to the intervals between
two consecutive zeros of a lattice random walk -- in other words to the lengths of the
{\it excursions}. 
Hence, as we will see below, the study of the ages of the records for a random walk
bears strong similarities with the excursion theory of the lattice random walk and Brownian motion.

As we discuss it in this section, 
the full statistics of the ages can be obtained from the renewal theory 
presented in section \ref{sec:renewal}, see (\ref{renewal.1}). A first rough and naive inspection of the
joint distribution of the ages in (\ref{renewal.1}) suggests that these ages are essentially independent (assuming
for the moment that the global constraint can be ignored) and also identical (except for
the last interval $\ell_M$ which is different). Therefore, if one is interested in the distribution $P(\ell_k|N)$
of the {\rm typical} age of a record, i.e., of $\ell_k$ with $k<M$, one naturally expects that 
\bea\label{age_k_rw}
P(\ell_k) = \lim_{N \to \infty} P(\ell_k|N) = f(\ell_k) \;,
\eea
where $f(\ell_k)$ is the first-passage probability (\ref{fl_def}). And this can be easily shown by an explicit calculation starting from (\ref{renewal.1}) (see e.g., \cite{GL2001}). Note that this result (\ref{age_k_rw}) holds for all $k$ (with $k < M$), which is quite different from the limiting distribution of the age of the $k$-th record for an i.i.d.~sequence in (\ref{Pell_explicit}), which depends explicitly on $k$. Furthermore, using that $f(\ell) \propto 1/\ell^{3/2}$ for large $\ell$ for a random walk (without drift), one obtains that the typical age $\ell_{\rm typ}$ behaves as $\ell_{\rm typ}= \langle \ell_k \rangle = \sum_{\ell=1}^N \ell \, f(\ell) \propto \sqrt{N}$. This behaviour can also be obtained 
by the simple following heuristic argument: given that the average number of records is $\langle M \rangle$, the typical age which is the typical time interval between two successive records is $\ell_{\rm typ} \sim N/\langle M\rangle \propto \sqrt{N}$, where we have used that 
$\langle M \rangle \propto \sqrt{N}$.

There are however rare records whose ages behave quite differently. A natural way to probe such atypical behaviours 
of the ages is to study the fluctuations of the largest $\ell_{\max,N}$ or the shortest lasting record $\ell_{\min, N}$. 
As already mentioned previously, the sequence of the ages of the records of a random walk are not all on the same footing, as the last record is still a record at step $N$ (see figure \ref{fig.rw}). 
This leads to different definitions of the longest (or shortest) age \cite{GMS2014} (see section \ref{sec:different}). 
Here we will mainly consider the somewhat simplest definition and define $\ell_{\max,N}$ and $\ell_{\min, N}$ as
\bea\label{def_lmax_lmin}
\ell_{\max,N} = \max\{\ell_1, \ell_2, \ldots, \ell_M\} \;\;\;, \;\;\; \ell_{\min,N} = \min\{\ell_1, \ell_2, \ldots, \ell_M\} \;.
\eea 

Besides, in order to characterize better the statistics of the last age $\ell_M$, following the definition introduced previously for i.i.d.~variables (see (\ref{eq:QN})),
a natural quantity to study is the probability $Q_N$ that the age of the last record is the longest one, or probability of record breaking for the sequence of ages,
\beq
Q_N=\Prob(\ell_M>\max(\ell_1,\dots,\ell_{M-1}))=\Prob(\ell_{\max,N}=\ell_M) \;.
\label{def_QN}
\eeq
It turns out that $Q_N$ is related to $\ell_{\max,N}$ as follows \cite{GMS2014,GMS2009}
\begin{eqnarray}\label{lmax_QN}
\langle \ell_{\max,N+1} \rangle = \langle \ell_{\max,N} \rangle + Q_N \;.
\end{eqnarray}
This relation (\ref{lmax_QN}) can be easily obtained if one considers the evolution of the random variable $\ell_{\max,N}$ as $N$ increases by one unit. Indeed, $\ell_{\max,N+1} = \ell_{\max,N} + 1$ if the last record is the longest one -- which by definition occurs with probability $Q_N$ (\ref{def_QN}) -- and otherwise it remains unchanged, $\ell_{\max,N+1} = \ell_{\max,N}$. Hence, on average, one obtains the relation in (\ref{lmax_QN}). 

As done in the i.i.d.~case [see above (\ref{cumulative distribution function_lmax_iid})], the cumulative distribution of $\ell_{\max,N}$, $F(\ell |N) = {\rm Prob}(\ell_{\max,N} \leq \ell)$,
is obtained by summing the joint distribution of the ages in (\ref{renewal.1}) over $\ell_k$ and $M$ such that $\ell_k \leq \ell$ for each $k$. As for the distribution of the record numbers (\ref{renewal.genf}), this summation is conveniently performed by considering the generating function of $F(\ell |N)$ with respect to $N$. It yields \cite{MZ2008}
\bea\label{cdf_F_RW}
\sum_{N \geq 0} F(\ell | N) z^N = \frac{\sum_{m=1}^\ell q(m) \, z^m}{1 - \sum_{m=1}^\ell f(m) \, z^m} \;,
\eea
where $q(m)$ and $f(m)$ are defined respectively in (\ref{ql_def}) and (\ref{fl_def}). From (\ref{cdf_F_RW}), one computes the $\GF$ of $\langle \ell_{\max,N} \rangle = \sum_{\ell \geq 1} [1 - F(\ell |N)]$ as
\bea\label{gf:lmax_av}
\sum_{N \geq 0} z^N \, \langle \ell_{\max,N} \rangle &=& \sum_{\ell \geq 0} \left[ \frac{1}{1-z} - \frac{\sum_{m=1}^{\ell} q(m) \, z^m}{1 - \sum_{m=1}^\ell f(m)\, z^m} 
 \right] \;.
\eea
 
Similarly, as done in the i.i.d.~case, one can compute the cumulative distribution function of $\ell_{\min,N}$, $G(\ell |N) = {\rm Prob}(\ell_{\min,N} \geq \ell)$ by summing the joint distribution in (\ref{renewal.1}) over $\ell_k$ and $M$, with $\ell_k \geq \ell$ for all values of $k$. Note that $\ell_{\min,N}$ as defined in (\ref{def_lmax_lmin}) takes values between 0 and $N$: indeed if there is a record at the last step, then $\ell_{\min,N} = \ell_M = 0$ and if there are no records beyond the first step, i.e., $M=1$, then $\ell_{\min,N} = \ell_1 = N$. The $\GF$ of $G(\ell |N)$ with respect to $N$ can then be obtained in a concise form \cite{MZ2008}
\bea\label{cdf_G_RW}
\sum_{N \geq 0} G(\ell | N) \, z^N = \frac{\sum_{m\geq \ell} q(m) \, z^m}{1 - \sum_{m\geq \ell} f(m) \, z^m} \;,
\eea
where we recall that $q(m)$ and $f(m)$ are given in (\ref{ql_def}) and (\ref{fl_def}) respectively. From (\ref{cdf_G_RW}), one obtains immediately the $\GF$ of the average value $\langle \ell_{\min,N} \rangle = \sum_{\ell \geq 1} G(\ell | N)$ as
\bea\label{gf:lmin_av}
\sum_{N \geq 0} \langle \ell_{\min,N} \rangle \, z^N = \sum_{\ell \geq 1} \frac{\sum_{m\geq \ell} q(m) \, z^m}{1 - \sum_{m\geq \ell} f(m) \, z^m} \;.
\eea
 
These formulae (\ref{gf:lmax_av}) and (\ref{gf:lmin_av}) show that $\langle \ell_{\max,N} \rangle$ and $\langle \ell_{\min,N} \rangle$ depend on the random walk under consideration, through $q(m)$ and $f(m)$. In particular, to obtain the large $N$ behaviour of these quantities, one needs to analyse their generating functions in equations (\ref{gf:lmax_av}) and (\ref{gf:lmin_av}) in the limit $z \to 1$. In this limit, it turns out that the discrete sum over $m$ is dominated by the large values of $m$, which thus depends on the large $m$ behaviour of the survival probability $q(m)$ (see table \ref{tab:cmu} above). Below we will discuss the behaviour for $\langle \ell_{\max,N} \rangle$ and $\langle \ell_{\min,N} \rangle$ in the large $N$ limit obtained from these general formulas (\ref{gf:lmax_av}) and (\ref{gf:lmin_av}) for a variety of random walks, with different jump distributions (continuous and discrete), both with and without drift.

\subsubsection{ Symmetric and continuous jump distribution.} 

In this case one can insert the explicit expression of $q(\ell)$ and $f(\ell)$ given respectively in equations (\ref{ql_gf}) and (\ref{fl_ql}) into (\ref{gf:lmax_av}) to obtain an exact expression for the $\GF$ of $\langle \ell_{\max,N} \rangle$, from which one can obtain in principle the exact value of $\langle \ell_{\max,N} \rangle$ for arbitrary $N$. For instance, one obtains $\langle \ell_{\max,N} \rangle = 0, 1, 3/2, 17/8, 11/4$ respectively for $N=0, 1, 2, 3, 4$ \cite{GMS2014}. The large $N$ behaviour of $\langle \ell_{\max,N} \rangle$ is obtained by analysing the behaviour of its $\GF$ (\ref{gf:lmax_av}) in the limit $z \to 1$, which yields \cite{MZ2008}
\beq
\langle \ell_{\max,N} \rangle \approx C \, N \;, \; C = \int_0^\infty \frac{1}{1+y^{1/2}\e^y \, \gamma(1/2,y)} \, dy = 0.626508\ldots \label{lmax_sym_cont} \;,
\eeq 
where 
\beq\label{def_g}
\gamma(\nu,x) = \int_0^{x} t^{\nu-1}\,\e^{-t}\,dt\;,
\eeq
is the lower incomplete gamma function. Hence, the longest age is much larger than the typical record age, which is of order ${\cal O}(\sqrt{N})$. Note that the constant $C$ also appears in the study of the longest excursion of Brownian motion \cite{GMS2009,PY97}. 
This is in line with the remark made in the introduction of section \ref{sec:age_rw}, where it was mentioned that
the study of the ages of the records for a random walk
bears strong similarities with the excursion theory of the lattice random walk and Brownian motion.

From this result (\ref{lmax_sym_cont}) together with (\ref{lmax_QN}), one obtains the large $N$ behaviour of the probability $Q_N$ that the last interval $\ell_M$ is the longest one \cite{GMS2014,GMS2009,GMS2015}
\bea\label{QN_asympt}
Q_N \to C = 0.626508\ldots \;, \; N \to \infty \;.
\eea
Similarly, by inserting the explicit expression of $q(\ell)$ (\ref{gf:lmin_av}) and $f(\ell)$ (\ref{fl_ql}) into (\ref{gf:lmin_av}), one obtains an explicit expression of the $\GF$ of $\langle \ell_{\min,N} \rangle$, from which the exact value of $\langle \ell_{\min,N} \rangle$ for arbitrary $N$ can be obtained, yielding $\langle \ell_{\min,N} \rangle = 0, 1/2, 1, 21/16, 51/32$ for $N=0, 1, 2, 3, 4$ \cite{GMS2014}. By analysing the $\GF$ in (\ref{gf:lmin_av}) in the limit $z \to 1$, one obtains the large $N$ behaviour of $\langle \ell_{\min,N} \rangle$ as \cite{MZ2008}
\bea
\langle \ell_{\min,N} \rangle \approx D \, \sqrt{N} \;, \; D = \frac{1}{\sqrt{\pi}} = 0.564190\ldots \;, \, \label{lmin_sym_cont}
\eea 
which is thus of the same order as the {\it typical} record age $\ell_{\rm typ}$, i.e., ${\cal O}(\sqrt{N})$. 

For symmetric and continuous jump distributions, one can investigate the full distribution of $\ell_{\max,N}$ and $\ell_{\min,N}$. The distribution of $\ell_{\min,N}$ turns out to be quite simple for large $N$ and given at leading order by 
\beq\label{pdf_lmin}
{\rm Prob}(\ell_{\min,N} = \ell) = \delta_{\ell,1} + {\cal O}(N^{-1/2}) \;.
\eeq
This shows that the average value of $\langle \ell_{\min,N} \rangle$ in (\ref{lmin_sym_cont}) is controlled by rare events. In fact, the main contribution to $\langle \ell_{\min,N} \rangle$ comes from the paths with a single record, $M=1$, occurring at $X_0=0$ \cite{GMS2014}. Indeed, the result in (\ref{lmin_sym_cont}) can be simply recovered by noting that a path with $M=1$ is such that it stays negative up to step $N$. Such paths occur with a probability $q(N) \approx 1/\sqrt{\pi N}$ and they contribute to a value of $\ell_{\min ,N} = N$, implying precisely the result in (\ref{lmin_sym_cont}). This shows explicitly that $\langle \ell_{\min,N} \rangle$ is dominated by rare events, such that the random walk never crosses the origin up to step $N$.

\begin{figure}
\centering
\includegraphics[width = 0.8\linewidth]{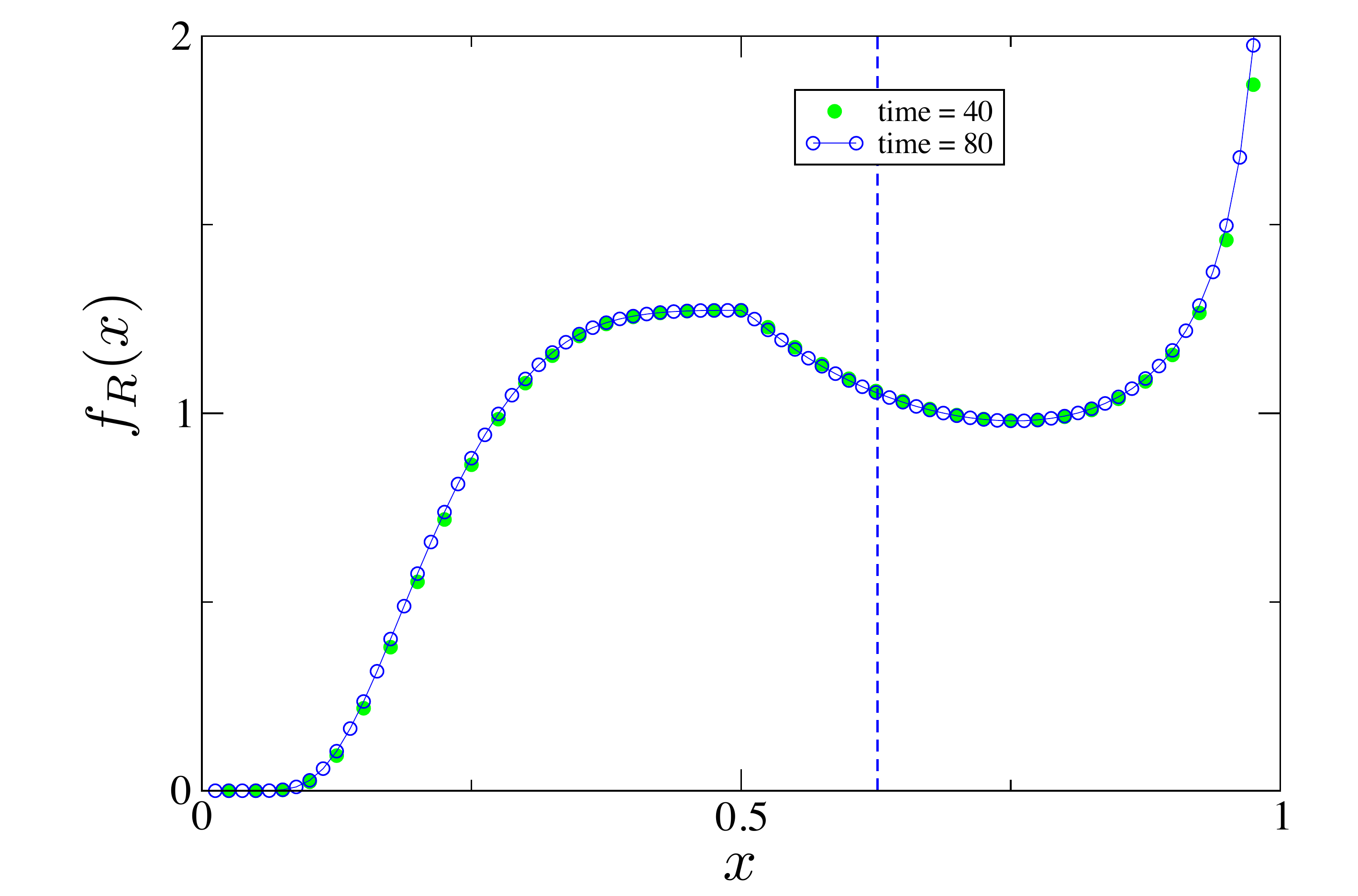}
\caption{Limiting distribution of the scaled random variable $R = \ell_{\max,N}/N$, see (\ref{def_fR}). It was obtained from the analytical expression of the generating function of $\ell_{\max,N}$ (\ref{cdf_F_RW}) for a random walk after $N=40$ steps (green full circles) and $N=80$ steps (blue empty circles), while the line is a guide to the eyes connecting the blue circles (see ref \cite{GMS2015} for more details). 
The good collapse of the data confirms the scaling form in (\ref{def_fR}).}
\label{fig:fR}
\end{figure}
The distribution of $\ell_{\max,N}$ has a much richer structure. As in the i.i.d.~case (\ref{pdf_lmax_iid}), one can show that the scaled random variable $R = \ell_{\max,N}/N$ reaches a limiting distribution in the large $N$ limit \cite{Lam61,GMS2015}
\bea\label{def_fR}
{\rm Prob}(\ell_{\max,N} = \ell) \rightarrow \frac{1}{N} f_R \left( \frac{\ell}{N}\right) \;,
\eea
where the function $f_{R}(x)$ is a piecewise continuous function on the interval $[0,1]$.
It is continuous on each interval of the form $[1/2,1]$, $[1/3,1/2]$, and so on, and exhibits singularities at the points $x_k = 1/k$ with $k=2,3, \ldots$ \cite{Lam61}. It turns out that the generating function of the random variable $V = 1/R$ has a rather simple explicit expression \cite{PY97,GMS2015,Lam61}, from which one obtains the asymptotic behaviours of $f_R(x)$ 
\begin{eqnarray}
f_R(x) \approx
\begin{cases}
&{2\,\alpha_0} \,x^{-2}\,\exp{\left(-{\alpha_0}/{x}\right)} \;, \; x \to 0\\
&\frac{1}{\pi}(1-x)^{-1/2} \;, \; x \to 1
\end{cases}
\end{eqnarray}
where $\alpha_0 = 0.854032\ldots$ is the only zero of the hypergeometric function $_1F_1(1,1/2,-x)$ on the real axis. In figure \ref{fig:fR}, we show a plot of the scaling function $f_R(x)$. 



\subsubsection{ Symmetric random walk on a lattice.}

In this case, the $\GF$ of the persistence probability $q(\ell)$ is given by (\ref{lw_recur.sol2}), yielding the large $\ell$ behaviour in (\ref{qrw_largel}), while $f(\ell)$ is given in (\ref{fl_ql}). In the large $N$ limit, one finds \cite{MZ2008}
\bea
\langle \ell_{\max,N} \rangle \approx C\, N \label{lmax_sym_dis} \;,
\eea
as in the continuous case (\ref{lmax_sym_cont}), despite the fact that the persistence probabilities differ by a factor $\sqrt{2}$ [see (\ref{ql.1}) and (\ref{qrw_largel})]. 
Similarly, the probability $Q_N$ also goes to the same constant $Q_N \rightarrow C$ as $N \to \infty$, as above (\ref{QN_asympt}).
However, this difference (by a factor $\sqrt{2}$) matters in the large $N$ behaviour of $\ell_{\min,N}$ which is given in this case by \cite{MZ2008} 
\bea
\langle \ell_{\min,N} \rangle \approx\sqrt{2}\,D \,\sqrt{N} \;, \label{lmin_sym_dis}
\eea
which is larger, by a factor $\sqrt{2}$, than its value for the continuous case (\ref{lmin_sym_cont}).
As for continuous jumps, the ages of the records bear strong similarities with the excursions between consecutive zero crossings of the discrete random walk, which have been extensively studied in the mathematical literature, see e.g., \cite{CH03}.

\subsubsection{Random walk in the presence of a constant drift.}

In this case, the random walk is characterized by two parameters which are
the L\'evy index $0 <\mu\leq 2$ and the constant drift $c$ [see (\ref{evol_drift.1}) and (\ref{smallk.1})]. As we emphasized it above, the large $N$ behaviours of $\langle \ell_{\max,N} \rangle$ and $\langle \ell_{\min,N} \rangle$ are governed by the asymptotic behaviour of the persistence probability $q(\ell)$ for large $\ell$, which depends strongly on the L\'evy index $0 <\mu\leq 2$ and the constant drift $c$, giving rise to five different regimes in the strip $(c, 0< \mu \leq 2)$ (see figure \ref{fig.phd}). In turn, both $\langle \ell_{\max,N} \rangle$ and $\langle \ell_{\min,N} \rangle$ depend on $\mu$ and $c$ and this dependence was studied in detail in ref \cite{MSW2012} (see also \cite{LDW09} for the case $\mu=1$). Without giving further details, we summarise in table \ref{tab:cmu_age} the main results for $\langle \ell_{\max,N} \rangle$ and $\langle \ell_{\min,N} \rangle$. Note that in this table all the amplitudes can be computed explicitly \cite{MSW2012}. 

\begin{table}[h]
\begin{center}
\begin{tabular}{|c||c|c|} 
\hline
${\rm regimes\,\, in\,\, figure\;} {\ref{fig.phd}}$&$\langle \ell_{\max,N}\rangle$&${\langle 
\ell_{\min,N}\rangle}$\\
\hline
I&$\approx C_I\, N$&$\approx D_I\, \sqrt{N}$\\
II&$\approx C_{II}\, N$&$\approx D_{II}\, N^{1-\theta(c)}$\\
III&$\approx C_{III}\, N^{1/\mu}$&$ D_{III}$\\
IV& $\approx C_{IV}\, \ln N 
$&$D_{IV} $\\
V&$\approx C_V\, N $&$ D_V \, N $\\
\hline
\end{tabular} 
\end{center} 
\caption{Asymptotic results for $\langle \ell_{\max,N}\rangle$ and $\langle \ell_{\min,N}\rangle$ for large $N$
in the five regimes in the $(c,0<\mu\le 2)$ strip in figure (\ref{fig.phd}). All the amplitudes can be computed explicitly \cite{MSW2012} (in particular $C_I = C$ and $D_I = D$ as given in equations (\ref{lmax_sym_cont}) and (\ref{lmin_sym_cont}) respectively), while the exponent $\theta(c)$ is given in (\ref{cauchy_mean.1}).}\label{tab:cmu_age}
\end{table}

\subsubsection{Continuous-time random walks.}

They are characterized by an exponent $0<\gamma\leq 1$ describing the power law tail of the time $\tau$ between two successive jumps [see (\ref{wtdist.1})]. The case $\gamma = 1$ corresponds to the discrete time random walk (and continuous jumps). The statistics of the longest and shortest lasting records for continuous-time random walks were studied in ref \cite{Sanjib2011} along the lines explained in section \ref{sec:record_nber}. In the limit of a large fixed time interval $[0,t]$, the average value of the longest time interval $\langle \ell_{\max}(t)\rangle$ grows linearly with $t$ with a non trivial amplitude $c(\gamma)$ \cite{Sanjib2011}
\bea
\langle\ell_{\max}(t)\rangle \approx c(\gamma) \,t \;, \; c(\gamma) = \int_0^\infty \frac{1}{1+ y^{\gamma/2} \e^y \gamma(1-\alpha/2,y)} \, dy \label{lmax_CTRW} \;.
\eea
As expected, for $\gamma = 1$ we recover the discrete-time result given in (\ref{lmax_sym_cont}), i.e., $c(1) = C$, given in equation (\ref{lmax_sym_cont}). On the other hand, the average shortest age is given, for large $t$, by \cite{Sanjib2011}
\bea
\langle \ell_{\min}(t)\rangle \approx \frac{\tau_0}{\Gamma(1-\frac{\gamma}{2})} \left( \frac{t}{\tau_0}\right)^{1-\frac{\gamma}{2}} \;, \label{lmin_CTRW} 
\eea
which, for $\gamma = 1$, yields back the result for the discrete time random walk given in~(\ref{lmin_sym_cont}), with the substitution $\ell_{\min,t} \to \tau_0 \, \ell_{\min,N}$ and $t \to N \tau_0$.

\subsection{Statistics of the record increments}\label{sec:increments}

\begin{figure}[ht]
\centering
\includegraphics[width = 0.8\linewidth]{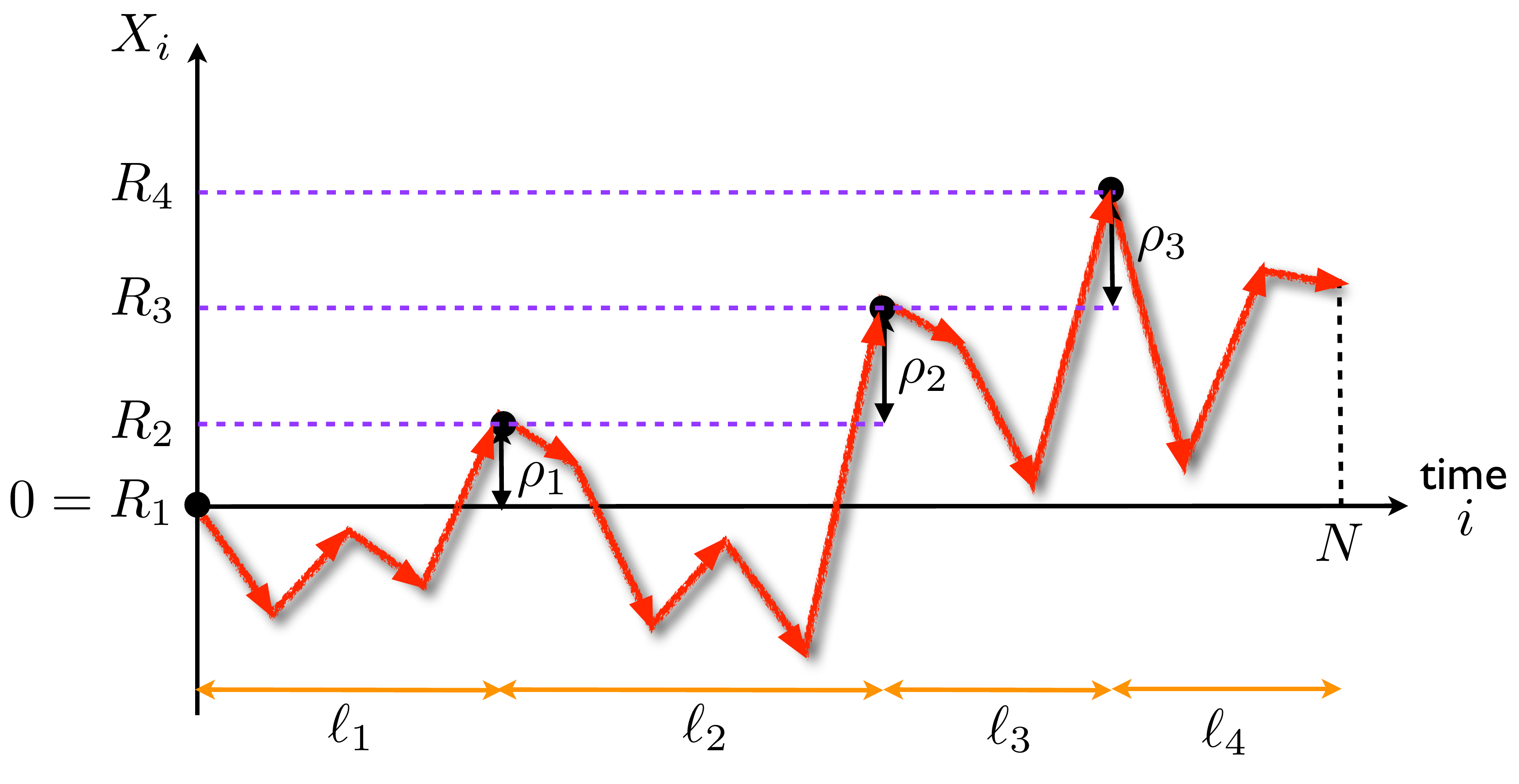}
\caption{Realization of a random walk trajectory of $N=15$ steps with $M=4$ records. The variables
$\ell_k$ denote the ages of the records, i.e., the intervals between successive records. The record values are noted
as $R_k$ and the increments between two successive record values are denoted by $\rho_k = R_{k+1}- R_k$.}\label{Fig:increments}
\end{figure}

Up to now, in the current section \ref{section:RW}, we have mainly focused on the 
number of records and on the ages of the records, for a given random walk of $N$ steps. 
The statistics of these observables have been obtained from the general renewal property
described in section \ref{sec:renewal} [see in particular (\ref{renewal.1})]. In this section we focus on random walks with a symmetric and continuous jump
distribution $\phi(\eta)$ and consider the statistics of
the record increments. Let us consider a particular realization of a random walk sequence with $M$ number of
records, as in figure \ref{Fig:increments}. We denote by $R_k$ the record values and by $\rho_k = R_{k+1}-R_k$ the
corresponding increments in this realization.
In this subsection, we focus on the joint $\PDF$ $P(\vec \rho,M|N)$ of the increments
$\vec{\rho} = (\rho_1, \rho_2, \dots, \rho_{M-1})$ and the number of records $M$ for a fixed number of steps. In particular, we are
interested in the large $N$ limit. 

To compute this joint $\PDF$, we first compute a more complicated object, which is the ``grand'' joint $\PDF$ $P(\vec{\rho}, \vec{\ell},M|N)$ 
of the record increments $\vec{\rho}$, the record ages $\vec{\ell}$ and the number of records $M$. The joint $\PDF$ $P(\vec{\rho},M|N)$ is then
obtained by integrating the age degrees of freedom $\vec{\ell}$ \cite{GMS2016}. 
To compute this grand joint $\PDF$ $P(\vec{\rho}, \vec{\ell},M|N)$ we need the three following quantities:
\begin{itemize}
\item{
The first one is the survival probability $q(\ell)$ (\ref{ql_def}), i.e., the probability that a random walk, starting at $x_0$, stays below
$x_0$ up to $\ell$ time steps, which is universal and given by the Sparre Andersen theorem (\ref{ql.1}).}
\item{The second is the first-passage probability $f(\ell)$ defined in (\ref{fl_def}), which is also universal and simply given by $f(\ell) = q(\ell-1) - q(\ell)$ [see (\ref{fl_ql})].}
\item
Finally, the third quantity we need is $J(\ell,\rho)$ (for a random walk starting at $x_0=0$), defined as
 \begin{equation}\label{def_Jl}
J(\ell,\rho) = {\rm Prob} (X_1 < 0, X_2<0, \ldots, X_{\ell - 1}<0, X_{\ell} = \rho > 0) \;.
\end{equation} 
\end{itemize} 

This denotes the probability that the walker, starting at the origin $x_0=0$, stays below the origin up to $\ell -1$ steps
and then jumps to the positive side, arriving at $\rho > 0$ at step $\ell$. If one integrates it over the final position $\rho$, one recovers the first
passage probability at step $\ell$, i.e., 
\begin{eqnarray}\label{eq:identity1}
\int_0^\infty J(\ell,\rho) \,d\rho = f(\ell) \;.
\end{eqnarray}
The probability $J(\ell,\rho)$ also appears in the study of the order statistics of random walks \cite{MMS13, MMS14} and its $\GF$ can be expressed in terms of the jump distribution $\phi(\eta)$ as follows (see ref \cite{GMS2016} for details). 
\begin{figure}[ht]
\centering
\includegraphics[width=0.8 \linewidth]{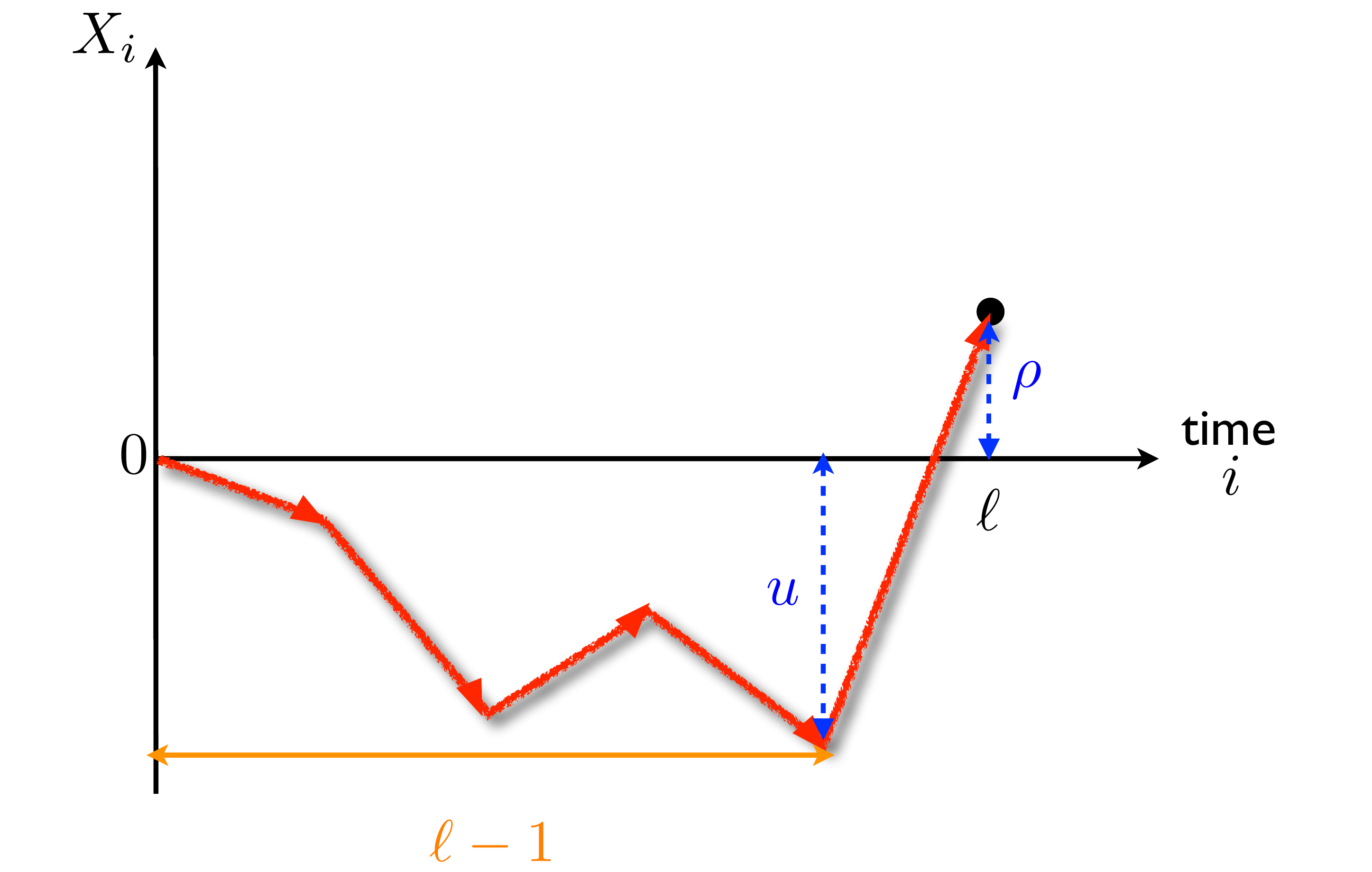}
\caption{A configuration of a random walk, starting at the origin $X_0=0$, that stays below the origin up to $\ell - 1$ steps and then jumps to $\rho>0$ at step $\ell$. We also use the notation $u$ for $ -X_{\ell -1}$, so that the last jump is of length $u+\rho$.}\label{fig_J}
\end{figure}
To compute $J(\ell,\rho)$, we first define $p_{\ell-1}(u)$ as the probability density for the walker to arrive at $u>0$ in $\ell-1$ steps, starting from the origin and staying above the origin up to $\ell -1$ steps. By symmetry $p_{\ell-1}(u)$ also denotes the probability density that the walker arrives at $-u$ in $\ell -1$ steps, staying negative up to $\ell -1$ steps. Note also that $p_{\ell-1}(u)$ is the probability density that a single walker reaches the level $u$ for the first time at step $\ell-1$, starting at the origin at step $0$. (This definition will be useful to study the record statistics of a multi-particle system in section \ref{sec:multi}.) 
Clearly one has
\begin{eqnarray}\label{expr_J}
J(\ell,\rho) = \int_0^\infty p_{\ell-1}(u) \phi(u+\rho) \, du \;,
\end{eqnarray} 
where $\phi(u+r)$ denotes the distribution of the last jump (see figure \ref{fig_J}). Consequently, the $\GF$ $\tilde J(z,\rho) =\sum_{\ell \geq 1} J(\ell,\rho) z^\ell$ is given by
\begin{eqnarray}\label{GF_J}
\tilde J(z,\rho) = \sum_{\ell \geq 0} z^{\ell +1} \int_0^\infty p_\ell(u) \phi(u+\rho) \, du \;,
\end{eqnarray} 
where we have shifted $\ell$ by 1, for convenience. It turns out that computing the constrained propagator $p_\ell(u)$ for arbitrary jump distribution $\phi(\eta)$ is rather nontrivial. Nevertheless there exists a fairly explicit formula \cite{Ivanov94} for the double Laplace transform of $p_\ell(u)$ which reads (for a recent review see \cite{MMS14,satya_leuven})
\begin{eqnarray}\label{eq:ivanov}
\int_0^\infty \sum_{\ell \geq 0} p_\ell(u) z^\ell \, \e^{-\lambda \, u} \, du = \psi(\lambda,z) \;.
\end{eqnarray}
The function $\psi(\lambda,z)$ is given by
\begin{eqnarray}\label{eq:def_phi}
\psi(\lambda,z) = 
\exp{\left(-\frac{\lambda}{\pi} \int_0^\infty 
\frac{\ln{[1-z\,\hat \phi(q)]}}{q^2+\lambda^2} \, d q \right)} \;,
\end{eqnarray} 
where $\hat \phi(q) = \int_{-\infty}^\infty \phi(\eta)\,\e^{i q\eta} \, d \eta$ is the Fourier transform
of the jump distribution. Thus the dependence of $p_\ell(u)$ on the jump distribution manifests itself
through its Fourier transform $\hat \phi(q)$. In general, it is very hard to compute explicitly $p_\ell(u)$ for any $\ell$ and $u$ from this relation (\ref{eq:def_phi}). However in the case of {\it a symmetric exponential jump distribution} $\phi(\eta) = 1/(2b)\,\e^{-|\eta|/b}$, the generating function of $p_\ell(u)$ with respect to $\ell$ can be computed explicitly, with the result 
\beq
\tilde p(u,z) = \sum_{\ell\ge1} z^\ell p_\ell(u) = \frac{1-\sqrt{1-z}}{b} \e^{-\frac{|u|}{b}\sqrt{1-z}} \;,\label{eq:expr_GG>_app}
\eeq
while $p_0(u) = \delta(u)$. Using this expression (\ref{eq:expr_GG>_app}) together with (\ref{GF_J}), one obtains
\beq\label{eq:integral1}
\hspace*{-1.cm}\sum_{\ell\ge1 } J(\ell,\rho)\, z^{\ell} = z \int_0^\infty d y \, \tilde p(y,z) \frac{1}{2b}\e^{-(y+\rho)/b} = \frac{1}{b}(1-\sqrt{1-z}) \e^{-{\rho}/{b}} \;.
\eeq
This equation (\ref{eq:integral1}) shows that the variables $\ell$ and $\rho$ decouple for the exponential jump distribution, yielding
\beq\label{eq:expr_TGF}
J(\ell,\rho) = \frac{1}{b}f(\ell) \, \e^{-\rho/b} \;, \; \sum_{\ell\ge1} f(\ell) z^\ell = 1-\sqrt{1-z} \;,
\eeq
which yields the expression of the coefficients $f(\ell)$ as
\beq\label{eq:expr_ck}
f(\ell) = (-1)^{\ell +1} \frac{\sqrt{\pi}}{2\,\Gamma(3/2-\ell)\Gamma(\ell+1)} \approx \frac{1}{2\sqrt{\pi} \ell^{3/2}} \;, \; {\rm as} \; \ell \to \infty \;.
\eeq
These formulae (\ref{eq:expr_TGF}), (\ref{eq:expr_ck}) will be useful in section \ref{sec:bridge} to study the records of a random walk bridge, with symmetric exponential jumps.


With these three quantities, one can express the grand joint $\PDF$, using again the renewal property of the random walk, i.e., the independence of the intervals between two successive records (see figure \ref{Fig:increments}). For $M\geq 2$ it reads
\begin{equation}\label{eq:full_joint}
P(\vec{\rho}, \vec{\ell},M|N) = \prod_{k=1}^{M-1} J(\ell_k,\rho_k) \, 
q(\ell_M) \delta\left(\sum_{k=1}^M \ell_k,N\right) \;,
\end{equation}
where the Kronecker delta ensures that 
the total number of steps is fixed to $N$. The factor $q(\ell_M)$ corresponds 
to the interval after the last record, {i.e.}, the probability that all the positions
$X_i$ after the last record stay below the last record value, which is 
given in (\ref{ql.1}). For $M=1$, only the starting point is a 
record, and the process stays negative during the entire time interval 
$N$. In this case, there is no record increment, but we set the record 
increment to be $\rho=0$ by convention and hence
\begin{equation}\label{eq:full_joint_M1}
P(\rho,\ell_1,M=1|N) = q(\ell_1) \delta(\ell_1,N) \delta(\rho) \;.
\end{equation}

 The joint $\PDF$ $P(\vec{\rho},M|N)$ is then obtained by summing $P(\vec{\rho}, \vec{\ell},M|N)$ 
in (\ref{eq:full_joint}) over $\ell_1, \ldots, \ell_{M-1}$ (each from 
$1$ to $\infty$) and $\ell_M$ (from $0$ to $\infty$). Hence the $\GF$ of
$P(\vec{\rho},M|N)$ with respect to $N$ reads, for $M \geq 2$
\begin{eqnarray}\label{eq:gf_jp}
\sum_{N \geq 0} P(\vec{\rho},M|N) z^N = 
\tilde q(z) \prod_{k=1}^{M-1} \tilde J(z,\rho_k) \;,
\end{eqnarray}
where $\tilde q(z)$ is given in (\ref{qz_exact}) and the $\GF$
$\tilde J(z,\rho)\equiv \sum_{\ell\geq 1} z^\ell J(\ell,\rho)$. From (\ref{eq:gf_jp}), it follows that $P(\vec \rho, M|N)$ is invariant under 
permutation of the labels of record increments, implying that the marginal 
distribution of $\rho_k$, $P(\rho_k|N)$, is independent of $k$. It can be computed by 
integrating $P(\rho,\rho_2,\ldots,\rho_{M-1},M|N)$ in (\ref{eq:gf_jp})
over $\rho_2,\ldots,\rho_{M-1}$ and 
then summing over $M$ (from $1$ to $+\infty$) (see \cite{GMS2016} for 
details). One gets 
\begin{eqnarray}\label{exact_gf_increments}
\sum_{N\geq 0} P(\rho|N) z^N = \frac{\tilde J(z,\rho)}{(1-z)} + 
\frac{\delta(\rho)}{\sqrt{1-z}} \;,
\end{eqnarray}
where we have used $\tilde q(z)=1/\sqrt{1-z}$ [see (\ref{qz_exact})] and 
$\tilde f(z) = 1- \sqrt{1-z}$ [see (\ref{fz_qz})]. 
As $z\to 1$, the right hand side of (\ref{exact_gf_increments}) 
behaves, to leading order, 
as $\tilde J(1,\rho)/(1-z)$, implying that in the large $N$ limit, 
\begin{eqnarray}
\lim_{N \to \infty} P(\rho|N) = p(r) = \tilde J(1,\rho) \;,
\end{eqnarray}
which shows that the increments have a 
stationary distribution as~$N~\to~\infty$.

For some jump distributions, $\tilde 
J(1,\rho)$ can be computed explicitly \cite{GMS2016} (see also \cite{Asm2003} for related results in the context of queuing theory). For instance, 
for $\phi(\eta) = 1/(2\,b) \e^{-|\eta|/b}$, one finds $p(\rho) = \e^{-\rho/b}/b$, with 
$\rho\ge 0$. Another exactly solvable case is $\phi(\eta) = 
1/(2\,b^2) |\eta|\,\e^{-|\eta|/b}$, for which one finds (with $\rho\ge 0$)
\begin{eqnarray}\label{eq:exact_linexp}
p(\rho) = \frac{1}{2\,b^2(1+\sqrt{3})}\,\e^{-\rho/b}\left({2b} \,(\sqrt{3}-1) 
+ 4\,\rho \right) \;.
\end{eqnarray}
Another interesting example is the case of L\'evy flights, corresponding to $\phi(\eta) \sim A \, |\eta|^{-1-\mu}$ 
with $0<\mu<2$. In this case one can obtain the tail of $p(\rho)$ exactly for large~$\rho$
\begin{eqnarray}\label{eq:asympt_heavy}
p(\rho) \approx B_\mu \, \rho^{-1 - \mu/2} \;, \; \rho \to \infty \;,
\end{eqnarray}
where $B_\mu$ is a computable constant (and depends both on $A$ and $\mu$). Interestingly, this result (\ref{eq:asympt_heavy}) decays more slowly than the jump distribution. 

These exact results for the grand joint $\PDF$ in (\ref{eq:full_joint}) or for the joint $\PDF$ of the record increments (\ref{eq:gf_jp}) given in this section are useful to compute many observables related to the records of random walk and its variants, and not only the marginal distribution of the increments $P(\rho|N)$ as discussed here. In the next section we will see that the grand joint $\PDF$ in (\ref{eq:full_joint}) is needed to study the record statistics of constrained random walks, like the random walk bridge. In section \ref{sec:other} we will further illustrate this by computing the probability ${\cal Q}(N)$ that the increments are monotonically decreasing up to step $N$.

\section{Record and age statistics for a constrained discrete time random walk}\label{sec:bridge}

\begin{figure}[ht]
\centering
\includegraphics[width=0.8\linewidth]{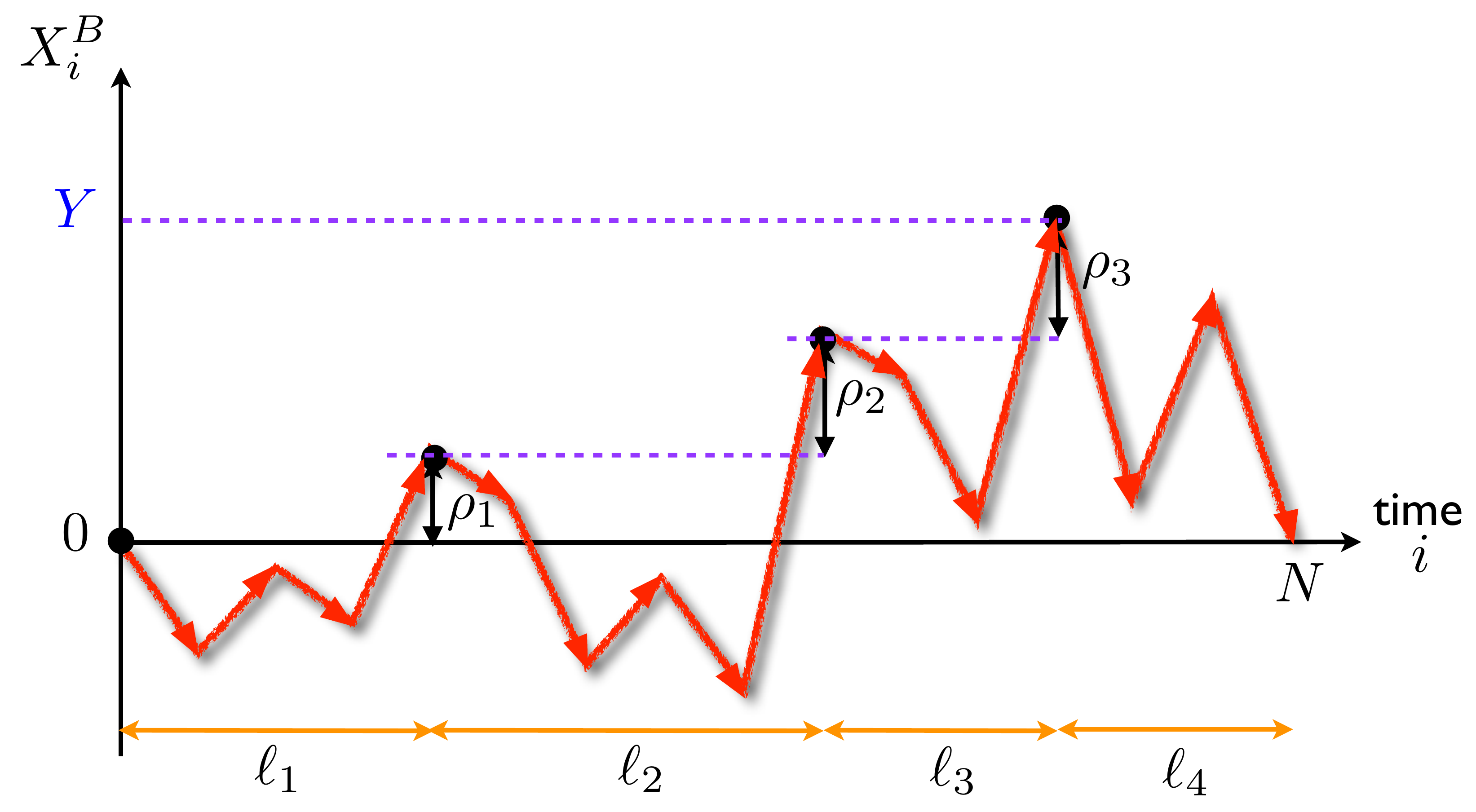}
\caption{A realization of a random walk bridge $X_i^B$ with $N=15$ steps. Here the number of records is $M=4$. The intervals $\ell_k$ denote the ages of the records and the $\rho_k$ are the increments between successive records while $Y$ denotes the value of the maximal value $X_{\max,N}^B$. The joint distribution of the random variables $\ell_k$, $\rho_k$, $M$ and $Y$, for a random walk bridge with symmetric exponential jumps is given in (\ref{eq:def_joint_pdf_exp_2}).}\label{Fig_bridge}
\end{figure}

As we have seen in the previous section, a remarkable feature of the record statistics of random walks with continuous jumps is that it is completely universal, i.e., independent of the jump distributions, even for a finite number of steps. It is thus natural to ask whether this universality still holds for constrained random walks. One of the most natural and interesting instance of such constrained random walks is the random walk bridge, which we mainly focus on here (see figure \ref{Fig_bridge}). 

As before we consider a time series $\{X_i\}, 0 \leq i \leq N$, starting from $X_0=0$ and evolving according to the Markovian rule in (\ref{evolrw.1})
\bea\label{def_RW_bridge}
X_i = X_{i-1} + \eta_i \;,
\eea 
where the jump variables $\eta_i$ are i.i.d.~random variables, drawn from the distribution $\phi(\eta)$. Here we restrict our analysis to the case where the jump distribution $\phi(\eta)$ is symmetric (no drift) but we will consider both the case of a discrete (the lattice random walk) and continuous jump distributions. In this section, we focus on the positions of the {\it random walk bridge} $\{X^B_i\}$, with $0 \leq i \leq N$, which is a random walk as defined in (\ref{def_RW_bridge}) conditioned to come back to the origin after $N$ time steps, $X^B_0 = X^B_N = 0$. Such a constrained random walk is relevant, for instance, to model periodic strongly correlated series (with $N$ being the period). 

The statistics of records for random walk bridges turn out to be rather different from the case of the free random walk. Technically, this constrained random walk is harder to analyse than the free random walk. Indeed, for free random walks, the computations require the full joint distribution of the ages of the records $\ell_1, \ell_2, \dots, \ell_{M}$ but there is no need to keep track of the actual value of the record at a given time step [see (\ref{renewal.1})]. The knowledge of the actual value of the record at a given time step is however required for bridges, where the random walk returns back to the origin after $N$ time steps. This is done here by considering the full joint distribution of the ages $\ell_k$ and of the record increments $\rho_k$ (which are the differences between two consecutive records), i.e., the grand joint $\PDF$ considered above in (\ref{eq:full_joint}) [see also figure \ref{Fig_bridge}]. Consequently, given this technical difficulty, less is known in the case of a bridge. Nevertheless, there are two special cases that can be analysed in detail: (i) the lattice random walk and (ii) the symmetric exponential jump distribution $\phi(\eta) = 1/(2b)\exp(-|\eta|/b)$, with $b>0$ \cite{GMS2015b}. The exact results obtained for these cases provide some insights 
on the record statistics for a bridge random walk with an arbitrary continuous jump distribution.

\subsection{Summary of the main results}

We first summarise the main results for the record of a random walk bridge, and refer the reader to ref \cite{GMS2015b} for more details. As in the case of the free random walk, discrete and continuous jump distributions yield different results. But in this case, for continuous distributions, the statistics of records (and of the ages) are not universal any longer and depend, for finite $N$, on the details of the jump distribution $\phi(\eta)$. Nonetheless, in the limit of large $N$, various observables characterizing the record statistics depend (at leading order for large $N$) only on the L\'evy index $\mu$ (\ref{smallk.1}) and not on further microscopic details of the jump distribution $\phi(\eta)$. 
We recall that the L\'evy index characterizes the small argument behaviour of the Fourier transform of the jump distribution $\hat \phi(q) = \int_{-\infty}^\infty d\eta\, \phi(\eta)\,\e^{i q \eta} \approx 1 - |l_\mu q|^\mu$, where $l_\mu$ is the characteristic length scale of the jumps.

Let us denote by $M$ the number of records for the random walk bridge after $N$ steps. For the lattice random walk, using the relation between $M$ and the maximum of the random walk bridge (i.e., the relation in (\ref{eq:record_max}) which can be straightforwardly generalised to the bridge) it is possible to compute exactly the full distribution of the record number. In particular, for large $N$, the mean record number still grows like $\sqrt{N}$ \cite{GMS2015b}
\beq\label{ampli_mu_discrete}
\langle M \rangle \approx \frac{\sqrt{\pi}}{2^{3/2}} \sqrt{N} \;,
\eeq
but with an amplitude which is smaller by a factor $\pi/4$ compared to the free random walk~(\ref{avg_rec_discrete}). In the large $N$ limit, the probability distribution of the random variable $M/\sqrt{N}$ converges to a stationary (i.e., $N$-independent) distribution given by
\bea\label{eq:gB}
P(M|N) \approx \sqrt{\frac{2}{N}} \, g_B\left( \frac{\sqrt{2}\,M}{\sqrt{N}}\right) \;, \; {\rm with} \; g_B(x) = 2 \,x\, \e^{-x^2} \Theta(x) \;,
\eea 
which is different from its counterpart for the free random walk (\ref{eq:scaling_lattice_RW}). Note that, as expected from (\ref{eq:record_max}), the limiting scaling function is the one of the maximum of the Brownian bridge on the unit time interval. 

For continuous jump distributions, the average number of records behaves as 
\beq\label{ampli_bridge_mu}
\langle M \rangle \approx A_B(\mu) \sqrt{N} \;,
\eeq
where the amplitude depends explicitly on $\mu$. The dependence on $\mu$ is quite involved and this amplitude can be evaluated explicitly only for $\mu=2$ with the result
\beq
A_B(\mu=2) = \frac{\sqrt{\pi}}{2} \label{ampli_mu_2} \;,
\eeq 
which, as for the lattice random walk, is also smaller by a factor $\pi/4$ compared to its continuous counterpart (\ref{avg_rec.2}). For an arbitrary continuous jump distribution, the analysis of the statistics of $M$, beyond the first moment, is quite difficult. However, exact results for the full distribution can be obtained for the symmetric exponential distribution, which is representative of the case $\mu=2$ [see (\ref{smallk.1})]. In this case, the distribution of the scaled variable $M/\sqrt{N}$ reaches a limiting distribution when $N \to \infty$ \cite{GMS2015b}
\beq
P(M|N) \approx \frac{1}{\sqrt{N}} g_B\left( \frac{M}{\sqrt{N}}\right) \;, \label{eq:gB_exp}
\eeq
where the scaling function $g_B(x)$ is the same as the one found for the lattice random walk bridge and given in (\ref{eq:gB}). 

On the other hand, for the record breaking probability $Q_N$ [see (\ref{def_QN})], exact results can be obtained only for the lattice random walk and for the random walk with symmetric exponential jump distribution. In both cases, $Q_N$ converges to the same constant, which can be expressed in terms of a non-trivial integral given by
\bea\label{eq:QN_bridge}
&&\hspace*{-2.cm}\lim_{N \to \infty} Q_N = \frac{2}{\sqrt{\pi}} \int_0^\infty \frac{d y}{\sqrt{y}} \e^{-y}\left[1-\sqrt{\pi\,y} F(y) \exp{[y F^2(y)]} {\rm erfc}\left[\sqrt{y} F(y) \right]\right] \nonumber \\
&&\hspace*{-1.8cm} {\rm where} \; \; F(y) = {\rm erf}(\sqrt{y}) + \frac{1}{\sqrt{\pi}} \frac{\e^{-y}}{\sqrt{y}} \;.
\eea
A numerical evaluation of the integral in (\ref{eq:QN_bridge}) yields, for the random walk bridge:
\bea\label{lim_QN_bridge}
\lim_{N \to \infty} Q_N = 0.6543037\ldots
\eea
which is different from, and slightly larger than, the one characterizing the free random walk and given in (\ref{QN_asympt}). 

On the other hand, for the lattice random walk and for the symmetric exponential jump distribution, the average age of the longest lasting record $\langle \ell_{\max,N}\rangle$ can be computed exactly in the large $N$ limit \cite{GMS2015b} 
\bea
\fl\lim_{N \to \infty} \frac{\langle \ell_{\max,N} \rangle}{N} &=& 4\int_0^\infty d y \left(\frac{1}{2} - \frac{F(y)\,\e^{-y+y\,F^2(y)}\,{\rm erfc}[\sqrt{y}\,F(y)]-\e^{-y}/\sqrt{\pi y} }{1-F^2(y)} \right) \nonumber \\
&=& 0.6380640\ldots \;, \label{lmax_bridge}
\eea 
which, at variance with the free random walk, is strictly smaller that the limiting value of $Q_N$ in (\ref{lim_QN_bridge}). Numerical simulations were performed in \cite{GMS2015b} to estimate numerically $Q_N$ as well as $\ell_{\max,N}$ and a very good agreement with the predictions in equations (\ref{eq:QN_bridge}) and~(\ref{lmax_bridge}) was found.

\subsection{Outline of the derivation of the main results}

In this section, we give the main ideas that lead to the results announced before for the random walk bridge and we refer to \cite{GMS2015b} for more details. 

\subsubsection{Mean number of records.}

To compute the mean number of records $\langle M \rangle$ we proceed as explained before for the i.i.d.~case in equations (\ref{def_sigma})--(\ref{exact_mean_iid}) and compute the record rate $r_{k}$, which is the probability that a record is broken at step $k$ -- for a random walk bridge of $N$ steps. One has indeed [see (\ref{average_rw_simple})]
\beq
\langle M \rangle = \sum_{k=0}^N r_{k} \label{av_R_bridge.1} \;. 
\eeq
Note that, at variance with the i.i.d.~or free random walk case, one expects that this record rate depends on {\it both} $k$ and $N$, as the random walk bridge must return to the origin after $N$ steps. To compute the record rate $r_{k}$, the two following quantities are required

\begin{itemize}
\item[$\bullet$]{The free Green's function (propagator) $G(x,x_0,\ell)$ that denotes the probability (for lattice random walk) or probability density (for continuous jump distribution) that a random walker starting at $x_0$ arrives at $x$ after $\ell$ steps.}
\item[$\bullet$]{The constrained Green's function $G_>(x,x_0,\ell)$ that denotes the probability (for lattice random walk) or probability density (for continuous distribution) that a random walker starting at $x_0$ arrives at $x$ after $\ell$ steps and staying strictly positive in between.} 
\end{itemize}

\begin{figure}
\centering
\includegraphics[width=0.8\linewidth]{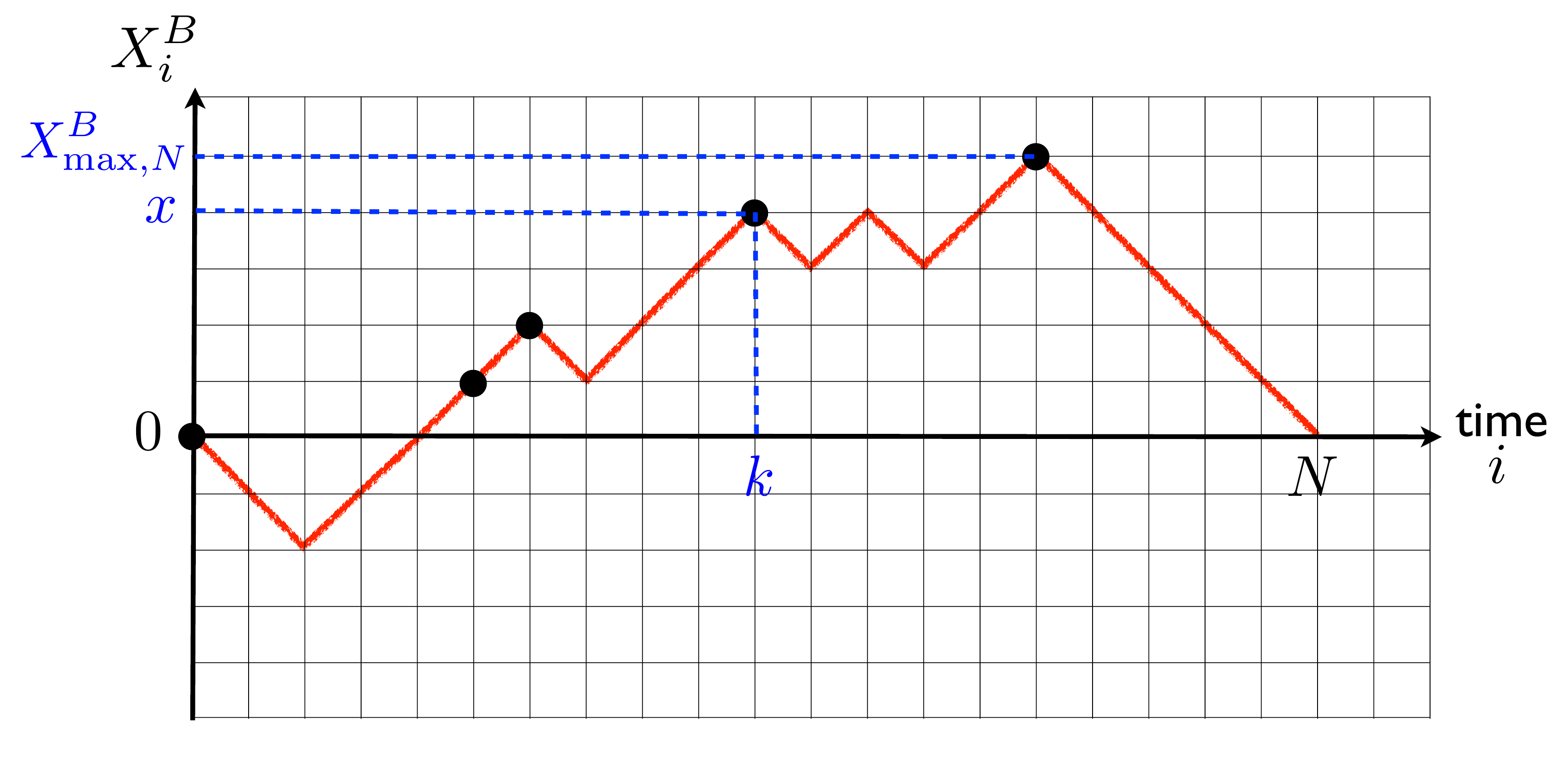}
\caption{A lattice random walk bridge of $N=20$ steps. Here the number of records is $M= X^B_{\max,20}+1 = 6$.}\label{Fig_bridge2}
\end{figure}
To compute $r_{k}$, let us suppose that a record happens at step $k$ with a record value $x$ (see figure \ref{Fig_bridge2}). This corresponds to the event that the walker, starting at the origin at step $0$, has reached the level $x$ for the first time at step $k$ and returns back to the origin after $N$ steps -- as we are considering random walk bridges. In the time interval $[0,k]$, the walker propagates from $0$ to $x$, being constrained to stay {\it strictly} below $x$. To compute the corresponding propagator, we take $x$ as the new origin of space and then reverse both the time and coordinate axes. Hence, we see that on the time interval $[0,k]$, the particle propagates with $G_{>}(x,0,k)$. On the other hand, between step $k$ and step $N$ (where the walker ends at the origin) the walker is free and thus propagates with $G(0,x,N-k) = G(x,0,N-k)$, as the jump distribution is symmetric. The record rate is then obtained by integrating the probability of this event over $x \geq 0$ as the record can take place at any level $x\geq 0$ (note that only the first record, i.e., $k=0$, is such that $x=0$). Using the statistical independence of the random walk in the time intervals $[0,k]$ and $[k,N]$ (being Markovian), one thus has, for $N \geq 1$:
\beq\label{expr_rm}
r_{k} = \frac{1}{G(0,0,N)}\int_0^\infty \, d x \, G_>(x,0,k) G(x,0,N-k) \, \;, \; 0 \leq k \leq N-1 \;,
\eeq
where we have divided by $G(0,0,N)$ as we are considering random walks that are conditioned to come back to the origin after $N$ time steps (bridges). Since for a bridge $X^B_N = X_0^B = 0$, a record can not be broken at the last step -- as a record is defined by a strict inequality [see (\ref{def_record})]. Note that in the case of a discrete random walk the integral over $x$ in (\ref{expr_rm}) has to be replaced by a discrete sum. 

The explicit computation of the record rate in (\ref{expr_rm}) for an arbitrary distribution and for arbitrary $k$ and $N$ is a very hard task, since the computation of the constrained propagator $G_>(x,x_0,k)$ can be carried out explicitly only in some special cases. Such exactly solvable cases include the lattice random walk, using the method of images, and the symmetric exponential jump distribution, using the so-called Hopf-Ivanov formula \cite{Ivanov94}. In these two cases, $\langle M \rangle$ can be computed explicitly for any $N$ \cite{GMS2015b} leading to the results in equations (\ref{ampli_mu_discrete}) and (\ref{ampli_mu_2}). For more general continuous jump distribution, although an exact calculation of $\langle M \rangle$ for any finite $N$ seems quite difficult, one can perform a large $N$ asymptotic analysis as we discuss it now. As we will see, the final large $N$ result depends only on the L\'evy-index $0\leq \mu \leq 2$ characterising the random walk (\ref{smallk.1}). 

We recall that the average number of records is given by the sum in (\ref{av_R_bridge.1}). This sum over $k$ is dominated by the values of $k \sim {\cal O}(N)$ which are thus large, when $N \gg 1$ \cite{GMS2015b}. Hence, to evaluate the record rates $r_{k}$ given in (\ref{expr_rm}) for large $k$ one can replace the propagators $G(x,0,N-k)$ and $G_>(x,0,k)$ by their scaling forms valid for $k,N \gg 1$, with $k/N$ fixed, and $x \gg 1$, with $x/N^{1/\mu}$ fixed. One has indeed
\begin{eqnarray}
&&G(x,0,N-k) \approx \frac{1}{l_\mu (N-k)^{1/\mu}} R\left(\frac{x}{l_\mu \, (N-k)^{1/\mu}} \right)\;, \label{eq:G_scaling} \\
&&G^{}_>(x,0,k) \approx \frac{1}{l_\mu \sqrt{\pi}k^{1/2+1/\mu}} R_+ \left(\frac{x}{l_\mu \, k^{1/\mu}} \right) \;, \label{eq:G>_scaling}
\end{eqnarray} 
where the scaling functions are normalised, i.e., $\int_{-\infty}^\infty d x \, R(x) = 1$ and $\int_0^\infty d x \, R_+(x) = 1$. We recall that $l_\mu$ in equations (\ref{eq:G_scaling}) and (\ref{eq:G>_scaling}) is the characteristic length scale of the jumps (\ref{smallk.1}). 
The scaling function $R(x)$ is a (symmetric) L\'evy stable distribution:
\beq\label{eq:stable_dist}
R(x) = \frac{1}{2 \pi} \int_{-\infty}^\infty d q \, \e^{-i q x} \e^{-|q|^\mu} \;,
\eeq
and in particular $R(0) = {\Gamma(1+1/\mu)}/{\pi}$. For $\mu = 2$, it corresponds to a Gaussian distribution while for $\mu=1$ this is the Cauchy distribution. On the other hand, there is no explicit expression for $R_+(x)$ for generic $\mu < 2$. For $\mu = 2$ one has $R_+(x) = 2 \,x \, \e^{-x^2}\Theta(x)$ and for $\mu=1$, it is also possible to write $R_+(x)$ explicitly as an integral \cite{Darling,GRS2011} (with $x>0$)
\begin{eqnarray}
&&R_+(x) = -\sqrt{x} \int_0^1 g\left(\frac{x}{v}\right)\,v^{-3/2}\,(1-v)^{-1/2} \, dv \nonumber \\
&&g(z) = \frac{d}{dz} \left[ \frac{1}{\pi} \frac{1}{(1+z^2)^{3/4}} \exp\left( -\frac{1}{\pi} \int_0^z \frac{\ln{u}}{1+u^2} \,du\right) \right] \label{R+_cauchy} \;.
\end{eqnarray}

With such a normalisation (\ref{eq:G>_scaling})
one can check in particular that by integrating $G_>(x,0,k)$ in (\ref{eq:G>_scaling}) over $x$ one recovers the survival probability $q(k)$, which is the probability that the walker, starting at the origin, stays positive up to step $k$:
\beq
\int_0^\infty {d x} \; G^{}_>(x,0,k) = q(k) \approx \frac{1}{\sqrt{\pi \,k }} \;, \; {\rm as} \; k \to \infty \;,
\eeq
in agreement with the Sparre Andersen theorem \cite{SA54}. By inserting these scaling forms (\ref{eq:G_scaling}, \ref{eq:G>_scaling}) into the expression for $r_{k}$ in (\ref{expr_rm}) one finds that for large $k$ and $N$ keeping $k/N = y$ fixed (with $0 \leq y \leq 1$): 
\bea\label{eq:scaling_rate}
r_{k} = \frac{1}{\sqrt{N}} H\left(y = \frac{k}{N}\right) \;, 
\eea
where the scaling function reads
\bea\label{eq:scaling_rate+}
\fl H(y) = \frac{\sqrt{\pi}}{\Gamma(1+1/\mu)} \frac{1}{\sqrt{y}(1-y)^{1/\mu}} \int_0^\infty d x \, R_+(x) R\left(\frac{x}{(y^{-1}-1)^{1/\mu}}\right) \;.
\eea
Finally, from this scaling form for the record rate (\ref{eq:scaling_rate}), one obtains 
\begin{eqnarray}\label{eq:av_rn_levy}
\langle M\rangle = \sum_{k=0}^{n} r^{\rm c}(k,n) \approx A_B(\mu )\sqrt{n} \;, \; A_B(\mu) = \int_0^1 d y \, H(y) \;.
\end{eqnarray} 
In particular, one can check that $A_B(\mu=2) = \sqrt{\pi}/2$, which coincides, as expected, with the result obtained in the exponential case. Note that a detailed analysis of this amplitude $A_B(\mu)$, as a function of $\mu$, has not been carried out, even numerically. 

\subsection{Joint distribution of the ages}

As we have discussed it in section \ref{sec:renewal} on the free random walk, the computation of the full statistics of most observables related to records (like the record number, the age of the longest lasting record $\ell_{\max,N}$ or the probability of record breaking $Q_N$) necessitates the knowledge of the joint distribution of the ages $\ell_1, \ell_2, \dots, \ell_M$ and the record number $M $, denoted by $P(\vec{\ell}, M|N)$ -- see (\ref{renewal.1}) for the free random walk. While for the free random walk this joint distribution can be computed for any jump density $\phi(\eta)$, for the random walk bridge, it is known for two special cases, the lattice random walk and the random walk symmetric exponential jumps, which we now discuss separately. 

\vspace*{0.5cm}
 
\noindent{\bf Lattice random walk bridge.} In this case the joint distribution of the set of the ages $\ell_1, \dots, \ell_M$ together with the number of records $M$ reads \cite{GMS2015b}
\bea\label{joint_discrete}
&&\hspace*{-0.cm}P(\ell_1, \dots, \ell_{M}, M|N) = \frac{P(\vec{\ell},M|N)_{(0)}}{G(0,0,N)} \;, \; 
\eea
where the numerator $P^{}(\vec{\ell},M|N)_{(0)}$ is given by
\beq\label{num_joint_discrete}
P^{}(\vec{\ell},M|N)_{(0)} = f^{}(\ell_1) \dots f^{}(\ell_{M-1})G^{}_\geq(M-1,0,\ell_M) \delta\left(\sum_{k=1}^{M}\ell_k ,N\right) \;,
\eeq
and $f(\ell)$ is the first-passage probability that the discrete random walk, starting from $x_0$, arrives at $x_0+1$ for the first time at step $\ell$.
In (\ref{num_joint_discrete}), $G_\geq(x,x_0,k)$ is the probability that the random walker, starting at $x_0$, arrives at $x$ after $k$ steps
 while staying non-negative (i.e., it may touch $0$ but not $-1$) in between. Note that this is $G^{}_\geq(M-1,0,\ell_M)$ which enters the expression in (\ref{num_joint_discrete}), and not $G^{}_>(M-1,0,\ell_M)$, since a record is defined by the strict inequality in (\ref{def_record}). This last block ensures that the random walk comes back to the origin, and is thus different from the last block entering the same joint distribution for the free random walk (\ref{renewal.1}), which in that case is simply the survival probability $q(\ell_M)$. 

The building blocks that enter into this joint probability in (\ref{num_joint_discrete}) can all be computed explicitly for the lattice random walk. First, since the random walk is invariant under translation, the first-passage probability $f(\ell)$ is independent of $x_0$ and for a discrete random walk, its generating function is given by (\ref{fz_qz}) and (\ref{lw_recur.sol2})
\beq\label{eq:gf_first_passage_rw}
\tilde f^{}(z) = \sum_{\ell\ge1} f^{}(\ell) z^\ell = \frac{1-\sqrt{1-z^2}}{z} \;,
\eeq
from which we deduce that
\beq\label{eq:first_passage_discrete}
f^{}(\ell) = 
\begin{cases}
&0 \;, \; \ell \; {\rm even} \;, \\
& (-1)^{(\ell-1)/2} \dfrac{\sqrt{\pi}}{2 \Gamma(1-\ell/2)\Gamma(3/2+\ell/2)}\;, \ell \; {\rm odd} \;.
\end{cases}
\eeq
Furthermore, the constrained propagator $G^{}_\geq(x,0,\ell)$ can be simply computed using the method of images with the result
\bea\label{eq:ggeq_app}
G_{\geq}(x,0,\ell) 
= 
\begin{cases}
&\frac{1}{2^\ell} \left({\ell \choose \frac{\ell+x}{2}} - {\ell \choose \frac{\ell+x}{2}+1} \right) \;, \; {\rm if} \; \ell+x \; {\rm is \; even} \\
&0 \;, \; {\rm if} \; \ell+x \; {\rm is \; odd}
\end{cases} \;.
\eea
From this joint probability (\ref{num_joint_discrete}) which is fully explicit in this case, using (\ref{eq:first_passage_discrete}) and (\ref{eq:ggeq_app}), the full statistics of the record number, the age of the longest lasting record $\ell_{\max,N}$ or the probability of record breaking $Q_N$ can be obtained, following the lines detailed in section \ref{sec:renewal}, and yielding the results given in equations (\ref{eq:gB}), (\ref{lim_QN_bridge}) and (\ref{lmax_bridge}). This joint probability (\ref{num_joint_discrete}) should be useful to compute any observable related to the ages of the lattice random walk bridge.

\vspace*{0.5cm}
 
\noindent{\bf Random walk bridge with symmetric exponential distribution.} For the symmetric exponential jump distribution $\phi(\eta) = 1/(2b)\e^{-|\eta|/b}$, the starting point of our analysis is the equivalent of the joint distribution given, for lattice random walks, in (\ref{joint_discrete}). However, because $\phi(\eta)$ is here a continuous distribution, this computation is more delicate than in the discrete case. Indeed, as we are considering random walk bridges, the weight of the last part of the paths, where the walker comes back to origin, i.e., the last segment of duration $\ell_M$ (see figure \ref{Fig_bridge}), involves the propagator $G_{\geq}(Y,0,\ell_M) = G_{>}(Y,0,\ell_M)$ (as there are no ties here since the jump distribution is continuous) where $Y=X^B_{\max,N}$ is the actual value of the last record, which coincides with the maximum of the random walk bridge after $N$ steps. For a lattice random walk the number of records $M$ and $X_{\max,N}^B$ are directly related through $X_{\max,N}^B= M-1$ but this relation does not hold for a continuous jump distribution. Consequently, we need to keep track both of the number of records and of the value of the last record. A convenient way to do so is to consider jointly the record increments $\rho_k$, which were introduced in section \ref{sec:increments} about the record increments of random walks [see (\ref{eq:full_joint})], as well as the value of the maximum. Hence, we introduce 
the joint distribution $P(\{\ell_k, \rho_k\}_{1 \leq k \leq M-1},\ell_M,M,Y|N)$ of the ages $\ell_k$, increments $\rho_k$, the number of records 
$M$ and $X_{\max,N}^B = Y$ (see figure \ref{Fig_bridge}):
\bea\label{eq:def_joint_pdf_exp_2}
\fl P(\{\ell_k, \rho_k\}_{1 \leq k \leq M-1},\ell_M,M,Y|N) =
\nonumber \\
\dfrac{\prod\limits_{k=1}^{M-1} J(\ell_k,\rho_k)
G_>(Y,0,\ell_M) }{G(0,0,N)}\, \delta\left(\sum_{k=1}^{M-1} \rho_k - Y\right) 
\delta\left(\sum_{k=1}^{M} \ell_k, N\right) \;.
\eea
The quantity $J(\ell,\rho)$ was introduced in section \ref{sec:increments} [see (\ref{def_Jl})] and will be further discussed below. 
The joint distribution of the $\ell_k$ and $M$, i.e., the equivalent of (\ref{joint_discrete}) for the discrete case, is obtained by integrating the formula in (\ref{eq:def_joint_pdf_exp_2}) over $\rho_k$ and $Y$: 
\bea\label{eq:def_joint_pdf_exp_partial_int}
P(\ell_1, \ell_2, \dots, \ell_{M}, M| N) = \frac{P(\vec{\ell},M|N)_{(0)}}{G(0,0,N)} \;, \; 
\eea 
where 
\bea\label{eq:num_exp1}
&&\hspace*{-.cm}P(\vec{\ell},M|N)_{(0)} = \prod_{k=1}^{M-1} \int_0^{\infty} d \rho_k \, J(\ell_k,\rho_k) \nonumber \\
&&\times \int_0^\infty d Y G_>(Y,0,\ell_M) \, \delta\left(\sum_{k=1}^{M-1} \rho_k - Y\right) \delta\left(\sum_{k=1}^{M} \ell_k , N\right) \;.
\eea
Note that this formula (\ref{eq:num_exp1}) is actually valid for any continuous jump distribution $\phi(\eta)$. However, its analysis is in general very hard to do, mainly because the constrained propagator $G^{}_>(x,0,n)$ does not have any explicit expression (see the discussion in section \ref{sec:increments}), which prevents one to perform the analysis of this multiple integral. Fortunately, such an explicit expression exists for the case of an exponential jump distribution $\phi(\eta) = 1/(2b) \e^{-|\eta|/b}$, which we now focus on.

In this case, the building block $J(\ell,\rho)$ has an explicit expression, given above in equations (\ref{eq:expr_TGF}) and (\ref{eq:expr_ck}). By injecting this explicit expression of $J(\ell,\rho)$ (\ref{eq:expr_TGF}, \ref{eq:expr_ck}) in (\ref{eq:num_exp1}), the joint probability distribution $P(\vec{\ell},M|N)_{(0)}$ can be written
\bea
&&P(\vec{\ell},M|N)_{(0)} = \prod_{k=1}^{M-1} f(\ell_k) \int_0^\infty d Y G_>(Y,0,\ell_M) \, \e^{-Y/b} \nonumber \\
&&\times\prod_{k=1}^{M-1} \int_0^\infty \frac{{\rm d} \rho_k}{b}\, \delta\left(\sum_{k=1}^{M-1} \rho_k - Y\right) \delta\left(\sum_{k=1}^{M} \ell_k, N\right) \;.
\eea 
Finally, using the identity
\beq\label{eq:identity}
\prod_{k=1}^{M-1} \int_0^\infty d \rho_k \, \delta\left(\sum_{k=1}^{M-1} \rho_k - Y\right) = \frac{Y^{M-2}}{(M-2)!} \;,
\eeq
which can be easily shown by taking the Laplace transform on both sides of (\ref{eq:identity}) with respect to $Y$, we obtain an expression for the joint probability of the $\ell_k$ and $M$ as
\bea
&&P(\vec{\ell},M|N)_{(0)} = \prod_{k=1}^{M-1} f^{}(\ell_k) q(M,\ell_M) \delta\left(\sum_{k=1}^{M} \ell_k, N\right) \;, \label{eq:num_expfinal} \\
&&q(M,\ell_M) = \frac{1}{(M-2)! b^{M-1}} \int_0^\infty d Y \e^{-Y/b} \, Y^{M-2} \, G_>(Y,0,\ell_M) \label{eq:qma_exp} \;,
\eea
which has thus a structure very similar to the one found in the discrete case (\ref{num_joint_discrete}), but with different building blocks. Furthermore, the $\GF$ of $q^{}(M,\ell_M)$ in (\ref{eq:qma_exp}) with respect to $\ell_M$ can be obtained explicitly as \cite{GMS2015b}
\beq\label{eq:GFq}
\tilde q(M,z) = \sum_{\ell\ge1} q^{}(M,\ell) \, z^{\ell} = \frac{1}{b}\frac{1-\sqrt{1-z}}{(1+\sqrt{1-z})^{M-1}} = \frac{(1-\sqrt{1-z})^M}{b\, z^{M-1}} \;.
\eeq

From this joint distribution (\ref{eq:num_expfinal}), together with equations (\ref{eq:expr_TGF}) and (\ref{eq:GFq}), it is possible to compute the statistics of all the observables related to the ages of the record of the random walk bridge with symmetric exponential jumps. In particular one obtains rather straightforwardly the results for the distribution of the number of records in (\ref{eq:gB_exp}), for the record breaking probability in (\ref{eq:QN_bridge}) or for $\langle \ell_{\max,N}\rangle$ in (\ref{lmax_bridge}).

\vspace*{0.5cm}
{\bf Conclusion and open questions.} These results for the random walk bridge in equations (\ref{ampli_mu_discrete})--(\ref{lmax_bridge}) show that the record statistics of constrained random walks are quantitatively different from their counterpart obtained for a free random walk. The computations in this case are technically much harder and, for most of the observables related to records, exact results are only available for the lattice random walk and for the random walk with symmetric exponential jump. One of the main differences with the free random walk is that the record statistics for the bridge is not universal and depends, for finite $N$, on the details of the jump distribution, while it is completely universal (for continuous jump distributions) for the free random walk. Nonetheless, one expects that in the large $N$ limit the record properties of a random walk bridge (with continuous jump distribution) are, to leading order for large $N$, solely determined by the L\'evy index $\mu$ in (\ref{smallk.1}). This implies in particular that the asymptotic results obtained in the exponential case should describe the record statistics in the large $N$ limit of any random walk with continuous jumps and $\mu =2$. The generalisation of these results to arbitrary value of the L\'evy index $0< \mu < 2$ remains a challenging open question. It is also interesting to notice that the limiting value of $\langle \ell_{\max,N}\rangle/N$ obtained for the exponential case in (\ref{lmax_bridge}) is much more complicated than its counterpart, the constant $C$, obtained for the free random walk (\ref{lmax_sym_cont}) -- and we refer the reader to \cite{GMS2015b} for the study of the full distribution of $\ell_{\max,N}/N$ for the bridge random walk with exponential jump distribution. In particular, it is not given by the theory of Poisson-Dirichlet distributions (see section \ref{sec:PD} below) and it will be very interesting to extend these results to other jump distributions, with different L\'evy index $0<\mu < 2$.

\section{Record statistics for multiple random walks on a line} \label{sec:multi}

\begin{figure}
\centering
\includegraphics[width=0.8\textwidth]{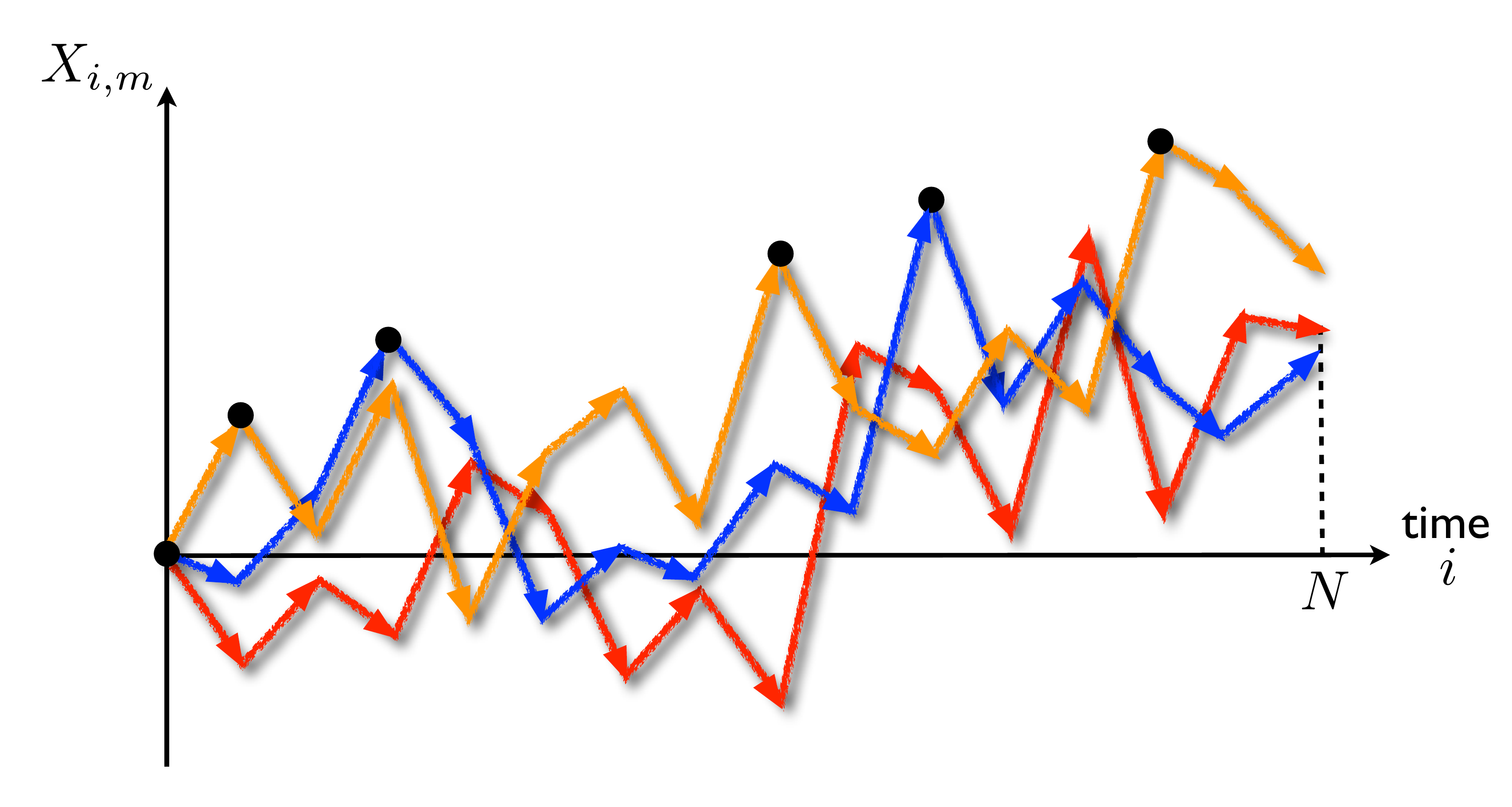}
\caption{Schematic trajectories of $K=3$ independent random walkers
up to step $N$, where $X_{i,m}$ denotes the position of the $m$-th walker
($m=1,2,\ldots, K$)
at step $i$, all of them starting at the origin at step $0$. A record 
happens at step $n$, if $X^{\rm max}_{n}>X^{\rm max}_{n'}$ for all 
$n'=0,1,2,\ldots, (n-1)$. The record values are shown by filled circles.}
\label{fig.multi}
\end{figure}

In the previous sections, we studied the statistics of the record number
in a sequence with entries $\{X_0=0, X_1,X_2,\ldots, X_N\}$ corresponding 
to the 
positions of a
{\em single} random walk at discrete times, starting at $X_0=0$. In this 
section, we will generalise some of these results for the single walker 
case to the case when one has $K\ge 1$ independent random walkers.
In this $K$-walker process, a record happens at an instant when the
maximum position of all the walkers at that instant exceeds all its 
previous values. The record statistics for this multiple walker case was
studied in great detail in ref \cite{WMS2012}, and it was found that
despite the fact that the walkers are independent, the record
statistics is rather rich, nontrivial, and partially universal even in this 
relatively simple model. Below we will describe the model precisely and 
outline the main results found in ref \cite{WMS2012}. For details of the 
computations, the reader may consult ref \cite{WMS2012}.

Consider $K\ge 1$ independent random walkers all starting at the origin 
at time $i=0$ [see figure (\ref{fig.multi})]. The position $X_{i,m}$ of the 
m-th walker ($m=1,2,\ldots, 
K$) at discrete time 
step $n$ evolves via the Markov evolution rule
\begin{equation}
X_{i,m}= X_{i-1,m} + \eta_{i,m} ,
\label{markov_K1}
\end{equation}
where $X_{0,m}=0$ for all $m=1,2,\ldots,K$ and the noise $\eta_{i,m}$ 
are i.i.d.~variables (independent from step to step and from walker to 
walker), each drawn from a symmetric distribution $\phi(\eta)$ as
in the previous section. We are interested in the record statistics of the
composite process. More precisely, consider at step $n$, the maximum
position of all the $K$ walkers
\begin{equation}
X^{\rm max}_{n}= {\rm max}\left[X_{n,1}, X_{n,2},\ldots, X_{n,K}\right]\, .
\label{max_K1}
\end{equation}
A record occurs at step $n$ if this maximum position at step $n$ is bigger 
than all previous maximum values, i.e., if $X^{\rm max}_{n}> X^{\rm 
max}_{n'}$ for all $n'=0,1,2,\ldots, (n-1)$ [see figure (\ref{fig.multi})]. 
In
other words, we are interested in the record statistics of the stochastic
discrete-time series $\{X^{\rm max}_n\}$, with the convention that
the initial position $X^{\rm max}_0=0$ is counted as a record. 
This new process, though derived from $K$ independent underlying Markov 
processes, is itself a rather complicated non-Markov process for $K>1$.
Consequently, for $K>1$, the simple renewal approach used before for the $K=1$ case
(which was valid since for $K=1$ the process is Markovian) breaks down and
one needs to find a new approach to compute the record statistics. We will see below
that while a new approach can be devised to compute the mean number of records for all $K\ge 1$, the
computation of the full distribution of the record number for $K>1$ is much more difficult
and remains partially an open problem. 

Let $M_{N,K}$ denote the number of number of records up to step $N$ of 
this composite $K$-walker process. 
Note that it is convenient
to use a notation that keeps track of the $N$-dependence of the number of records. 
In this section we are interested  in 
the statistics of $M_{N,K}$. 
For a single $K=1$ walker, we recall from the
previous section that the statistics of the record number $M_{N,1}$ is 
completely universal for all $N$, i.e., independent of the jump 
distribution $\phi(\eta)$ for symmetric and continuous $\phi(\eta)$.
In particular, the statistics is identical for Gaussian walkers as well as 
for L\'evy flights with index $0<\mu<2$. 
It turns out that for $K>1$, the statistics of $M_{N,K}$ is no
longer universal for all $N$ \cite{WMS2012}. However, for large $N$, a 
different
sort of universality emerges in the limit of large number of walkers
$K \gg 1$ \cite{WMS2012}, that we summarise below.

\vskip 0.5cm

\noindent{\bf {Summary of the main results:}}
In the large $N$ and large $K$ limit, there are essentially two universal
asymptotic behaviours of 
$M_{N,K}$, depending on whether the second moment $\sigma^2= 
\int_{-\infty}^{\infty} \eta^2\, \phi(\eta)\, d\eta$ of the
jump distribution is finite or divergent. 
For example, for Gaussian, exponential, uniform jump distributions
$\sigma^2$ is finite. In contrast, for L\'evy flights where 
$\phi(\eta)\sim
|\eta|^{-\mu-1}$ for large $\eta$ with the L\'evy index $0<\mu<2$, the 
second
moment $\sigma^2$ is divergent. In these two cases, the
following behaviours for the record statistics have emerged \cite{WMS2012}.

\vspace{0.3cm}

\noindent {\bf Case I ($\sigma^2$ finite):} Consider
first jump distributions $\phi(\eta)$ that are symmetric with
a finite second moment $\sigma^2=\int_{-\infty}^{\infty} \eta^2\, 
\phi(\eta)\,
d\eta $. In this case, the Fourier transform of the jump distribution 
${\hat \phi}(q)= \int_{-\infty}^{\infty} \phi(\eta)\,\e^{iq\eta}\, d \eta$ 
behaves, 
for small $q$, as
\begin{equation}
{\hat \phi}(q) = 1- \frac{\sigma^2}{2}\, q^2 + \ldots
\label{fourier1}
\end{equation}
Examples include the Gaussian jump distribution,
$\phi(\eta)=\sqrt{a/\pi}\,\e^{-a\,\eta^2} $, exponential jump distribution
$\phi(\eta)= 1/(2 b)\, \exp[-|\eta|/b] $, uniform jump distribution over 
$[-1/2,1/2]$, etc. For such jump distributions,
it was found \cite{WMS2012} that 
for large number of walkers $K$, the mean number of
records grows asymptotically for large $N$ and large $K$ as
\begin{equation}
\langle M_{N,K}\rangle \xrightarrow[N\to \infty]{K\to \infty} 
2\, \sqrt{\ln K}\,
\sqrt{N} \;.
\label{meanrecord1}
\end{equation}
Note that this asymptotic behaviour is universal in the sense that it does 
not
depend explicitly on $\sigma$ as long as $\sigma$ is finite.

Moreover, it was argued \cite{WMS2012} that for large $K$ and large $N$, 
the scaled random 
variable
$M_{N,K}/\sqrt{N}$ converges, in distribution, to the Gumbel form, i.e,
\begin{equation}
{\rm Prob}\left( \frac{M_{N,K}}{\sqrt{N}}\le x\right) \xrightarrow[N\to
\infty]{K\to
\infty} F_{1}\left[\left(x-2\,
\sqrt{\ln K}\right)\,\sqrt{\ln K}\right], \
{\rm with} \
F_1(z)=\exp\left[-\e^{-z}\right].
\label{gumbel.1}
\end{equation}
Indeed, for large $N$ and large $K$, the scaled variable
$M_{N,K}/\sqrt{N}$ converges, in distribution, to the maximum of $K$
independent random variables
\begin{equation}
\frac{M_{N,K}}{\sqrt{N}}\xrightarrow[N\to \infty]{K\to
\infty} M_K\, \quad {\rm where}\quad M_K= {\rm max}(y_1,y_2,\ldots, y_K)
\label{gumbel.2}
\end{equation}
where $y_i\ge 0$ are i.i.d.~non-negative random variables each drawn from the
distribution $p(y)= \frac{1}{\sqrt{\pi}}\, \e^{-y^2/4}$ for $y\ge 0$ and
$p(y)=0$ for $y<0$.

\vspace{0.3cm}
\noindent {\bf Case II ($\sigma^2$ divergent ):} Let us next consider
the opposite case, i.e., 
jump
distributions $\phi(\eta)$ such that the second moment $\sigma^2$ is 
divergent.
In this case, the Fourier transform ${\hat \phi}(q)$ of the noise 
distribution
behaves, for all
$g$, as
\begin{equation}
{\hat \phi}(q) = 1- |a\,q|^{\mu} + \ldots
\label{fourier2}
\end{equation}
where $0<\mu<2$.
Examples include L\'evy flights where $\phi(\eta)\sim |\eta|^{-\mu-1}$ 
with 
the 
L\'evy index $0<\mu< 2$. For the noise distribution in 
(\ref{fourier2}), it turns out \cite{WMS2012}, quite amazingly, that
in the large $N$ and large $K$ limit, the record statistics is (i) 
completely
universal, i.e., independent of $\mu$ and $a$,
(ii) more surprisingly and unlike in case I (corresponding to finite $\sigma$), the record statistics also
becomes independent of $K$ as $K\to \infty$. For example, it was proved 
that
for large $K$, the mean number of records grows asymptotically with $N$
as
\begin{equation}
\langle M_{N,K}\rangle \xrightarrow[N\to \infty]{K\to \infty}
\frac{4}{\sqrt{\pi}}\, \sqrt{N} \;,
\label{meanrecord2}
\end{equation}
which is exactly twice that of one walker, i.e., $\langle M_{N,K\to
\infty}\rangle = 2\, \langle M_{N,1}\rangle$ for large $N$.
Similarly, it was shown \cite{WMS2012} that the scaled variable 
$M_{N,K}/\sqrt{N}$, for large 
$N$
and large $K$, converges to a universal distribution
\begin{equation}
{\rm Prob}\left( \frac{M_{N,K}}{\sqrt{N}}\le x\right) \xrightarrow[N\to
\infty]{K\to
\infty} F_2(x) \;,
\label{distri2}
\end{equation}
which is independent of the L\'evy index $\mu$ as well as of the scale $a$
in (\ref{fourier2}).
While this universal distribution $F_2(x)$ was numerically computed 
rather
accurately \cite{WMS2012}, deriving it analytically
remains a challenging open problem. Numerically, it seems that $F_2(x)$ can be fitted very well with a Weibull form:
$F_2(x) \approx 1- \exp\left[-(b\,x)^{\gamma}\right]$, where the fitting parameters
$b\approx 0.89$ and $\gamma\approx 2.56$ \cite{WMS2012}. This means that the $\PDF$ $F_2'(x) \sim x^{\gamma-1}\, \exp\left[-(b\,x)^{\gamma}\right]$
for large $x$, indicating a faster than Gaussian tail as $x\to \infty$. 

\vskip 0.5cm

\noindent{\bf {Outline of the derivation for the mean number of records:}}
Let us briefly outline below the main idea behind the calculation of at least the mean number of records $\langle M_{N,K}\rangle$, and referring
the readers to \cite{WMS2012} for the derivation of the full distribution of $M_{N,K}$. 
Let $M_{n,K}$ be the number of records up to step $n$ for $K$ random walkers,
i.e., for the maximum process $X^{\rm max}_n$. Let us write (following equations (\ref{def_sigma}) and~(\ref{def_rate}))
\begin{equation}
M_{n,K}= M_{n-1,K}+ \sigma_{n} \;,
\label{recur.1} 
\end{equation} 
starting from $M_{0,K}=1$. Here $\sigma_{n}$ is a 
binary random 
variable taking values $0$ or $1$.
The variable $\sigma_{n}=1$ if a record happens at step $n$ and
$\sigma_{n}=0$
otherwise. Clearly, the total number of records up to step $N$ is
\begin{equation}
M_{N,K}= 1+ \sum_{n=1}^N \sigma_{n} \, .
\label{recordnumber.1}
\end{equation}
So, the mean number of records up to step $N$ is
\begin{equation}
\langle M_{N,K} \rangle = 1+ \sum_{n=1}^N \langle \sigma_{n}\rangle= 1+ 
\sum_{n=1}^N
r_{n,K} \;,
\label{mean.1}
\end{equation} 
where $r_{n,K}= \langle \sigma_{n}\rangle$ is just the record rate, 
i.e., the
probability that a record happens at step $n$ for 
the maximum process $X^{\rm max}_n$ of $K$ independent walkers. To compute 
the mean number
of records, we will first evaluate
the record rate $r_{n,K}$ and then sum over $n$, as in (\ref{mean.1}).

\begin{figure}
\centering
\includegraphics[width=0.7\textwidth]{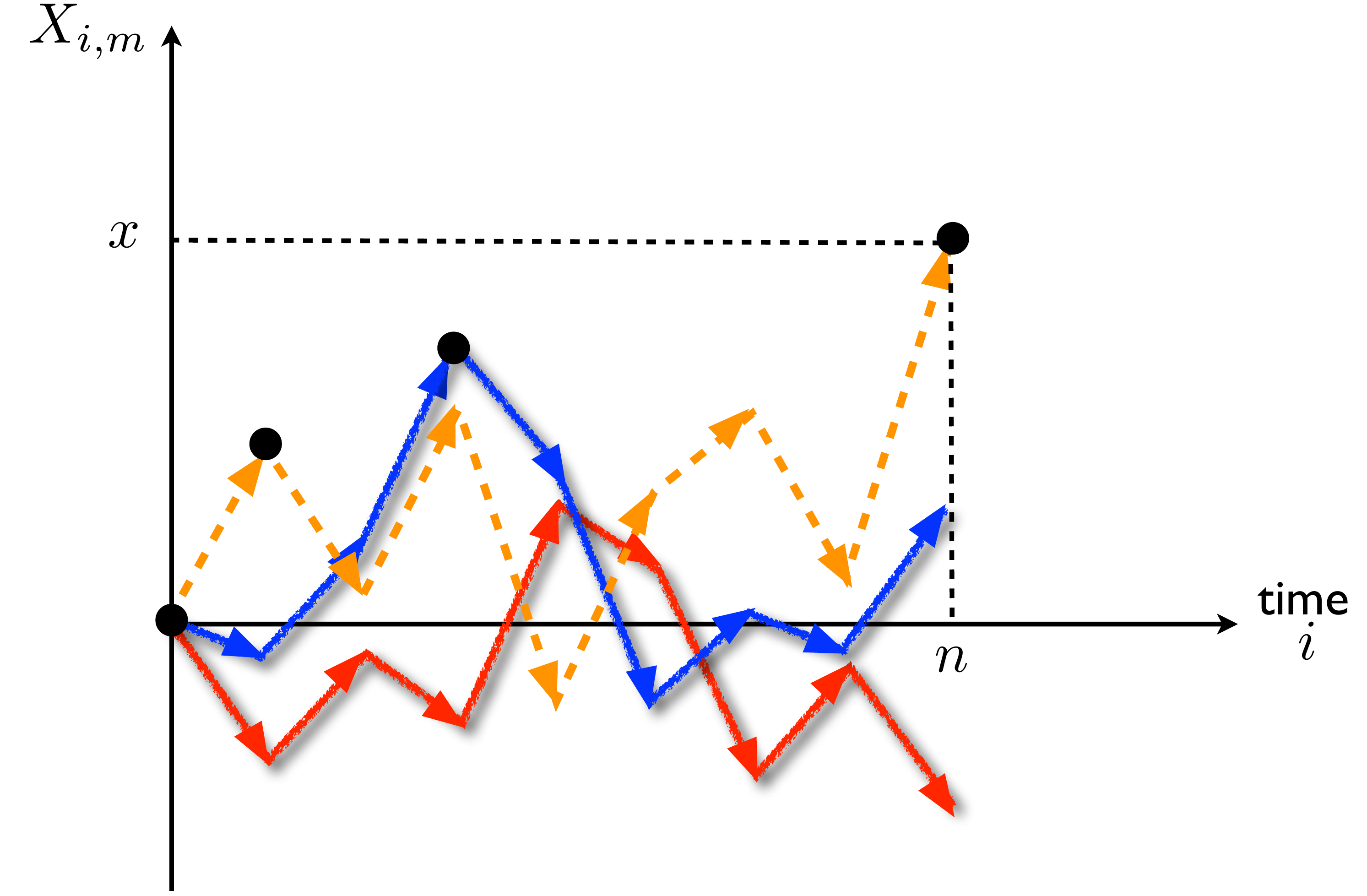}
\caption{A record happens at step $n$ with record value $x$ for $K=3$ walkers,
all starting at the origin (the index $m$ thus runs over $m=1, \cdots, K=3$).
This event corresponds to one walker (the dashed line) reaching the level $x$
for the first time at step $n$, while the other walkers stay below the level 
$x$ up to step $n$.}
\label{fig.event}
\end{figure}

To compute $r_{n,K}$ at step $n$, we need to sum
the probabilities of all trajectories that lead to a record event at step $n$.
Suppose that a record happens at step $n$ with the record value $x$ (see figure
\ref{fig.event}). This corresponds to the event that one of the $K$ walkers
(say the dashed trajectory
in figure \ref{fig.event}), starting at the origin at step $0$, has reached
the level $x$ for the first time
at step $n$, while the rest of the $K-1$ walkers, starting at the origin at step
$0$, have all stayed below the level $x$ until the step $n$. Also, the walker
that actually reaches $x$ at step $n$ can be any of the $K$ walkers. 
Finally this event can take place at any level $x>0$ and one needs to
integrate over the record value $x$.
Using the independence of $K$ walkers and taking into account the event
detailed above, one can
then write
\begin{equation}
r_{n,K} = K\, \int_0^{\infty} p_n(x)\, \left[q_n(x)\right]^{K-1}\, dx \;,
\label{rate.1}
\end{equation}
where $q_n(x)$ denotes the probability that a single walker, starting at the
origin, stays below the level $x$ up to step $n$ and $p_n(x)$ is the
probability density that a single walker reaches the level $x$ for the
first time at step $n$, starting at the origin at step $0$.
Note that $q_n(0)$ is precisely the probability that a single walker, starting
at $0$, stays below the level $0$ up to step $n$, and is identical to the
persistence probability $q(n)$ defined in (\ref{ql_def}). Hence $q_n(0)$, by Sparre Andersen theorem discussed before,
is completely universal (independent of $\phi(\eta)$ for symmetric and continuous $\phi(\eta)$)
\begin{equation}
q_n(0)= {{2n} \choose {n}}\, 2^{-2\,n}\, .
\label{qn0.1}
\end{equation}
Furthermore, it is easy to see, by the reflection principle, that the following identity holds~\cite{WMS2012}
\begin{equation}
\int_0^{\infty} p_n(x)\, dx= q_n(0)= {{2n} \choose {n}}\, 2^{-2\,n}\, .
\label{pn_qn}
\end{equation}

Fortunately, the generating functions of these two quantities $p_n(x)$ and $q_n(x)$
are known exactly for arbitrary jump distributions $\phi(\eta)$ (for a detailed discussion see \cite{WMS2012}).
They are given by the pair of formulae:
\begin{eqnarray}
\sum_{m\ge0}s^m\, \int_0^{\infty} p_m(x)\, \e^{-\lambda x}\, dx &= &
\psi(s,\lambda) \label{pm.2} \\
\sum_{m\ge0} s^m\, \int_0^{\infty} q_m(x)\, \e^{-\lambda x}\, dx &= & 
\frac{1}{\lambda \sqrt{1-s}}\, \psi(s,\lambda)\, .
\label{qm.2}
\end{eqnarray}
where the function $\psi(s,\lambda)$ is given explicitly by 
\begin{equation}
\psi(s,\lambda)= \exp\left[-\frac{\lambda}{\pi}\,\int_0^{\infty} \frac{\ln [1-s
{\hat \phi}(q)]}{\lambda^2+q^2}\, dq\right] \quad {\rm where}\quad {\hat \phi}(q)= 
\int_{-\infty}^{\infty} \phi(\eta)\, \e^{iq\eta}\, d\eta \;.
\label{ivanov.2}
\end{equation}
The formula in (\ref{qm.2}) is known in the literature as the
celebrated Pollaczek-Spitzer formula \cite{Pollaczek,Spitzer} and has
been used in a number of works to derive exact results on the maximum of a
random jump process \cite{Darling,GRS2011,CM2005,KMR2011}. Interestingly,
this formula has also been useful to compute
the asymptotic behaviour of
the flux of particles to a spherical trap in three
dimensions \cite{MCZ2006,ZMC2007,ZMC2009}, as well as in the exact computation of 
the order and gap statistics for random walks in a recent series of papers \cite{MMS13,MMS14,SM2012,MSM16,MS16}.
The formula in (\ref{pm.2}) can be derived from a more general formula derived by Ivanov~\cite{Ivanov94} (see
ref \cite{WMS2012} for a detailed discussion), and it was used previously in the study of record increments in section \ref{sec:increments} [see (\ref{eq:ivanov})].

Let us first remark that by making a change of variable $\lambda x=y$ on
the left side of (\ref{qm.2}) and taking $\lambda
\to \infty$, one recovers the universal Sparre Andersen result for all $m$
\begin{equation}
\sum_{n\ge0} q_n(0)\, s^n = \frac{1}{\sqrt{1-s}} \Longrightarrow
q_n(0)= {{2n}\choose n}\frac{1}{2^{2n}} \;.
\label{sa1}
\end{equation}
In particular, for large
$m$, $q_n(0)\approx 1/\sqrt{\pi n}$.
Hence, for the case of a single walker $K=1$, it follows from
 (\ref{rate.1}) that the record
rate at step $n$ is simply given by
\begin{equation}
r_{n,1}= \int_0^{\infty}p_n(x)\, dx= q_n(0)= {{2n}\choose n}\frac{1}{2^{2n}}
\xrightarrow{n\to \infty} \frac{1}{\sqrt{\pi n}} \;,
\label{K1.1}
\end{equation}
where we used the identity in (\ref{pn_qn}).
Consequently, one recovers from (\ref{mean.1}), for $K=1$, the universal result for the mean number of records mentioned
in (\ref{avg_rec.1}) and in particular, its large $N$ asymptotic limit in (\ref{avg_rec.2})
\begin{equation}
\langle M_{N,1}\rangle \xrightarrow{N\to \infty} \frac{2}{\sqrt{\pi}}\, \sqrt{N}\, . 
\label{K1.2}
\end{equation}

In contrast, for $K>1$, we need the full functions $p_n(x)$ and $q_n(x)$
to compute the record rate in (\ref{rate.1}). This is hard to compute
explicitly for all $n$. However, one can make progress in computing the
asymptotic behaviour of the record rate
$r_{n,K}$ for large $n$ and large $K$ \cite{WMS2012}.
In turns out that for large $n$, the integral in (\ref{rate.1}) is dominated
by the asymptotic scaling behaviour of the two functions $p_n(x)$ and $q_n(x)$
for large $n$ and large $x$. These asymptotics can be derived explicitly \cite{WMS2012} starting
from the two generating functions
in equations (\ref{pm.2}) and (\ref{qm.2}) respectively.
The next step is to use these asymptotic expressions in the main formula
in (\ref{rate.1}) to determine the record rate $r_{n,K}$ at step $n$ for
large $n$ and large $K$. Here, we will skip all the details and just use the
main results for the asymptotics derived in ref \cite{WMS2012} to
derive the results announced in equations (\ref{meanrecord1}) and (\ref{meanrecord2}). 
The asymptotic behaviour of $p_n(x)$ and $q_n(x)$ depend on whether
$\sigma^2=\int_{-\infty}^{\infty} \eta^2 \, \phi(\eta)\, d\eta$ is finite or
divergent. This gives rise to the two cases mentioned in section II.

\vskip 0.3cm

\noindent {\bf Case I ($\sigma^2$ finite):} In this case, it was shown
in \cite{WMS2012} that in the scaling limit $x\to \infty$, $n\to \infty$ but keeping the
ration $x/\sqrt{n}$ fixed, $p_n(x)$ and $q_n(x)$ approach the following
scaling behaviour
\begin{eqnarray}
p_n(x)&\to & \frac{1}{\sqrt{2\sigma^2}\, n}\,
g_1\left(\frac{x}{\sqrt{2\,\sigma^2\,
n}}\right)\,,
\quad
{\rm where}\quad g_1(z)= \frac{2}{\sqrt{\pi}}\,z\, \e^{-z^2} \;, \label{pm1scaling} \\
q_n(x) & \to & h_1\left(\frac{x}{\sqrt{2\,\sigma^2\,
n}}\right)\,,
\quad
{\rm where}\quad h_1(z)= {\rm erf}(z) \;,\label{qm1scaling}
\end{eqnarray}
where ${\rm erf}(z)= \frac{2}{\sqrt{\pi}}\, \int_0^z \e^{-u^2}\, du$. Note that
$dh_1(z)/dz= g_1(z)/z$.

\vskip 0.3cm

\noindent {\bf Case II ($\sigma^2$ divergent):} In this case the
Fourier
transform of the jump distribution ${\hat \phi}(q)$ has the small $q$ behaviour
as in (\ref{fourier2}), and it was shown \cite{WMS2012} that in the scaling limit
when $x\to \infty$, $n\to \infty$, but keeping the ratio $x/n^{1/\mu}$ fixed,
\begin{eqnarray}
p_n(x)&\to & \frac{1}{n^{1/2+1/\mu}}\, g_2\left(\frac{x}{n^{1/\mu}}\right) \:, \label
{pm2scaling} \\
q_n(x) & \to & h_2\left(\frac{x}{n^{1/\mu}}\right). \label
{qm2scaling}
\end{eqnarray}
While it is hard to obtain explicitly the full scaling functions $g_2(z)$
and $h_2(z)$ for all $z$, one can compute the large $z$ asymptotic
behaviour and obtain
\begin{eqnarray}
g_2(z) &\underset{z \to \infty}{\approx} & \frac{A_{\mu}}{z^{1+\mu}} \;,\label{g2largez} \\
h_2(z) & \underset{z \to \infty}{\approx} & 1- \frac{B_{\mu}}{z^{\mu}} \;, \label{h2largez}
\end{eqnarray}
where the two amplitudes are
\begin{eqnarray}
A_{\mu} &= & \frac{2\mu}{\sqrt{\pi}}\, \beta_{\mu} \;, \label{amu1} \\
B_{\mu} &=& \beta_{\mu} \;, \label{bmu1}
\end{eqnarray}
with the constant $\beta_{\mu}$ given by \cite{WMS2012}
\begin{eqnarray}\label{unif_beta}
\beta_{\mu} = \frac{a^\mu \Gamma(\mu) \sin{(\frac{\mu \pi}{2})}}{\pi} \; \quad {\rm for}\quad 0<\mu<2\; .
\end{eqnarray}

Next we use these asymptotic behaviour of $p_n(x)$ and $q_n(x)$
in (\ref{rate.1}) to deduce the large $n$ behaviour of the record rate.
Noting that for large $n$, the integral is dominated by the scaling regime,
we substitute in (\ref{rate.1}) the scaling forms of $p_n(x)$ and $q_n(x)$
found in equations (\ref{pm1scaling}), (\ref{qm1scaling}), (\ref{pm2scaling}) and
(\ref{qm2scaling}). This gives, for large $n$,
\begin{equation}
r_{n,K}\approx \frac{K}{\sqrt{n}}\, \int_0^{\infty} g(z)\,
[h(z)]^{K-1}\, dz \;,
\label{rate.2}
\end{equation}
where $g(z)=g_{1,2}(z)$ and $h(z)=h_{1,2}(z)$ depending on the two cases (I and II).
So, we notice that in all cases the record rate decreases as
$n^{-1/2}$ for large $n$, albeit with different $K$-dependent prefactors in
the two cases. Hence, the mean number of records $\langle M_{N,K}\rangle $ up to
step $N$ grows, for large $N$, as
\begin{equation}
\langle M_{N,K}\rangle \approx \alpha_K\, \sqrt{N} \,, \quad {\rm where}\quad
\alpha_K= 2 K\,
\int_0^{\infty} g(z)\, [h(z)]^{K-1}\, dz \;.
\label{avgrec.1}
\end{equation}

The constant $\alpha_K$ can be estimated for large $K$. From (\ref{avgrec.1}), the
constant $\alpha_K$ can be re-expressed as
\begin{equation}
\alpha_K= 2\, \int_0^{\infty} \frac{g(z)}{h'(z)}\, \frac{d}{dz} \{[h(z)]^K\}\, dz \;,
\label{alphaN}
\end{equation}
where $h'(z)= dh/dz$. Noticing that $h(z)$ is an increasing function of
$z$ approaching $1$ as $z\to \infty$, the dominant contribution to the
integral for large $K$ comes from the large $z$ regime. Hence, we need
to estimate how the ratio $g(z)/h'(z)$ behaves for large $z$. Let us
again consider the two cases separately.

\vskip 0.3cm

\noindent {\bf Case I ($\sigma^2$ finite): } In this case, we have
explicit expressions for $g_1(z)$ and $h_1(z)$ respectively in equations (\ref{pm1scaling}) and (\ref{qm1scaling}). Hence we get
\begin{eqnarray}
\alpha_K &=& 2\, \int_0^{\infty} dz\, z\, \frac{d}{dz} [{\rm erf}(z)]^K \\
&=& \int_0^{\infty} dy\, y\, \frac{d}{dy} [{\rm erf}(y/2)]^{K}.
\label{alphaN1.case1}
\end{eqnarray}
The right hand side of (\ref{alphaN1.case1}) has a nice interpretation.
Consider $K$ i.i.d.~positive random variables $\{y_1,\,y_2,\ldots, y_K\}$, each
drawn from the
distribution: $p(y)= \frac{1}{\sqrt{\pi}}\,\e^{-y^2/4}$ for $y\ge 0$
and $p(y)=0$ for $y<0$. Let $Y^{\rm max}_K$
denote their
maximum. Then the cumulative distribution function of the maximum is given by
\begin{equation}
{\rm Prob}(Y^{\rm max}_K\le y)= \left[\int_0^{y} p(y')\,dy'\right]^K= [{{\rm erf}(y/2}]^K\,.
\label{maximum.1}
\end{equation}
The probability density of the maximum is then given by: $\frac{d}{dy} [{{\rm
erf}(y/2}]^K$. Hence, the right hand side of (\ref{alphaN1.case1}) is just the average
value $\langle Y^{\rm max}_K \rangle $ of the maximum. This gives us an identity
for all $K$
\begin{equation}
\alpha_K= \langle Y^{\rm max}_K\rangle \;.
\label{maximum.2}
\end{equation}
From the standard extreme value analysis of i.i.d.~variables \cite{gumbel,SM_review}, it is easy to show 
that to leading order for large $K$, $\langle Y^{\rm max}_K\rangle \approx 2\sqrt{\ln K}$
which then gives, via (\ref{avgrec.1}), the leading asymptotic
behaviour of the mean record number announced in (\ref{meanrecord1})
\begin{equation}
\langle M_{N,K} \rangle \xrightarrow[N\to \infty]{K\to \infty} 2 \sqrt{\ln K}\, \sqrt{N} \;.
\label{avgrec.case1}
\end{equation}

\vskip 0.3cm

\noindent {\bf Case II ($\sigma^2$ divergent): } To evaluate $\alpha_K$
in (\ref{alphaN}), we note that when $\sigma^2$ is divergent, unlike in case
I,
we do not have the full explicit form of the scaling functions $g_2(z)$
and $h_2(z)$. Hence evaluation of $\alpha_K$ for all $K$ seems difficult, since we do not have
explicit forms of these scaling functions for all $z$.
However, we can make progress for large $K$. As mentioned before, for large $K$,
the dominant contribution to the integral in (\ref{alphaN}) comes
from large $z$. For large $z$, using the asymptotic expressions in equations (\ref{g2largez}) and (\ref{h2largez}), we get
\begin{equation}
\frac{g_2(z)}{h_2'(z)}\xrightarrow{z\to \infty} \frac{A_\mu}{\mu\,
B_{\mu}}=\frac{2}{\sqrt{\pi}} \;,
\label{ratiolargez}
\end{equation}
where we have used (\ref{amu1}) and (\ref{bmu1}) for the expressions
of $A_\mu$ and $B_{\mu}$. We next substitute this asymptotic constant
for the ratio $g_2(z)/h_2'(z)$ in the integral on the right hand side of (\ref{alphaN}).
The integral can then be performed trivially and we get, for large $K$, 
\begin{equation}
\alpha_K \xrightarrow{K\to \infty} \frac{4}{\sqrt{\pi}} \;.
\label{alphaN.case2}
\end{equation}
From (\ref{avgrec.1}) we then get, for the mean record number up to $N$ steps, the result mentioned
in (\ref{meanrecord2}), i.e., 
\begin{equation}
\langle M_{N,K} \rangle \xrightarrow[N\to \infty]{K\to \infty}
\frac{4}{\sqrt{\pi}}\, \sqrt{N} \, .
\label{avgrec.case2}
\end{equation}
In contrast to case I in (\ref{avgrec.case1}), here
the mean record number becomes independent of $K$ for large $K$.

\vskip 0.5cm

\noindent {\bf Full distribution of $M_{N,K}$ for $K>1$:} While for the mean record number, a fairly complete
analysis can be done for all $K\ge 1$ \cite{WMS2012}, the corresponding result for the full distribution of $M_{N,K}$ is much 
less complete for $K>1$. In ref \cite{WMS2012}, it was argued that in case I when $\sigma^2$ is finite,
$M_{N,K}$ approaches a Gumbel variable asymptotically [see (\ref{gumbel.1}) and (\ref{gumbel.2})].
Intuitively this result derives from the fact that the record number $Y_{N,K}$ statistically becomes
identical (up to a constant scale factor) to the global maximum of all the $K$ walkers up to step $m$.
In contrast, in case II when $\sigma^2$ is divergent, the asymptotic scaling function $F_2(x)$ in
 (\ref{distri2}) is known only numerically. In this case, there is no correspondence
to the global maximum. Moreover, the fact that this scaling function $F_2(x)$ is completely
independent of $0<\mu<2$ is rather intriguing. For more details on the distribution of $M_{N,K}$ for $K>1$, the reader may consult
ref \cite{WMS2012}.

\vskip 0.5cm

\noindent {\bf Open problems:} The record statistics for multiple, independent random walkers is a fascinating problem
where many questions are still very much open. Even though the effective process (the maximum process $X^{\rm max}_n$)
 for $K>1$ walkers is highly non-Markovian, some results can still be derived analytically as we discussed above.
Still there are many questions which seem solvable (tantalizingly), but still remain wide open. 
For example,
as mentioned above, determining analytically the $\mu$-independent scaling function $F_2(x)$
associated with the distribution of $M_{N,K}$ for
L\'evy walks (with a divergent variance of the jump distribution) remains a challenging open problem. Even the fact that
$F_2(x)$ decays faster than Gaussian for large $x$ has not
been proved, but only observed numerically. Finally, 
we have not discussed at all the statistics of record ages $\{\ell_1,\,\ell_2,\ldots, \ell_M\}$ for $K>1$ walkers.
While we have full knowledge of the age statistics for $K=1$, so far there have been no studies on the
age statistics for $K>1$. It would be extremely interesting to know, e.g., how the maximal or the minimal
age behave for~$K>1$.

\section{Generalisations and extensions}
\label{sec:generalise}

In this section we give several natural generalisations and extensions to the questions addressed in the bulk of the present review. 

\subsection{The longest excursion}
\label{sec:longest}

As mentioned in the previous sections, the study of the ages of the records for a general random walk bears strong similarities with the excursion theory of the lattice random walk and Brownian motion.
The joint distribution of these excursions has the same renewal structure as in
(\ref{renewal.1}) with a corresponding distribution of the individual ages $f(\ell_k) \sim \ell_k^{-3/2}$ for $\ell_k \gg 1$ and $k<M$. 

It is then natural to consider more general renewal processes with a generic $f(\ell)$~\cite{GL2001} and address similar extreme value questions concerning $\ell_{\max, N}$ or $Q_N$ \cite{PY97,GMS2009,GMS2015}. 
Renewal processes have found a wide range of applications in probability theory \cite{Feller,Cox1962} and in statistical physics, including phase ordering kinetics \cite{GL2001,BBDG99}, blinking quantum dots \cite{blinking}, persistence properties \cite{MBCS1996,DHZ1996}, etc. In many of these applications, time is a continuous variable and we denote by $t$, instead of $N$, the duration of the process. 
As before, the lengths of time, $\ell_1, \ell_2, \ldots, \ell_{M-1}$ are identical, while $\ell_M$ is different from the others, 
however these variables are not independent, due to the global constraint that fixes their sum to be exactly equal to $t$. Nevertheless it can be shown that if $f(\ell)$ decays faster than $1/\ell^2$ for large $\ell$, i.e., if $f(\ell)$ admits a first moment, then this constraint is unimportant in the large $t$ limit, as far as the extreme-value statistical properties are concerned. Consequently, 
the limiting distribution of $\ell_{\max}(t)$, properly centred and scaled, is given by the classical theory of extreme value statistics for i.i.d.~random variables \cite{GMS2015}. However, if $f(\ell) \sim \ell^{-1-\alpha}$ for large $\ell$ with $0<\alpha<1$, the scaled variable $\ell_{\max}(t)/t$ converges to a non-trivial distribution when $t \to \infty$. The exponent $\alpha$ is called the persistence exponent \cite{Satya_review,Bray_review,derrida94,bray94,AS2015}.
For $\alpha = 1/2$, one recovers the result found for Brownian motion [see (\ref{lmax_sym_cont}) and (\ref{def_fR})] but, for arbitrary $\alpha \in (0,1)$, the limiting distribution depends continuously on $\alpha$. In particular, the first moment is given by \cite{GMS2009,PY97,GMS2015}
\bea\label{CI_theta}
\lim_{t \to \infty} \frac{\langle\ell_{\max}(t) \rangle}{t} = C(\alpha) \;, \; C(\alpha) = \int_0^\infty \frac{1}{1+y^{\alpha}\e^y \, \gamma(1-\alpha,y)} \, dy \;.
\eea
%
An important outcome of this study is to show that, for $0<\alpha<1$, there is universality of the results with respect to the distribution of intervals $f(\ell)$ \cite{GMS2009,GMS2015}. 
Note that the result obtained for the CTRW in (\ref{lmax_CTRW}) corresponds to $\alpha = \gamma/2$, i.e., $c(\gamma) = C(\gamma/2)$, where $c(\gamma)$ is defined in (\ref{lmax_CTRW}).

A similar generalisation can be made for the ages of the records of i.i.d.~variables \cite{GL2008}.
Starting from (\ref{eq:ratios}), a natural generalisation consists in considering the times $t_k$ as representing the locations of the zeros of a multiplicative process in continuous time $t$ such that the variables $U_k = t_{k-1}/t_k$ have the common distribution $\rho(u)=\theta u^{\theta-1}$.
This yields
\beq\label{eq:goltheta}
\lim_{t \to \infty} \frac{\langle\ell_{\max}(t) \rangle}{t} = Q(\theta)=\int_0^\infty d s\,\e^{-s-\theta E(s)},
\eeq
which gives back~(\ref{c1}) for $\theta=1$ -- we recall that $E(s) = \int_s^\infty d y \, \e^{-y}/y$.

In principle, the longest excursion $\ell_{\max}(t)$ can be defined for any stochastic process, not only for renewal processes or multiplicative processes. 
An interesting instance in the context of coarsening systems is the case where the process is the magnetisation (local or global) of a ferromagnet and in this case, the intervals $\ell_k$ denote the times between two consecutive sign changes of the magnetisation. In many situations, it was shown numerically \cite{GMS2009} that the longest excursion $\langle\ell_{\max}(t) \rangle$ grows linearly with time $t$ (for $t \gg 1$) and with an amplitude which, rather remarkably, is well approximated by $C(\alpha)$ in (\ref{CI_theta}), $\alpha$ being the associated persistence exponent of the process \cite{derrida94,bray94,Bray_review}. 
Likewise, comparisons of the theoretical prediction~(\ref{eq:goltheta}) to the equivalent quantities measured numerically on various approximately multiplicative processes can be found in \cite{GMS2009}.

This observable $\langle\ell_{\max}(t) \rangle$ was also computed numerically for the fractional Brownian motion with Hurst exponent $H$ \cite{GRS2010}. For $H=1/2$, it corresponds to Brownian motion but for $H \neq 1/2$, it is a non-Markovian process. Nevertheless the persistence exponent $\alpha$ is known exactly for any value of $H$, and it is given by $\alpha = 1 - H$ \cite{Mol1999,KKMCBS1997}. Numerical simulations show that $\langle\ell_{\max}(t) \rangle$ also grows linearly with time $t$ (for $t \gg 1$) and, except for $H = 1/2$, the amplitude $\langle\ell_{\max}(t) \rangle/t$ shows a clear deviation from the renewal result (\ref{CI_theta}) with $\alpha = 1-H$. This is one of the rare observables for fractional Brownian motion that clearly exhibits its non-Markovian nature \cite{GRS2010}.

\subsection{Different definitions of the longest age or the longest excursion}
\label{sec:different}

As noted previously, for a random walk, the last record does not stand on an equal footing with the others. To probe the effects of this last record on various observables associated to the ages, ref \cite{GMS2014} studied distinct sequences of the ages of random walks differing only by their last element. For instance, to avoid the ambiguity of the age of the last record, one may simply discard $\ell_M$ and consider the restricted list of ages $\{\ell_1, \ell_2, \dots, \ell_{M-1}\}$, which is a set of identically distributed random variables (though not independent since their sum is constrained to be less than $N$). This set is a rather natural choice as a toy model for the statistics of avalanches close to the depinning transition of elastic manifolds in random media \cite{LDW09}. In this context, $\ell_k$ with $k<M$ corresponds precisely to the size of the $k$-th avalanche, while the quantity $\ell_M$ in this context does not have a direct physical meaning. The study performed in \cite{GMS2014} showed that observables such as $\ell_{\max, N}, \ell_{\min,N}$ or $Q_N$ are actually quite sensitive to this last record, even in the limit $N \to \infty$. The mechanism behind this high sensitivity is that these observables associated to the ages are dominated by rare events, whose statistics is controlled, to a large extent, precisely by the last record duration. 
This study extends to the case of excursions as well as to more general renewal processes~\cite{GMS2015} (see also \cite{Sch95} in the mathematical literature).

\subsection{Joint distribution of the ranked ages: Poisson-Dirichlet distributions}
\label{sec:PD}

In both cases discussed previously, i.e., in the i.i.d.~case as well as in the random walk case, one can study the full order statistics of the ages of the records, 
$ \ell^{(1)}_N > \ell^{(2)}_N > \cdots > \ell^{(M)}_N$, with $ \ell^{(1)}_N \equiv\ell_{\max,N}$. In the limit of large $N$, one can show that $\ell^{(k)}_N$ grows linearly with $N$, for any fixed $k$, and that the joint distribution of the scaled ranked ages $\ell^{(1)}_N/N, \ell^{(2)}_N/N, \ldots$ converges to a limiting distribution, which depends on two real parameters $0 \leq \alpha\leq 1$ and $\theta > - \alpha$, which are known under the name of Poisson-Dirichlet distributions, denoted by ${\rm PD}(\alpha,\theta)$. The distribution with parameters ${\rm PD}(0,1)$ describes the statistics of the (scaled and ranked) ages of the records for the i.i.d.~sequence, while ${\rm PD}(1/2,0)$ describes the statistics of the (scaled and ranked) ages of the records for random walks. 

The family of distributions ${\rm PD}(0,\theta)$, with $\theta>0$, was initially introduced by Kingman in ref \cite{Kin75}. They describe the statistics of the (scaled and ranked) time intervals between successive zeros of the multiplicative process, indexed by $\theta$, and discussed 
in the paragraph above (\ref{eq:goltheta}). These distributions naturally arise in the study of asymptotic distributions of random ranked relative frequencies in various contexts ranging from number theory \cite{Ver1986} and combinatorics \cite{VS1977} to Bayesian statistics \cite{Fer93} or population genetics \cite{Ewe88} (for reviews see \cite{Pitman_StFlour,Fen10}). 
This one-parameter family of distribution was later generalised by Pitman and Yor to a two-parameter family denoted by ${\rm PD}(\alpha,\theta)$, with $0\leq \alpha \leq 1$ and $\theta > - \alpha$, in order to study the ranked statistics of excursions of Brownian motion and Bessel processes \cite{PY97}. In this framework, the distribution PD$(\alpha,0)$ describes the (scaled and ranked) statistics of the intervals between successive zeros of a renewal process with a corresponding distribution of the individual ages that decays algebraically as $f(\ell) \sim \ell^{-1-\alpha}$, with $0 \leq \alpha \leq 1$, which is the renewal process discussed in the paragraph above (\ref{CI_theta}). Hence, using the aforementioned correspondence for a random walk between the record breaking events and the zeros of a lattice random walk, we see indeed that the joint distribution of the (scaled and ranked) ages of the records of random walk $(\ell_N^{(1)}/N, \ell_N^{(2)}/N, \dots, \ell_N^{(M)}/N)$ converges, in the large $N$ limit, to PD$(1/2,0)$ \cite{SV2016}.

There is no simple explicit expression for the Poisson-Dirichlet distributions PD$(\alpha,\theta)$ but 
ref \cite{PY97} provided various probabilistic interpretations and constructions of this joint law. 
In particular, they gave a nice description of ${\rm PD}(\alpha,\theta)$, in terms of stick-breaking processes that generalises the multiplicative process described in the paragraph above (\ref{eq:goltheta}) (for a review see \cite{Pitman_StFlour}). For instance, this construction allows to compute the average value of the $k$-th longest age of 
the records for i.i.d.~random variables [corresponding to PD$(0,1)$] and for the random walk case [corresponding to PD$(1/2,0)$]. For the i.i.d.~case, one finds
\beq\label{eq:lambdak}
\langle \ell^{(k)}_N \rangle = \lambda^{(k)}\, N + {\cal O}(1) \;, \; \lambda^{(k)} = \frac{1}{\Gamma(k)} \int_0^\infty ds \, \e^{-s} E(s)^{k-1} \e^{-E(s)} \;,
\eeq
where the function $E(s)$ is defined below (\ref{eq:goltheta}). In particular, by setting $k=1$ in (\ref{eq:lambdak}), one recovers $\lambda^{(1)} = \lambda$, where $\lambda$ is the Golomb-Dickman constant given in (\ref{c1}). The first values can be evaluated numerically, yielding $\lambda^{(2)}=0.20958\ldots$, $\lambda^{(3)} = 0.08831\ldots$.
One can easily check from (\ref{eq:lambdak}) that $\sum_{k\geq 1} \lambda^{(k)} = 1$. On the other hand, in the case of a random walk, one finds \cite{PY97,SV2016}
\bea\label{average_lk}
\langle \ell^{(k)}_N\rangle \approx C^{(k)} \, N \;, \; C^{(k)} = \frac{1}{2^{k-1}} \int_0^\infty \frac{y^{-1/2}\,\e^{-y} \Gamma(-1/2,y)^{k-1}}{(y^{-1/2} \e^{-y} + \gamma(1/2,y))^k} \, dy \;,
\eea 
where
\bea
\Gamma(\nu,x) = \int_x^\infty t^{\nu-1}\,\e^{-t} \, dt \;,
\eea
is the upper incomplete Gamma function. In particular, by setting $k=1$ in (\ref{average_lk}), one recovers $C^{(1)} = C$, where $C$ is given in (\ref{lmax_sym_cont}). The first values can be evaluated numerically, yielding $C^{(2)}=0.14301\ldots$, $C^{(3)} = 0.06302\ldots$ \cite{SV2016}. Here also one can check that $\sum_{k\ge1} C^{(k)} = 1$, as expected.

\subsection{Excursions for the tied-down random walk, the Brownian bridge and related renewal processes}
The probability distribution of the longest interval between two consecutive zeros
of a lattice random walk starting and ending at the origin, and of its continuum
limit, the Brownian bridge, is another related subject of interest.
This problem was first addressed by Wendel \cite{Wen64}, then revisited in several works.
In \cite{CGtd} the problem is revisited and extended to renewal processes with the ``tied-down'' condition, i.e., the last interval drawn with the common distribution $f(\ell)$ (as defined in section \ref{sec:longest}) exactly terminates at time $t$.
Interestingly, the corresponding situation for the records of random walks is when one imposes the condition that the last record of the random walk occurs exactly at $N$, which is the {\it fixed} number of steps of the random walk \cite{CGtd}, or,
otherwise stated, when the maximum of the random walk occurs exactly at the last step $N$.
Extension of this study to the statistics of the ranked longest intervals can also be performed \cite{CGtd}.

Related studies have been addressed recently in the context of mixed-order phase transitions and we refer the reader to ref \cite{BMSM2016} for more details.

\section{Other related problems and open questions}
\label{sec:other}

In this section, we discuss related works or various questions related to records that have been recently studied 
in the literature.

\subsection{Effects of measurement error and noise}

In all the previous studies, a record happens at step $k$ if the $k$-th entry exceeds all the previous
entries [see (\ref{def_record})]. However, to apply these results to real data one needs to reconsider 
the definition of a record in a more pragmatic way. Indeed, in many applications, the observations of the data $X_i$ are subject to 
uncertainty, due to instrument error $\delta$ or noise $\xi$. For instance, $\delta$ can be the assurance 
limit of the detector, while $\xi$ can describe white noise from an instrument reading. It is then natural to ask how the presence of measurement error $\delta$ or noise $\xi$ affects the records statistics, in particular the growth of the average record number $\langle M \rangle$ with the size of the sample $N$. Related questions were raised in the statistics literature, e.g., in the context of $\delta$-exceedance records \cite{BBP96,GLS12}, and more recently in the physics literature \cite{rounding,edery,PK16}.

Here, we first discuss the presence of a (fixed) measurement error $\delta$. We define $X_k$ to be a record breaking event, called a $\delta$-record for short, if it exceeds all previous values in the sequence, by at least $\delta$, i.e., if
\begin{eqnarray}\label{delta_record}
X_k > \max \{ X_1, \ldots, X_{k-1}\} + \delta \,,
\end{eqnarray}
where, here, $\delta > 0$ (in the case $\delta < 0$, $X_k$ is sometimes called a {\it near record} \cite{BPS05}). In fact, most of the results related to this problem (except \cite{edery} that we discuss below) have been obtained for the case of i.i.d.~random variables. In this case, it was shown~\cite{rounding} that an immediate consequence of introducing an error parameter $\delta >0$ is that the strong universality of the record statistics for i.i.d.~[as in (\ref{rate_iid})] is lost and replaced by an explicit dependence on the right tail of the parent distribution of the variables $X_i$ --reminiscent of the different universality classes existing for the extreme value statistics for i.i.d.~random variables. We refer the reader to \cite{gregor_review} for a review of these results for i.i.d.~random variables and focus here instead on the case of strongly correlated variables, for which much less is know.

Following ref \cite{edery} we thus consider a random walker that starts at $X_0 =0$ and evolves according to (\ref{evolrw.1}) with a continuous and symmetric jump distribution $\phi(\eta)$. If one denotes by $r_k \equiv r_k(\delta)$ the probability that a record is broken at step $k$, the mean number of record is simply given by (\ref{average_rw_simple}), i.e., 
\begin{equation}\label{average_rw_delta}
\langle M \rangle = \sum_{k=0}^N r_k(\delta) \;.
\end{equation}
By definition $X_0=0$ is a record and thus $r_0 = 1$ and, for $k\geq 1$, $r_k(\delta)$ is defined by
\beq\label{rk_delta_1}
r_k(\delta) = {\rm Prob}[x_k - \delta > \max(X_0, \cdots, X_{k-1})] \;.
\eeq
Thus $r_k(\delta)$ is the probability of the event that the walker arrives in $x_k-\delta$ for the first time at time $k$ while staying below $x_k - \delta$ at all intermediate steps between 0 and time~$k$ (and where one needs to finally integrate over $x_k \geq \delta$). To compute this probability, it is convenient to change variables and define $y_i = x_k-x_{k-i}$, i.e., observe the sequence $\{y_i\}$ with respect to the last position and measure time backwards, as explained for the random walk without error, i.e., for $\delta = 0$, in figure \ref{fig:record_survival} -- where, here, in addition the $y$-axis is also reversed. Then, $r_{k}(\delta)$ is the probability that the ``new'' walker $y_i$, starting at the new origin at $i=0$, makes a jump $\geq \delta$ at the first step and then subsequently, up to $k$ steps, stays above $\delta$, i.e.,
\beq\label{rk_delta_2}
r_k(\delta) = {\rm Prob}[y_1 \geq \delta, \cdots, y_{k} \geq \delta | y_0 = 0] \;.
\eeq
To compute the probability in (\ref{rk_delta_2}), we decompose the corresponding event into the first step where the walker, starting in $y_0 = 0$ jumps to $y_1 =\delta + u$ where $u \geq 0$ and the $k-1$ subsequent steps during which the random walk stays above $\delta$. Hence, writing $y_i = \delta + u_i$, we can re-express $r_k(\delta)$ as \cite{edery} 
\beq\label{rk_delta_3}
r_k(\delta) = \int_0^\infty du \,\phi(\delta+ u) q_{k-1}(u) \;, 
\eeq
where $q_n(u)$ is the probability that the random walk, starting at $u \geq 0$, stays positive up to step $n$. This probability $q_n(u)$ was studied in detail before, see (\ref{qm.2}) and below it. In particular, from (\ref{qm.2}), one can show \cite{edery} that for large $n$, keeping $u$ fixed,
\beq\label{rk_delta_4}
q_n(u) \approx \frac{h(u)}{\sqrt{\pi\, n}} \;\;, {\rm with} \;\; \tilde h(\lambda) = \int_0^\infty du \, \e^{-\lambda u} h(u) = \frac{\psi(1,\lambda)}{\lambda} \;,
\eeq 
where the function $\psi(z,\lambda)$ is given in (\ref{ivanov.2}) and depends explicitly on the jump distribution $\phi(\eta)$. Finally, combining equations (\ref{average_rw_delta}) and (\ref{rk_delta_4}), one finds that for $N \gg 1$ the average number of records $\langle M \rangle$ behaves has \cite{edery}
\beq
\langle M \rangle \approx S(\delta) \sqrt{N} \;, \; S(\delta) = \frac{2}{\sqrt{\pi}} \int_0^\infty du\, \phi(u+\delta) h(u) \;.
\eeq
Hence for an arbitrary jump distribution the average record number grows universally as $\sqrt{N}$ (as in the case $\delta = 0$) while the prefactor $S(\delta)$ depends explicitly on the jump distribution \cite{edery}. Computing explicitly $S(\delta)$ for an arbitrary distribution is a very hard task and exact results exist only in very special cases. For instance, for a symmetric exponential jump distribution $\phi(\eta) = 1/(2b) \e^{-|\eta|/b}$, one finds that $S(\delta) = (2/\sqrt{\pi}) \e^{-\delta/b}$ \cite{edery}. On the other hand, for jump distributions with a power law tail $\phi(\eta) \sim |\eta|^{-1-\mu}$, with $\mu > 0$, one finds that $S(\delta)$ decays algebraically for large $\delta$, $S(\delta)\sim \delta^{-\mu + \alpha}$ with $\alpha = \mu/2$ for $\mu \leq 2$ while $\alpha = 1$ for $\mu \geq 2$.

The influence of the measurement noise $\xi$ was also studied in ref \cite{edery}. To quantify the effects of the noise, one considers that a record is registered at step $k$ if 
\beq
X_k + {\cal N}(0,\xi) \Delta x > \max\{X_0, \cdots, X_{k-1}\} \;, \label{def_record_gamma}
\eeq
where ${\cal N}(0,\xi)$ is a Gaussian random variable of zero mean and standard deviation $\xi$, while $\Delta x$ is the characteristic length scale of the jump. Hence, in (\ref{def_record_gamma}), the term ${\cal N}(0,\xi) \Delta x$ mimics the effects of noise measurement. In that case, it was found numerically that, for random walk, the mean number of records still grows like $\sqrt{N}$, i.e., $\langle M \rangle \approx T(\xi)\sqrt{N}$ with an amplitude $T(\xi)$ which is an increasing function of $\xi$ for all $\xi$. Hence in this case, the noise $\xi$ leads to an erroneous counting of the records, rendering an apparent mean number of records $\langle M \rangle$ larger than the actual one. We refer the reader to ref \cite{edery} for more details on $\xi$-records, in particular for a possible use of $T(\xi)$ to infer ``signal-to-noise'' ratio in diffusion-type experiments.

\subsection{Statistics of superior records}\label{sec:superior}

Let us consider a time series generated by $N$ i.i.d.~random variables $X_1, X_2\dots, X_N$ with a continuous density $p(X)$. We denote by 
$X_{\max,n}$ the value of the last record after $n$ time steps, i.e., the value
of the running maximum:
\beq\label{def_xmax}
X_{\max,n}=\max(X_1, X_2\ldots,X_n),
\eeq
and we denote its average by
\beq\label{def_max_av}
\langle X_{\max,n}\rangle=\mu_n \;.
\eeq
Note that we use the subscript $n$ (and not $N$) to emphasize that this is a {\it running} maximum. The study is thus restricted to distributions with finite average.
A superior sequence $\{X_1, X_2\dots, X_N\}$ is such that the running maximum is always above its average, i.e., $X_{\max,n}>\mu_n$ for all $n\le N$ \cite{BK2013}.
The probability $S_N$ of this event is found to decay~as
\beq\label{eq:Sn}
S_N\sim N^{-\beta},
\eeq
where the exponent $\beta$ is the root of an integral equation \cite{BK2013}.
This exponent is non-universal and depends on the choice of the distribution $p(X)$.
For instance for a uniform distribution, $\beta\approx 0.450$, while for an exponential distribution, $\beta\approx 0.621$.
This latter value turns out to be an upper bound for this exponent, whatever the choice of distribution~$p(X)$. 

Similarly, the probability that a sequence is inferior (that is, with running maximum always below its average) also decays with a power law
\beq\label{eq:In}
I_N\sim N^{-\alpha},
\eeq
where the exponent $\alpha$ is computable explicitly and depends on the parent distribution $p(X)$. For instance, for the uniform distribution $\alpha = 1$, while $\alpha = \e^{-\gamma_E} = 0.561459\ldots$, where $\gamma_E$ is the Euler gamma constant, for the exponential distribution. These results were compared to real earthquake data in ref \cite{BK2013}, to which we refer for more details. 

In a subsequent work \cite{BK2014}, these results were generalised to a strongly correlated time series, namely when the $X_i$ correspond to the position of a symmetric random walk after $i$ steps (\ref{evolrw.1}). While the problem is well defined for any type of random walk, including L\'evy flights, analytical results are known only for jumps $\eta_i$ with mean zero and a finite variance, such that the random walk converges after a large number of steps to Brownian motion. In this case, 
the average running maximum (\ref{def_max_av}) is known to grow as $\langle X_{\max,n}\rangle\approx\sqrt{2/\pi} \sqrt{n}$ for $n \gg 1$.
The behaviours~(\ref{eq:Sn}) and (\ref{eq:In}) are again found to hold \cite{BK2014}.
The exponent $\beta\approx 0.382$ and $\alpha\approx 0.241$ are the roots of parabolic cylinder functions
\beq
D_{2\beta+1}(\sqrt{2/\pi})=0,\quad D_{2\alpha}(\sqrt{-2/\pi})=0.
\eeq
Note the close similarity of this problem with the problem of survival of a diffusing particle in the presence of an absorbing moving boundary whose position grows like $\propto \sqrt{t}$ (see e.g., \cite{Bray_review,KR1996}). We conclude this section by mentioning that 
the study of these questions related to superior and inferior records for L\'evy flights remains a challenging open problem.

\subsection{Scaling exponents for ordered maxima}\label{sec:ordered}

Consider now $N$ i.i.d.~random variables $\{X_1,\ldots,X_N\}$ with a common distribution $p(X)$. 
A plot of the running maximum $X_{\max,n}$ against $n$ is a staircase with jumps at the successive occurrences of records, as in figure \ref{fig_record}.
Consider now $K \geq 1$ such sequences.
These sequences are said to be perfectly ordered if the corresponding staircases do not cross~\cite{BKL2015}.
The probability of this event has a power-law decay \cite{BKL2015}: 
\beq
P_{N,K}\sim N^{-\sigma_K},
\eeq
where the exponents $\sigma_K$ are known analytically only for $K=2,3$, 
\beq
\sigma_2=1/2,\quad\sigma_3\approx 1.302931 \;,
\eeq
where $\sigma_3$ is the root of some transcendental equation. For the two latter cases the probability $P_{N,K}$ is universal, i.e., it does not depend on the distribution $p(X)$.
The property is conjectured to hold for $K>3$. Bounds upon the exponents demonstrate that $\sigma_K$ should grow as $K$ but an explicit computation of $\sigma_K$ remains unknown. 

As above in section \ref{sec:superior}, the same problem can be generalised to $K$ random walks \cite{BK2014b}.
Likewise, $P_{N,K}$ is the probability that the maxima of the positions of $K$ independent random walkers are ordered up to step $N$
\beq
P_{N,K}\sim N^{-\nu_K},
\eeq
as demonstrated by numerical simulations \cite{BK2014b}.
The only analytical result concerns two random walks for which
\beq
\nu_2=\frac{1}{4}.
\eeq
An interesting connection between the case of i.i.d.~random variables and the case of random walks is given in \cite{BKL2015} where the relation $\nu_K\approx\sigma_K/2$ is observed (numerically) to be a good approximation.

\subsection{Incremental records}

Other interesting questions concern the sequence of record increments, which were discussed earlier in the context of random walks in section \ref{sec:increments}. We recall that, if one denotes the record values of a time series by $R_k$, the increments $\rho_k$ are defined, for $k \geq 1$, by $\rho_k = R_{k+1}-R_k$, as depicted in figure \ref{Fig:increments}. Intuitively, one expects that the sequence of increments $\{\rho_1, \rho_2, \cdots, \rho_{M-1}\}$ is typically decreasing. Indeed, as time goes on, the value of the current record is growing and it seems rather unlikely that the next record improves upon it by a large amount. Motivated by this intuition, Miller and Ben-Naim asked the following question \cite{MBN13}: what is the probability ${\cal Q}_N$ that the sequence of increments is monotonically decreasing up to step $N$? Such records with monotonically decreasing increments are called ``incremental records''.

This probability ${\cal Q}_N$ was first investigated in the case of i.i.d.~random variables whose parent distribution $p(X)$ has a finite support, $p(X) = \mu (1-X)^{\mu-1}$, for $0\leq X \leq 1$, and $p(X) =0$ otherwise. Numerical simulations showed that ${\cal Q}_N$ decreases algebraically for large $N$
\beq\label{eq:Q_incremental.1}
{\cal Q}_N \sim N^{-\nu} \;, \; N \gg 1 \;,
\eeq
with a non-trivial exponent $\nu$, which in addition depends on $\mu$ \cite{MBN13}. Computing this exponent $\nu$ turns out to be quite difficult and an exact computation was possible only for the case $\mu = 1$, which corresponds to a uniform distribution of the 
variables $X_i$. In this case, $\nu$ is given by the solution of an ``eigenvalue'' equation, and it can be evaluated numerically with high precision, yielding $\nu = 0.317621 \ldots$. No analytical solution exists for other values of $\mu$ nor for other types of distribution of the variables $X_i$. But the existing results already suggest that, for i.i.d.~random variables, ${\cal Q}_N$ is a rather non-trivial observable which is quite sensitive to the parent distribution $p(X)$.

In a subsequent work \cite{GMS2016}, this probability ${\cal Q}_N$ was studied in the case where the variables $X_i$ are the positions of a random walk as in (\ref{evolrw.1}) with a continuous and symmetric jump distribution $\phi(\eta)$. To compute ${\cal Q}_N$, it is convenient to write it as ${\cal Q}_N = 
\sum_{M\geq 1}{\cal Q}_N(M)$ where ${\cal Q}_N(M)$ is the joint probability that an 
$N$-step random walk sequence has exactly $M$ records and that the record 
increments are monotonically decreasing. This probability ${\cal Q}_N(M)$ is 
obtained from the joint probability of the increments $\rho_k$ and the number of records $P(\rho_1, \dots, \rho_{M-1},M|N)$ studied in section \ref{sec:increments} [see (\ref{eq:gf_jp})], by integrating it over $\rho_1>\rho_2>\cdots>\rho_{M-1}>0$. It turns out that this $(M-1)$-dimensional nested integral can be computed exactly \cite{GMS2016}, which allows to obtain the $\GF$ of ${\cal Q}_N(M)$ with respect to $N$ in a quite simple form, valid for all $M\geq 1$
\begin{eqnarray}\label{GF_inter}
\sum_{N\geq 0} z^N {\cal Q}_N(M) = \tilde q(z) \frac{1}{(M-1)!} \left[\tilde f(z) \right]^{M-1} \;,
\end{eqnarray}
in terms of the $\GF$ $\tilde q(z)$ and $\tilde f(z)$ of the survival probability (\ref{ql_def}) and of the first-passage probability (\ref{fl_def}) respectively. Quite remarkably, for continuous and symmetric jump distributions $\phi(\eta)$, this $\GF$ in (\ref{GF_inter}) is completely universal, as $\tilde q(z)$ and $\tilde f(z)$ are themselves 
universal, thanks to the Sparre Andersen theorem (\ref{qz_exact}). By summing up this formula (\ref{GF_inter}) over $M$ from 1 to $\infty$, one obtains the $\GF$ of ${\cal Q}_N$ as \cite{GMS2016} 
\beq\label{GF_Q_increment}
\sum_{N \geq 0} z^N {\cal Q}_N = \tilde q(z)\, \e^{\tilde f(z)} = \frac{1}{\sqrt{1-z}} \,\e^{1 - \sqrt{1-z}} \;.
\eeq
From (\ref{GF_Q_increment}), ${\cal Q}_N$ can be computed explicitly, with the result \cite{GMS2016}
\begin{equation}\label{eq:explicit_qn}
{\cal Q}_N = \e\,\sqrt{\frac{2}{\pi}} \, 
K_{N+1/2}(1) \frac{2^{-N}}{N!} = 
\sum_{j=0}^N {{N+j} \choose N} \frac{2^{-N-j}}{(N-j)!} \;,
\end{equation}
where $K_\nu(x)$ is the modified Bessel function of index $\nu$. 
For instance, ${\cal Q}_1 = 1$, ${\cal Q}_2 = 7/8$, ${\cal Q}_3 = 37/48$, etc. 
For large $N$, we find that ${\cal Q}_N$ decays as a power law
\begin{eqnarray}\label{eq:Qn_asympt}
{\cal Q}_N \sim \frac{{\cal A}}{\sqrt{N}} \;, \; 
{\cal A} = \frac{\e}{\sqrt{\pi}} = 1.53362 \ldots \; ,
\end{eqnarray}
which holds for any random walk with a continuous and symmetric jump distribution $\phi(\eta)$, hence even for L\'evy flights. Therefore this universal result found for random walks is quite different from the results for i.i.d.~random variables where, despite the fact that ${\cal Q}_N$ also decays algebraically, ${\cal Q}_N \sim N^{-\nu}$ (\ref{eq:Q_incremental.1}), it is much more sensitive (including the exponent $\nu$) to the distribution of the variables $X_i$. 

\section{Conclusion}

In this review we have presented various aspects of the record statistics of a time series with stochastic entries. While this topic has been a subject of study since the early fifties, most of the results were derived in the case where the entries are i.i.d.~random variables. This i.i.d.~case has been covered in detail both in the mathematics literature (see e.g., the textbooks \cite{Nevzorov,ABN1992,Res1987}) and, more recently, in the physics literature \cite{gregor_review,GL2008,SM_review}, where the study of records received a renewed interest. In this review we recalled the main results for the record statistics in the i.i.d.~case. In particular, this part also contains some detailed results on the statistics of the ages of records, for which it is hard to find explicit results in the previous surveys. We also note that, even in the i.i.d.~case, there remain some non-trivial open problems, notably concerning the record increments (see section \ref{sec:other}).

The main focus of the present review has been on the case of the time series whose entries correspond to the positions of a discrete-time random walker/L\'evy flight on a line. This is a natural example of a time series with strongly correlated entries. The computation of record statistics for a strongly correlated time series is, in general, very hard and challenging. However for the random walk case, many questions concerning record statistics can be addressed analytically as reviewed in this article.


The reason for solvability in this case can be traced back to the renewal structure of the underlying Markov process (see section \ref{sec:renewal}). As emphasized in this review, calculating various observables associated to the record statistics of this time series, makes very interesting links to first-passage properties as encoded in the Sparre Andersen theorem~(\ref{SA.1}), as well as to extreme value statistics of random walks, as captured by 
the rather sophisticated results of Pollaczeck-Spitzer (\ref{pm.2}) and Hopf-Ivanov (\ref{qm.2}). 
These tools turn out to be extremely useful to analyse the records for a variety of random walk models, including random walks and L\'evy flights with a linear drift, constrained random walks like the random walk bridge, continuous time random walks, as well as multiple random walks.

We hope that the analytical methods presented in this review will be useful to study the record statistics of
other models of strongly correlated time series, including the challenging issue of non-Markovian processes. For instance, in a recent paper, the record statistics for the number of distinct sites of a random walker on a fully connected lattice has been studied analytically \cite{Tur2015}. Even though the evolution of the position of the random walker is Markovian, the temporal evolution of the number of distinct sites visited is strongly history dependent and hence is a non-Markovian process. Amongst other non-Markovian models, one can cite the random acceleration process \cite{Bur14} (or the integrated random walk in discrete time), which evolves according to $X_{i+1}-2 X_i + X_{i-1} = \eta_i$, where $\eta_i$ are i.i.d.~random variables. Although in this case $X_i$ is a non-Markovian process, the two-dimensional process $(X_i,V_i)$, where $V_i = X_{i}-X_{i-1}$ is the velocity, is Markovian. Hence, it may be possible to generalise the renewal structure in phase space in order to study the record statistics of the random acceleration process. For more general non-Markovian processes, like the fractional Brownian motion for instance, such a renewal structure does not exist. Nevertheless, first-passage properties as well as extreme value statistics might provide a useful guideline and framework to study the record statistics of such non-Markovian processes.

\section*{Acknowledgments}

We thank A. Bar, E. Ben-Naim, B. Berkowitz, J.-P. Bouchaud, Y. Edery, J. Franke, R. Garcia-Garcia, A. Kostinski, P. Krapivsky, J. Krug, A. Kundu, J.-M. Luck, I.~Marzuoli, Ph. Mounaix, D. Mukamel, L. Palmieri, J. Pitman, A. Rosso, S. Redner, S. Sabhapandit, W. Tang, G. Wergen and R. M. Ziff, for collaboration and useful discussions.

\section*{References}


\begin{thebibliography}{9}


\bibitem{Cha1952}
K. N. Chandler, The Distribution and Frequency of Record Values, J. Roy. Statist. Soc., Ser. B {\bf 14}, 220 (1952).

\bibitem{hoyt}
D. V. Hoyt, Weather records and climatic change, {Climatic Change} {\bf 3}, 243 (1981).

\bibitem{basset}
G. W. Basset, Breaking recent global temperature records, {Climatic Change} {\bf 21}, 303 (1992).


\bibitem{SZ1999}
B. Schmittmann and R. K. Zia, ``Weather'' records: Musings on cold days after a long hot Indian summer, Am. J. Phys. {\bf 67}, 1269 (1999).



\bibitem{benestad}
R. E. Benestad, How often can we expect a record event?, {Climate Res.} {\bf 25}, 1 (2003).

\bibitem{RP2006}
R. Redner and M. R. Petersen, Role of global warming on the statistics of record-breaking temperatures, {Phys. Rev. E} {\bf 74}, 061114 (2006).

\bibitem{WK2010}
G. Wergen and J. Krug, Record-breaking temperatures reveal a warming climate, {Europhys. Lett.} {\bf 92}, 30008 (2010).

\bibitem{AB2010}
A. Anderson and A. Kostinski, Reversible Record Breaking and Variability: Temperature Distributions across the Globe, {J. Appl. Meteor. Clim.} {\bf 49}, {1681} (2010). 





\bibitem{WHK2013}
G. Wergen, A. Hense and J. Krug, Record occurrence and record values in daily and monthly temperatures, Clim. Dynam. {\bf 22}, 1 (2013).



\bibitem{records_finance}
G. Barlevy and H. N. Nagaraja, Characterization in a random record model with a non-identically distributed initial
Record, {J. Appl. Prob.} {\bf 43}, 1119 (2006); G. Barlevy, Identification of search models using record statistics, {Rev. Econ. Stud.} {\bf 75}, 29 (2008).

\bibitem{WBK2011}
G. Wergen, M. Bogner and J. Krug, Record statistics for biased random walks, with an application to financial data, {Phys. Rev. E} {\bf 83}, 051109 (2011). 

\bibitem{SL2014}
B. Sabir and M. S. Lanthanum, Record statistics of financial time series and geometric random walks, Phys. Rev. E {\bf 90}, 032126 (2014).


\bibitem{records_hydrology}
N. C. Matalas, Stochastic Hydrology in the Context of Climate Change, {Climatic Change} {\bf 37}, 89 (1997); R. M. Vogel, A.
Zafirakou-Koulouris, and N. C. Matalas, Frequency of record-breaking floods in the United States, {Water Res. Research} 
{\bf 37}, 1723 (2001).

\bibitem{Gembris}
D. Gembris, J. G. Taylor and D. Suter, Sports statistics: Trends and random fluctuations in athletics, {Nature} {\bf 417}, 506 (2002). 

\bibitem{sports}
E. Ben-Naim, S. Redner and F. Vazquez, Scaling in Tournaments, {Europhys. Lett.} {\bf 77}, 30005 (2007).


\bibitem{FWK2012}
J. Franke, G. Wergen and J. Krug, Correlations of record events as a test for heavy-tailed distributions, 
Phys. Rev. Lett. {\bf 108}, 064101 (2012).





%
%
%
%





\bibitem{glick}
N. Glick, Breaking records and breaking boards, {Amer. Math. Monthly} {\bf 85}, 2 (1978).

\bibitem{gregor_review}
G. Wergen, {Records in stochastic processes -- 
Theory and applications}, {J. Phys. A: Math. Th.} {\bf 46}, 223001 (2013).

\bibitem{gumbel}
E. J. Gumbel, {Statistics of Extremes}, Dover (1958).

\bibitem{Gal87} J. Galambos, The Asymptotic Theory of Extreme Order Statistics (R.E. Krieger Publishing Co., Malabar, 1987).


\bibitem{Redner_book}
S. Redner, {A guide to first-passage processes}, Cambridge University Press, Cambridge, (2001).

\bibitem{Satya_review} S. N. Majumdar, {Persistence in nonequilibrium systems}, {Curr.\ Sci.}\ {\bf 77}, pp. 370--375 (1999).

\bibitem{Bray_review}
A. J. Bray, S. N. Majumdar and G. Schehr, Persistence and first-passage properties in non-equilibrium systems, {Adv. Phys.} {\bf 62}, 225 (2013).

\bibitem{jensen}
J. H. Jensen, Evolution in Complex Systems: Record Dynamics in Models of Spin Glasses, Superconductors and Evolutionary Ecology, {Adv. Solid State Phys.} {\bf 45}, 95 (2005).

\bibitem{sibani}
P. Sibani, G. F. Rodriguez and G. G. Kenning, Intermittent quakes and record dynamics in the thermoremanent magnetisation of a spin-glass, {Phys. Rev. B} {\bf 74} 224407 (2006); P. Sibani, Linear response in aging glassy systems, intermittency and the Poisson statistics of record fluctuations, {Eur. Phy. J. B} {\bf 58}, 483 (2007).

\bibitem{oliveira}
L. P. Oliveira, H. J. Jensen, M. Nicodemi and P. Sibani, Record dynamics and the observed temperature plateau in the magnetic creep-rate of type-II superconductors, {Phys. Rev. B} {\bf 71}, 104526 (2005). 

\bibitem{fisher}
D.~S. Fisher, Collective transport in random media: from
superconductors to earthquakes, {Phys. Rep.} {\bf 301}, 113 (1998).

\bibitem{sibani_littlewood}
P. Sibani and P. B. Littlewood, Slow dynamics from noise adaptation, {Phys. Rev. Lett.} {\bf 71}, 1482 (1993). 

\bibitem{ABBM}
B. Alessandro, C. Beatrice, G. Bertotti and A. Montorsi, Domain-wall dynamics and Barkhausen effect in metallic ferromagnetic materials. I. Theory, {J. Appl. Phys.} {\bf 68}, 2901--2907 (1990).

\bibitem{LDW09}
P. Le Doussal and K. J. Wiese, {Driven particle in a random landscape: Disorder correlator, avalanche distribution, and extreme value statistics of records}, {Phys. Rev. E} {\bf 79}, 051105 (2009).

\bibitem{sibani_fitness}
P. Sibani, M. Brandt and P. Alstrom, Evolution and extinction dynamics in rugged fitness landscapes, {Int. J. Mod. Phys. B} {\bf 12}, 361 (1998).


\bibitem{krugjain}
J. Krug and K. Jain, Breaking records in the evolutionary race, {Physica A} {\bf 358}, 1 (2005).

\bibitem{franke}
J. Franke, A. Kl\"ozer and J. Arjan G. M. de Visser and J.~Krug, Evolutionary accessibility of mutational pathways, {PLos Comp. Biol.} {\bf 7}, e1002134 (2011). 

\bibitem{RBSY2016}
M. Robe, S. Boettcher, P. Sibani and P. Yunker, Record Dynamics: Direct Experimental Evidence from Jammed Colloids, 
EPL {\bf 116}, 38003 (2016).


\bibitem{PRLKM2016}
G. Pak, F. Raischel, S. Lennartz-Sassinek, F. Kun and I. G. Main, Record breaking bursts during the compressive failure of porous materials, Phys. Rev. E {\bf 93}, 033006 (2016). 



\bibitem{GL2008}
C. Godr\`eche and J.-M. Luck, A record-driven 
growth process, {J. Stat. Mech.} P11006 (2008).


\bibitem{SLJ2013}
S. C. L. Srivastava, A. Lakshminarayan and S. R. Jain, Record statistics in random vectors and quantum chaos, EPL {\bf 101}, 10003 (2013).


\bibitem{SL2015}
S. C. L. Srivastava, A. Lakshminarayan, Records in the classical and quantum standard map, 
Chaos, Solitons \& Fractals {\bf 74}, 67 (2015).





\bibitem{ZCDS1998}
S. Zapperi, P. Cizeau, G. Durin and H.~E. Stanley, Dynamics of a ferromagnetic domain wall: Avalanches, depinning transition, and the Barkhausen effect,, Phys. Rev. B {\bf 58}, 6353 (1998). 

\bibitem{Nevzorov}
V. B. Nevzorov, {Records: Mathematical Theory}, Am. Math. Soc. (2004). 

\bibitem{ABN1992}
B. C. Arnold, N. Balakrishnan and H. N. Nagaraja, Records (New York: Wiley), (1998).




\bibitem{Res1987}
S. I. Resnick, {Extreme Values, Regular Variation, and Point Processes}, Springer, New York, (1987). 





\bibitem{Kru2007}
J. Krug, Records in a changing world, J. Stat. Mech. P07001 (2007).


\bibitem{FWK2010}
J. Franke, G. Wergen and J. Krug, Records and sequences of records from random variables with a linear trend, J. Stat. Mech. P10013 (2010). 

\bibitem{WFK2011}
G. Wergen, J. Franke and J. Krug, Correlations between record events in sequences of random variables with a linear trend, J. Stat. Phys. {\bf 144}, 1206 (2011).




\bibitem{MZ2008}
S. N. Majumdar and R. M. Ziff, Universal record statistics of random walks and L\'evy flights, {Phys. Rev. Lett.} {\bf 101}, 050601 (2008). 


\bibitem{satya_leuven} S. N. Majumdar, {Universal first-passage properties of discrete-time random walks and 
L\'evy flights on a line: Statistics of the global maximum and records}, {Physica A} {\bf 389}, 4299 (2010).


\bibitem{SM_review}
G. Schehr and S. N. Majumdar, Exact record and order statistics of random walks via first-passage ideas, in "First-Passage Phenomena and Their Applications", Eds. R. Metzler, G. Oshanin, S. Redner. World Scientific (2014), arXiv:1305.0639.



\bibitem{Riordan}
J. Riordan, {Introduction to combinatorial analysis}, Dover, New-York (2002). 

\bibitem{rounding}
G. Wergen, D. Volovik, S. Redner and J. Krug, Rounding Effects in Record Statistics, {Phys. Rev. Lett.} {\bf 109}, 164102 (2012).

\bibitem{Schorrock}
R. W. Schorrock, {On record values and record times}, J. Appl. Prob. {\bf 9}, 316 (1972).

\bibitem{Ahsanullah}
M. Ahsanullah, {Record Values Theory and Applications}, New York, University Press of America
Inc., (1995). 

\bibitem{Lorenzo}
L. Palmieri, S. N. Majumdar and G. Schehr, in preparation.

\bibitem{Neuts}
M. F. Neuts, {Waiting times between record observations}, J. Appl. Prob. {\bf 4}, 206 (1967).

\bibitem{Finch_book}
S. R. Finch, \emph{Mathematical constants}, Cambridge University Press, pp. 284--292 (2003). 

\bibitem{MMS09}
S. N. Majumdar, K. Mallick and S. Sabhapandit, Statistical properties of the final state in one-dimensional ballistic aggregation, {{Phys. Rev. E}} {\bf 79}, 021109 (2009).

\bibitem{Gou96}
X. Gourdon, {Combinatoire, Algorithmique et G\'eom\'etrie des Polyn\^omes}, PhD thesis, Ecole Polytechnique, (1996). 

\bibitem{PY97}
J. W. Pitman and M. Yor, The two-parameter Poisson-Dirichlet distribution derived from a stable subordinator, {Ann. Probab.} {\bf 25}, 855 (1997). 

\bibitem{SP1966}
L. A. Shepp and S. P. Lloyd, Ordered cycle lengths in a random permutation, {{Trans. Amer. Math. Soc.}} {\bf 121}, 340 (1966). 

\bibitem{PLDW2009}
P. Le Doussal and K. J. Wiese, Driven particle in a random landscape: disorder correlator, avalanche distribution and extreme value statistics of records, Phys. Rev. E {\bf 79}, 051105 (2009).

\bibitem{MSW2012}
S. N. Majumdar, G. Schehr and G. Wergen, {Record statistics and persistence for a random walk with a drift}, {J. Phys. A: Math. Theor.} {\bf 45}, 
355002 (2012). 

\bibitem{Feller} W. Feller, {Introduction to Probability Theory and 
Its Applications} (Wiley, New York, 1966), Vol.~2.


\bibitem{Cox1962} D. R. Cox, {Renewal theory} (London: Methuen) (1962).

\bibitem{GL2001}
C. Godr\`eche and J.-M. Luck, Statistics of the occupation time of renewal processes, J. Stat. Phys. {\bf 104}, 489 (2001). 

\bibitem{SA54}
 E. Sparre Andersen, {On the fluctuations of sums of random variables I}, { Math.\ Scand.} 
{\bf 1}, 263 (1953); {On the fluctuations of sums of random variables II}, { Math. Scand.} {\bf 2}, 195 (1954).

\bibitem{WMS2012}
G. Wergen, S. N. Majumdar and G. Schehr, Record statistics for multiple random walks, {Phys. Rev. E} {\bf 86}, 011119 (2012). 

\bibitem{BG1990} J.-P. Bouchaud and A. Georges, Anomalous diffusion in disordered media: Statistical mechanisms, models and physical applications, Phys. Rep. {\bf 195}, 127 (1990).

\bibitem{MK2000} R. Metzler and J. Klafter, The random walk's guide to anomalous diffusion: a fractional dynamics approach, Phys. Rep. {\bf 339}, 1 
(2000).

\bibitem{BGL1999} M. Bauer, C. Godr\`eche and J.-M. Luck, Statistics of persistent events in the binomial random walk: Will the drunken sailor hit the sober man?, J. Stat. Phys. {\bf 96}, 963 (1999).

\bibitem{GLS2015}
R. Gout, F. J. Lopez and G. Sanz, Records from stationary observations subject to a random trend, Adv. in Appl. Probab. {\bf 47}, 1175 (2015).

\bibitem{Cha15a}
D. Chalet, {One- and two-sample nonparametric tests for the signal-to-noise ratio based on record statistics}, arXiv:1502.05367, (2015).

\bibitem{Cha15b}
D. Chalet, {Sharper asset ranking from total drawdown durations}, arXiv:1505.01333, (2015).

\bibitem{Cha_url}
D. Challet, \url{https://cran.r-project.org/web/packages/sharpeRratio/index.html}.


\bibitem{Sanjib2011} S. Sabhapandit, Record Statistics of Continuous Time Random Walk, Europhys. Lett. {\bf 94}, 20003 
(2011).

\bibitem{MW1965} E. W. Montroll and G. H. Weiss, Random Walks on Lattices II, J. Math. Phys. {\bf 6}, 
167 (1965).

\bibitem{GMS2014} C. Godr\`eche, S. N. Majumdar and G. Schehr, Universal statistics of longest lasting records of random walks and L\'evy flights, J. Phys. A: Math. Theor. {\bf 47}, 255001 (2014).

\bibitem{GMS2009}
C. Godr{\`e}che, S.~N. Majumdar and G. Schehr, The longest excursion of stochastic processes in nonequilibrium systems, {Phys. Rev. Lett.} {\bf 102}, 240602 (2009). 


\bibitem{GMS2015}
C. Godr\`eche, S. N. Majumdar and G. Schehr, Statistics of the longest interval in renewal processes, J. Stat. Mech. P03014 (2015).

\bibitem{Lam61}
J. P. Lamperti, A contribution to renewal theory, Am. Math. Soc. {\bf 12}(5), 724 (1961). 

\bibitem{CH03}
E. Cs\'aki and Y. Hu, Lengths and heights of random walk excursions, Discrete Math. Theo. Comput. Sci. AC, 45 (2003)


\bibitem{GMS2016}
C. Godr\`eche, S. N. Majumdar, G. Schehr, Exact statistics of record increments of random walks and L\'evy flights, Phys. Rev. Lett. {\bf 117}, 010601 (2016).



\bibitem{MMS13}
S. N. Majumdar, Ph. Mounaix and G. Schehr, Exact statistics of the gap and time interval between the first two maxima of random walks, {Phys. Rev. Lett.} {\bf 111}, 070601 (2013).

\bibitem{MMS14}
S.~N. Majumdar, Ph. Mounaix and G. Schehr, On the Gap and Time Interval between the First Two Maxima of Long Random Walks, J. Stat. Mech. P09013 (2014).

\bibitem{Ivanov94} V. V. Ivanov, Resolvent method: exact solutions of half-space transport problems by elementary means, Astron. Astrophys. {\bf 286}, 328 (1994).



\bibitem{Asm2003}
S. Asmussen, {Applied Probability and Queues}, (Springer, New York, 2003)


\bibitem{GMS2015b}
C. Godr\`eche, S. N. Majumdar and G. Schehr, {Record statistics for random walk bridges}, J. Stat. Mech. P07026 (2015).

\bibitem{Darling}
D. A. Darling, The maximum of sums of stable random variables, Trans. Am. Math. Soc. {\bf 83}, 164 (1956).

\bibitem{GRS2011} R. Garcia-Garcia, A. Rosso and G. Schehr, L\'evy flights on the half line, Phys. Rev. E {\bf 86}, 011101 (2012).


\bibitem{Pollaczek} F. Pollaczek, Fonctions caract\'eristiques de certaines r\'epartitions d\'efinies au moyen de la notion d'ordre, Comptes rendus {\bf 234}, 2334 (1952).

\bibitem{Spitzer} F. Spitzer, The Wiener-Hopf equation whose kernel is a probability density, Duke Math. J. {\bf 24}, 327 (1957).

\bibitem{CM2005} A. Comtet and S.N. Majumdar, Precise Asymptotics for a Random Walker's Maximum, J. Stat. Mech.: Theo. Exp. {P06013}, (2005).

\bibitem{KMR2011} M. Kwasnicki, J. Malecki and M. Ryznar, Suprema of L\'evy processes, Ann. Probab. {\bf 41}, 2047 (2013).


\bibitem{MCZ2006} S. N. Majumdar, A. Comtet and R. M. Ziff, Unified Solution of the Expected Maximum of a Random Walk and the Discrete Flux to a Spherical Trap, J. Stat. Phys. {\bf 122}, 833 (2006).

\bibitem{ZMC2007} R. M. Ziff, S. N. Majumdar and A. Comtet, General flux to a trap in one and three dimensions, J. Phys. C: Cond. Matter {\bf 19}, 065102 (2007).

\bibitem{ZMC2009} R. M. Ziff, S. N. Majumdar and A. Comtet, Capture of particles undergoing discrete random walks., J. Chem. Phys. {\bf 130}, 204104 (2009).

\bibitem{SM2012}
G. Schehr and S.~N. Majumdar, Universal Order Statistics of Random Walks, Phys. Rev. Lett. {\bf 108}, 040601 (2012).

\bibitem{MSM16} Ph. Mounaix, G. Schehr and S.~N. Majumdar, On the Gap and Time Interval between the First Two Maxima of Long Continuous Time Random Walks, J. Stat. Mech. P013303 (2016).

\bibitem{MS16} Ph. Mounaix and G. Schehr, First Gap Statistics of Long Random Walks with Bounded Jumps, arXiv:1609.03202. 

\bibitem{BBDG99}
A. Baldassarri, J.P. Bouchaud, I. Dornic and C. Godr\`eche, Generalized persistence exponents: an exactly soluble model, Phys. Rev. E {\bf 59}, R20 (1999).

\bibitem{blinking} X. Brokmann, J.-P.~Hermier, G. Messin, P. Desbiolles, J.-P. Bouchaud and M.~Dahan, Statistical Aging and Non Ergodicity in the Fluorescence of Single Nanocrystals, {Phys. Rev. Lett.} {\bf 90}, 120601 (2003); G. Margolin and E. Barkai, Aging Correlation Functions for Blinking Nano-Crystals, and Other On -- Off Stochastic Processes, {J. Chem. Phys.} {\bf 121}, 1566 (2004);
F.~D.~Stefani, J.-P.~Hoogenboom and E. Barkai, Beyond quantum jumps: Blinking nano-
scale light emitters, {Phys. Today} {\bf 62}, 34 (2009).

\bibitem{MBCS1996}
S. N. Majumdar, A. J. Bray, S. J. Cornell and C. Sire, Nontrivial Exponent for Simple Diffusion, Phys. Rev. Lett. {\bf 77}, 2867 (1996).

\bibitem{DHZ1996} 
B. Derrida, V. Hakim and R. Zeitak, Persistent spins in the linear diffusion approximation of phase ordering and zeros of stationary gaussian processes, Phys. Rev. Lett. {\bf 77}, 2871 (1996).

\bibitem{derrida94} B. Derrida, A. J. Bray and C. Godr{\`e}che, Non-trivial exponents in the zero temperature dynamics of the 1D Ising and Potts models, {J. Phys. A} {\bf 27}, L357 (1994).

\bibitem{bray94} A. J. Bray, B. Derrida and C. Godr{\`e}che, Non-Trivial Algebraic Decay in a Soluble Model of Coarsening, {Europhys. Lett.} {\bf 27}, 175 (1994).

\bibitem{AS2015}
F. Aurzada and T. Simon, Persistence probabilities and exponents in {L\'evy matters V}, p. 183?221, Lecture Notes in Math., 2149, Springer, (2015).

\bibitem{GRS2010}
R. Garc{\'i}a-Garc{\'i}a, A. Rosso and G. Schehr, The longest excursion of fractional Brownian motion: numerical evidence of non-Markovian effects, 
{Phys. Rev. E} {\bf 81}, 010102(R) (2010).

\bibitem{Mol1999}
G. M. Molchan, Maximum of a Fractional Brownian Motion: Probabilities of Small Values, Comm. Math. Phys. {\bf 205}, 97 (1999). 

\bibitem{KKMCBS1997}
J. Krug, H. Kallabis, S.~N. Majumdar, S.~J. Cornell, A. J. Bray and C. Sire, Persistence exponents for fluctuating interfaces, Phys. Rev. E {\bf 56}, 2702 (1997).

\bibitem{Sch95}
C. L. Scheffer, The rank of the present excursion, {Stoch. Proc. Appl.} {\bf 55}, 101 (1995). 

\bibitem{Kin75}
J. F. C. Kingman, Random discrete distributions, J. Roy. Statistic. Soc. Ser. B {\bf 37}, 1 (1975).

\bibitem{Ver1986}
A. M. Vershik, {The asymptotic distribution of factorizations of natural numbers into prime divisors}, Soviet. Math. Dokl. {\bf 34}, 57 (1986).

\bibitem{VS1977}
A. M. Vershik and A. Schmidt, {Limit measures arising in the theory of groups, I}, Theory Probab. Appl. {\bf 22}, 79 (1977).

\bibitem{Fer93}
T. Ferguson, {A Bayesian analysis of some nonparametric problems}, Ann. Statist. {\bf 1}, 209 (1973).

\bibitem{Ewe88}
W. Ewens, {Population genetics theory -- the past and the future}, in {Mathematical and Statistical Problems in Evolution} (S. Lessard, ed.). Univ. Montreal Press.

\bibitem{Pitman_StFlour}
J. W. Pitman, {Combinatorial stochastic processes}, Ecole d'\'et\'e de probabilit\'es de Saint-Flour, Lecture Notes Math., Springer (2002).

\bibitem{Fen10}
S. Feng, {The Poisson-Dirichlet Distribution and Related Topics -- Models and Asymptotic Behaviors}, Springer Science \& Business Media, (2010).

\bibitem{SV2016}
R. Szab{\'o}, B. Vet{\"o}, {Ages of records in random walks}, J. Stat. Phys. {\bf 165}, 1086 ((2016).



\bibitem{Wen64}
J. Wendel, Zero-free intervals of semi-stable Markov processes, Math. Scand. {\bf 14}, 21 (1964).

\bibitem{CGtd} C. Godr\`eche, {Longest interval between zeros of the tied-down random walk, the Brownian bridge and related renewal processes}, arXiv:1611.01434 (2016).

\bibitem{BMSM2016}
A. Bar, S. N. Majumdar, G. Schehr and D. Mukamel, Exact extreme value statistics at mixed order transitions, Phys. Rev. E {\bf 93}, 052130 (2016).

\bibitem{BBP96}
N. Balakrishnan, K. Balasubramanian and S. Panchapakesan, $\delta$-exceedance records, J. Appl. Stat. Sci. {\bf 4}, 123 (1997).

\bibitem{GLS12}
R. Gouet, F. J. Lopez and G. Sanz, Central Limit Theorem for the Number of Near-Records, Comm. Stat. Their. Methods {\bf 41}, 309 (2012).

\bibitem{edery}
Y. Edery, A. B. Kostinski, S. N. Majumdar and B. Berkowitz, Record-breaking statistics for random walks in the presence of measurement error and noise, {Phys. Rev. Lett.} {\bf 110}, 180602, (2013).





\bibitem{PK16}
S. C. Park, J. Krug, $\delta$-exceedance records and random adaptive walks, J. Phys. A: Math. Theor. {\bf 49}, 315601 (2016).

\bibitem{BPS05}
N. Balakrishnan, A. Pakes, A. Stepanov, On the number and sum of near-record observations, Adv. Appl. Probab. {\bf 37}, 765 (2005).

\bibitem{BK2013}
E.~Ben-Naim, P.~L.~Krapivsky, Statistics of Superior Records, Phys. Rev. E {\bf 88}, 022145 (2013). 

\bibitem{BK2014}
E.~Ben-Naim, P.~L.~Krapivsky, Persistence of Random Walk Records, J. Phys. A {\bf 47}, 255002 (2014).

\bibitem{KR1996}
P. L. Krapivsky, S. Redner, Life and Death in a Cage and at the Edge of a Cliff, Am. J. Phys. {\bf 64}, 548 (1996).

\bibitem{BKL2015}
E. Ben-Naim, P. L. Krapivsky and N. W. Lemons, Scaling Exponents for Ordered Maxima, Phys. Rev. E {\bf 92}, 062139 (2015). 

\bibitem{BK2014b}
E.~Ben-Naim, P.~L.~Krapivsky, Slow Kinetics of Brownian Maxima, Phys. Rev. Lett. {\bf 113}, 030604 (2014). 

\bibitem{MBN13}
P.~W. Miller and E. Ben-Naim, Scaling Exponent for Incremental Records, {J. Stat. Mech.} P10025 (2013).

\bibitem{Tur2015}
L. Turban, Records for the number of distinct sites visited by a random walk on the fully-connected lattice, J. Phys. A {\bf 48}, 445001 (2015). 

\bibitem{Bur14}
T. W. Burkhardt, {First Passage of a Randomly Accelerated Particle} in {First-Passage Phenomena and Their Applications}, edited by R. Metzler, G. Oshanin and S. Redner (World Scientific, 2014), arXiv:1603.07017. 








\end{thebibliography}
\end{document}